%% file: main_loco.tex
\documentclass[a4paper,aps,preprintnumbers,showpacs,twocolumn,superscriptaddress,nofootinbib,amsmath,amssymb]{revtex4-1}

\usepackage{graphicx}
\usepackage{amssymb}
\usepackage{bm}
\usepackage{color}
\usepackage{cancel}
\usepackage{soul}

\input kigou.tex

\newcommand{\GA}{\alpha}
\newcommand{\GB}{\beta}
\newcommand{\GG}{\gamma}
\newcommand{\GD}{\delta}
\newcommand{\GE}{\epsilon}

\newcommand{\GL}{\lambda}
\newcommand{\GR}{\rho}

\newcommand{\GC}{\psi}

\newcommand{\GO}{\omega}

\newcommand{\GP}{\phi}

\newcommand{\GU}{\theta}

\newcommand{\pd}{\partial}

\newcommand{\be}{\begin{equation}}
\newcommand{\ee}{\end{equation}}

\usepackage[]{amsmath}
\usepackage[]{amsfonts}
\usepackage[]{amssymb}
\usepackage[]{mathrsfs}
\usepackage{times}
\usepackage{hyperref}
\hypersetup{
  colorlinks=true,        
  linkcolor=blue,         
  citecolor=cyan,         
}
\newcommand{\cf}{cf.,~}
\newcommand{\ie}{i.e.,~}

\begin{document}

\title{Initial-data contribution to the error budget of gravitational waves
  from neutron-star binaries}


\author{Antonios Tsokaros}
\affiliation{Institute for Theoretical Physics, Max-von-Laue-Strasse 1,
  60438 Frankfurt am Main, Germany}
\author{Bruno C. Mundim}
\affiliation{Institute for Theoretical Physics, Max-von-Laue-Strasse 1,
  60438 Frankfurt am Main, Germany}
\author{Filippo Galeazzi}
\affiliation{Institute for Theoretical Physics, Max-von-Laue-Strasse 1,
  60438 Frankfurt am Main, Germany}

\author{Luciano Rezzolla} 
\affiliation{Institute for Theoretical Physics, Max-von-Laue-Strasse 1,
  60438 Frankfurt am Main, Germany} 
\affiliation{Frankfurt Institute for Advanced Studies,
  Ruth-Moufang-Str. 1, D-60438 Frankfurt am Main, Germany}

\author{K\=oji Ury\=u}
\affiliation{Department of Physics, University of the Ryukyus, Senbaru,
  Nishihara, Okinawa 903-0213, Japan}

\date{\today}

\begin{abstract}
As numerical calculations of inspiralling neutron-star binaries reach
values of accuracy that are comparable with those of binary black holes,
a fine budgeting of the various sources of error becomes increasingly
important. Among such sources, the initial data is normally not accounted
for, the rationale being that the error on the initial spacelike
hypersurface is always far smaller than the one gained during the
evolution. We here consider critically this assumption and perform a
comparative analysis of the gravitational waveforms relative to
essentially the same physical binary configuration when computed with two
different initial-data codes, and then evolved with the same evolution
code. More specifically, we consider the evolution of irrotational
neutron-star binaries computed either with the pseudo-spectral code
\lorene{}, or with the newly developed finite-difference code \cocal{};
both sets of initial data are subsequently evolved with the high-order
evolution code \whiskythc{}. In this way we find that despite the initial
data shows global (local) differences that are $\lesssim 0.02\%\ (1\%)$,
the gravitational-wave phase at the merger time differs by $\sim 0.5$
radians after $\sim 3$ orbits, a surprisingly large value. Our results
highlight the highly nonlinear impact that errors in the initial data can
have on the subsequent evolution and the importance of using exactly the
same initial data when comparative studies are done.
\end{abstract}

\maketitle

\section{Introduction}
\label{sec:intro}

With the first direct detection of gravitational waves from a merging
system of black holes \cite{Abbott2016a}, the long awaited
gravitational-wave astronomy has finally become a reality in which a
series of advanced interferometers such as LIGO, GEO, Virgo, KAGRA, ET
\cite{Abramovici92, Accadia2011_etal, Kuroda2010, Aso:2013, Punturo:2010}
is eagerly operating to unveil that part of the universe that can be
observed in terms of gravitational radiation. Neutron star binary systems
are prime actors of this universe and have received enormous attention
over the last ten years.

In addition, neutron-star binaries are leading candidates for the engine
of short gamma ray bursts \cite{Narayan92, Eichler89, Rezzolla:2011,
  Bartos:2012, Berger2013b}, and possible sites for the production of the
heaviest elements in the universe \cite{Lattimer74, Li:1998, Tanvir2013,
  Berger2013, Tanaka2013, Rosswog2014a, Sekiguchi2015,
  Radice2016}. Starting from the first successful simulations of binary
neutron-star mergers \cite{Shibata99d} and the first complete description
of this process from the inspiral down to the formation of an accreting
black-hole--torus system \cite{Baiotti08}, considerable progress has been
done, so that it is now possible to consider rather realistic scenarios
involving nuclear-physics equations of state \cite{Sekiguchi2011b,
  Takami2015}, neutrino cooling \cite{Sekiguchi2011, Galeazzi2013,
  Sekiguchi2015, Radice2016} and magnetohydrodynamics \cite{Anderson2008,
  Etienne08, Giacomazzo:2009mp, Kiuchi2014, Dionysopoulou2015}.

Obviously, any simulation of neutron-star binaries needs initial data to
get started and this is carefully crafted through standalone codes like
\cocal{} \cite{Uryu:2012b,Tsokaros2015}, \lorene{} \cite{lorene_web},
\kadath{} \cite{Grandclement09}, \scrid{} \cite{Tichy:2009, Tichy12}, or
through the elliptic solvers of the evolution codes like \spec{}
\cite{spec_web}, Princeton's \cite{East2012d}, or \bam{}
\cite{Bruegmann:2006at}. Although the first initial data for
neutron-star binaries has been computed for corotating systems
\cite{Baumgarte98b}, the large majority of the simulations performed to
date has used irrotational configurations, since neutron star viscosity
is believed to be too small to tidally lock the two stars prior to merger
\cite{Kochanek92,Bildsten92}. At the same time, the most advanced efforts
over the last couple of years have been concentrated on approaches to
reduce the eccentricity of the orbits or to produce binary systems with
arbitrary neutron star spins \cite{Kyutoku2014, Moldenhauer2014,
  Bernuzzi2013, Tichy12, Kastaun2013, Tsatsin2013, Tsokaros2015,
  Dietrich:2015b, Tacik15}.

In the past, the \cocal{} code has been used to compute quasi-equilibrium
sequences for binary black holes \cite{Uryu2012,Uryu:2012b,Tsokaros2012},
and a pointwise comparison was made with the spectral code \kadath{},
both for the gravitational fields, as well as for global quantities like
the ADM mass and angular momentum, finding excellent agreement. More
recently, the \cocal{} code has been used to compute quasi-equilibrium
sequences neutron-star binaries that are irrotational or spinning, with
spins aligned with the orbital angular momentum \cite{Tsokaros2015}; also
in this case, the comparison with the \lorene{} code for irrotational
sequences has shown excellent agreement. Overall, both sets of studies
show that when considering close binaries of compact objects, be it black
holes or neutron stars, the use of \cocal{} has led to agreements in the
global quantities to less than $0.03\%$, while for the individual metric
components the differences were less than $1\%$.

In this work we focus on neutron-star binaries and perform a close
comparison with another spectral code, \lorene{}, not only for the data
on the initial slice, but also for its subsequent evolution.  More
specifically, given irrotational binaries of neutron stars produced by
either \lorene{} or \cocal{}, we consider the same physical initial data
in terms of gravitational mass, rest mass, orbital frequency, and evolve
both sets of initial data with the high-order code \whiskythc{}
\cite{Radice2013b, Radice2013c, Radice2015}\footnote{We note that this is
  also the first time that evolutions are carried out using initial data
  of any type produced with the \cocal{} code.}. The evolutions are
performed at a number of resolutions, the highest of which have spacings
of $\GD x =0.1\,M_{\odot} \simeq 150\,{\rm m}$ and represent a major
computational cost, which has been reported before for one binary only in
\cite{Radice2015}, where it is referred to as ``very high''. Across all
simulations, we have monitored in detail the violations of the constraint
equations and we have performed a gravitational-wave analysis with
respect to the phase of the $\ell=m=2$ mode of the Weyl scalar
$\Psi_4$. Although the initial data between \cocal{} and \lorene{} shows
global (local) differences that are $\lesssim 0.02\%\ (1\%)$, or that the
waveforms have only very small differences, and that even the convergence
properties of the gravitational-wave signal are almost identical for the
two sets of initial data, we find that the Richardson-extrapolated phases
differ by an order of magnitude, \ie of about one radian, at the merger
time, after $\sim 3$ orbits. These results highlight therefore the highly
nonlinear impact that errors in the initial data can have on the
subsequent evolution, so that extra care needs to be employed when
computing waveforms of neutron-star binaries spanning tens of
orbits. More importantly, because this is the first time that evolutions
from different initial-data solvers is presented, our results issue an
important warning signal about the importance of using exactly the same
initial data when comparative studies of neutron star binary evolutions,
such as the ones carried out in \cite{Baiotti:2010ka, Read2013}, are
performed.

The plan of the paper is as follows. In Section \ref{sec:equations} we
provide a review of the quasiequilibrium equations and present the
\cocal{} driver to the \cactus{}~\cite{cactus_web} infrastructure, while
in Section \ref{sec:comparison} we describe the techniques developed to
import the initial data produced by \cocal{} in an evolution code,
performing a global and local close comparison of an
irrotational binary as computed with \lorene{} and with \cocal{}. Section
\ref{sec:comp_evol} is instead dedicated to the detailed comparison of
the evolution of the two sets of initial data for the various
configurations considered and to the presentation of the corresponding
convergence properties. Finally, our conclusions are presented in
\ref{sec:conclusions}. As complementary material, we present in Appendix
\ref{sec:comp_corot} a short study for corotating initial data produced
by \cocal{} and \lorene{}, again at $45\,{\rm km}$, mostly as a benchmark
of future arbitrary spinning binaries.

Hereafter, spacetime indices running from $0$ to $3$ will be indicated
with Greek letters, while spatial indices running from $1$ to $3$ with
Latin letters. The metric has signature $(-,+,+,+,)$, and we use a set of
geometric units in which $G=c=M_\odot=1$, unless stated otherwise (we
recall that in these units $1\,M_\odot=4.92674\,{\rm \mu s}=1.477\,{\rm km}$).

\section{Review of the quasiequilibrium equations}
\label{sec:equations}

In this Section we only state the basic definitions and equations that
are solved while we refer to \cite{Tsokaros2015} and references within
for more details. The spacetime metric in a $3+1$ decomposition is
written as 
\begin{equation}
ds^2 = -\GA^2 dt^2 + \GG_{ij}(dx^i+\GB^i dt)(dx^j+\GB^j dt)\,,
\label{eq:4metric}
\end{equation}
where $\GA,\ \GB^i,\ \GG_{ij}$ are, respectively, the lapse function, the
shift vector, and the three-metric on some spacelike slice $\Sigma_t$,
which is taken to be conformally flat
\begin{equation}
\GG_{ij}=\GC^4 \GD_{ij}\,.
\label{eq:3metric}
\end{equation}
Here we use the Cartesian components of the shift. The extrinsic
curvature is defined as $K_{\GA\GB}:=-\frac{1}{2}
\mathcal{L}_{\boldsymbol{n}}\GG_{\GA\GB}$, where
$\mathcal{L}_{\boldsymbol{n}}$ is the Lie derivative along the (timelike)
unit vector normal $\boldsymbol{n}$ to $\Sigma_t$. The assumption of
stationarity, $\pd_t \GG_{\GA\GB}=0$, yields $K_{ij} = \frac{1}{2\GA}
\mathcal{L}_{\boldsymbol{\beta}}\GG_{ij}$, while assuming maximal slicing
the conformally rescaled trace-free part of the extrinsic curvature
becomes
\begin{equation}
\tilde{A}^{ij} = \frac{1}{2\GA} \left(\pd^i\GB^j + \pd^j\GB^i
-\frac{2}{3}\GD^{ij}\pd_k
\GB^k\right)=\frac{1}{2\GA}(\widetilde{\mathbb{L}}\GB)^{ij} \,.
\label{eq:Lijcf}
\end{equation}
Note that $\tilde{A}_i^{\ j}=A_i^{\ j}$. The last term in
Eq. \eqref{eq:Lijcf} is the longitudinal operator and the tilde symbol
denotes the fact that it is related to the conformally flat geometry.

With the help of Eq. (\ref{eq:Lijcf}), the constraint equations and the
spatial trace of the time derivative of the extrinsic curvature (assuming
$\pd_t K=0$), result in five elliptic equations for the conformal factor
$\GC$, the shift $\GB^i$, and the lapse function $\GA$
\begin{eqnarray}
&& \nabla^2 \GC  = -\frac{\GC^5}{32\GA^2}(\widetilde{\mathbb{L}}\GB)^{ab}(\widetilde{\mathbb{L}}\GB)^{ij}\GD_{ia}\GD_{jb} 
                    - 2\pi E\GC^5 \label{eq:hamcon}\,, \\
&& \nabla^2 (\GA\GC)  = \frac{7\GC^5}{32\GA}(\widetilde{\mathbb{L}}\GB)^{ab}(\widetilde{\mathbb{L}}\GB)^{ij}\GD_{ia}\GD_{jb} 
                        + 2\pi\GA\GC^5(E+2S)\,,  \label{eq:trdotkij} \\
&& \nabla^2 \GB^i  = -\frac{1}{3}\pd^i\pd_j \GB^j + \pd_j\ln\left(\frac{\GA}{\GC^6}\right)(\widetilde{\mathbb{L}}\GB)^{ij} 
                    + 16\pi\GA\GC^4 j^i\,,  \qquad\label{eq:momcon}
\end{eqnarray} 
where the matter sources are $E:=n_\GA n_\GB T^{\GA\GB}$,
$S:=\GG_{\GA\GB} T^{\GA\GB}$, and $j^i := -\GG^i{}_\GA n_\GB
T^{\GA\GB}$. The boundary conditions for the equations above are dictated
by asymptotic flatness, \ie $\lim_{r\rightarrow\infty} \GC = 1$,
$\lim_{r\rightarrow\infty} \GA = 1$, and $\lim_{r\rightarrow\infty} \GB^i
= 0$.

For the stress-energy tensor we assume a perfect fluid with
\begin{equation}
T_{\GA\GB}=(\GE+p)u_\GA u_\GB + pg_{\GA\GB}=\GR h u_\GA u_\GB +
pg_{\GA\GB}\,, 
\label{eq:set}
\end{equation}
where $u^\GA$ is the four-velocity of the fluid and $\GR, \GE, h$, and
$p$ are, respectively, the rest-mass density, the total energy density,
the specific enthalpy, and the pressure as measured in the rest frame of
the fluid (see \cite{Rezzolla_book:2013} for details). The specific
internal energy $e$ is related to the enthalpy through
$h:=({\GE+p})/{\GR}=1+e+{p}/{\GR}$. The 4-velocity is decomposed as
$u^\GA=u^t(t^\GA+v^\GA)$ or $u^\GA=u^t(k^\GA+V^\GA)$ which correspond to
an inertial frame or the corotating frame decomposition, respectively. 
For the fluid variables we assume helical symmetry,
\begin{equation}
\mathcal{L}_{\mathbf{k}}(hu_\GA)=\mathcal{L}_{\mathbf{k}}\GR=0 \,,
\end{equation}
where
\begin{equation} 
k^\mu := t^\mu + \Omega\GP^\mu\,,
\label{eq:hkv}
\end{equation} 
is the helical Killing vector, and, without loss of generality,
\begin{equation}
\label{eq:phii}
\GP^i=(-y,x,0)\,,
\end{equation}
is the rotational generator. For corotating binaries, $V^\GA=0$, and the
Euler equation results into a first integral
\begin{equation}
\frac{h}{u^t}=C,\quad\mbox{with}\quad u^t = \frac{1}{\sqrt{\GA^2 - \GO_i
    \GO^i}}\,.
\label{eq:corot} 
\end{equation}
where $\GO^i:=\GB^i+\Omega \GP^i$ is the corotating shift. For
irrotational binaries $hu_\GA = \nabla_\GA \Phi$, $\Phi$ being the fluid
velocity potential, and the first integral of the Euler equation is
\begin{equation}
\frac{h}{u^t} + V^j D_j\Phi = C,\quad
h=\sqrt{\GL^2/\GA^2 - D_i\Phi D^i\Phi}\,,
\label{eq:irrot}
\end{equation}
with $\GL:=C+\GO^i D_i\Phi$. The fluid potential $\Phi$ is determined from
conservation of rest mass, $\nabla_\GA(\GR u^\GA)=0$, which yields
\begin{eqnarray}
\nabla^2\Phi & = & -\frac{2}{\GC}\pd_i\GC\pd^i\Phi + \GC^4 \GO^i\pd_i(hu^t)  \nonumber \\
             & \phantom{=} &+   [\GC^4hu^t \GO^i-\pd^i\Phi]\pd_i\ln\left(\frac{\GA\GR}{h}\right) \,,
\label{eq:cbm4}
\end{eqnarray}
with boundary condition on the star surface
\begin{equation}
\left[\left(\GC^4hu^t \GO^i-\pd^i\Phi\right)
       \pd_i\GR \right]_{{\rm surf.}} =0\,. 
\label{eq:bcphi}
\end{equation}
This condition is derived either from Eq.~\eqref{eq:cbm4} assuming that
the baryon density vanishes on the stellar surface, or by demanding that
the fluid velocity is tangent to the stellar surface in the corotating
frame $[V^\mu\nabla_\mu\GR]_{\rm surf.} =0$. Equations
(\ref{eq:hamcon})--(\ref{eq:momcon}) will be solved together with
(\ref{eq:corot}) for corotating motion or (\ref{eq:irrot}) and
(\ref{eq:cbm4}) for irrotational motion, and the two involving constants
$\Omega,\ C$ will be determined in the process. Details about the methods
we use in \cocal{} to solve these equations are described in
\cite{Tsokaros2015}.

\begin{figure*}
  \includegraphics[scale=0.7]{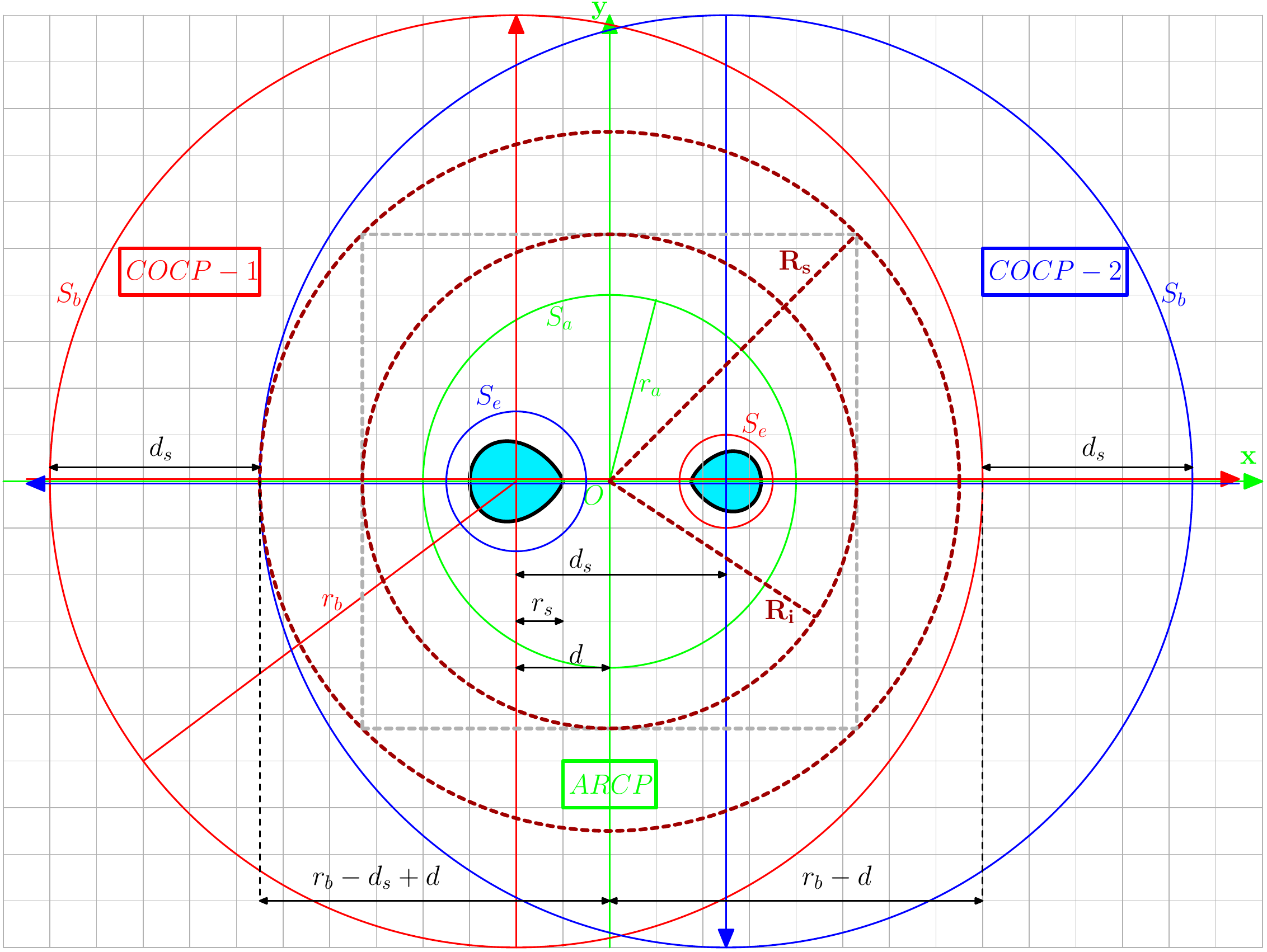}
   \caption{Structure of a two-dimensional cross-section of the \cocal{}
     grids (colored spherical coordinates) overlaid on a Cartesian
     coordinate system used for the evolution of the initial data. Here
     we assume that the $z=0$ plane of the evolutionary grid coincides
     with the corresponding \cocal{} plane. Evolution gridpoints
     $P(x_p,y_p,z_p)$ inside the sphere of radius $R_i$ are interpolated
     from the coordinate patches COCP-1 or COCP-2 depending whether
     $x_p\leq 0$ or $x_p>0$. Points outside that sphere are interpolated
     from the asymptotic region patch ARCP. Note that the figure is not
     in scale; in particular, the size of the sphere of radius $R_i$ is
     much larger than the size of the inner boundary $S_a$ of ARCP. The
     outer boundary of ARCP is not shown here and extends to very large
     values when compared to the compact-object sizes. Typical values are
     $r_b=100$, $d_s=2.5$, $d=1.25$, $R_s=98.75$, $R_i=69.125$,
     $r_a=5.0$. The point where the neutron-star's surface intersects the
     positive $x$-axis of the COCP takes values $r_s\leq 1$ and is in
     general different for the two stars.}
	\label{fig:coctocac}
\end{figure*}

\section{Initial data import and comparison}
\label{sec:comparison}

\cocal{} uses finite differences on spherical coordinates to compute the
various field variables. Importing the initial data into an evolution
code involves interpolating from the \cocal{} grid to the one used by the
evolution code, which in most cases is in Cartesian coordinates. In this
Section we describe the \coctocac{} driver, which interpolates the
\cocal{} grid variables to the \etoolkit{}~\cite{loeffler_2011_et,
  EinsteinToolkit:web}. The full description of the coordinate systems
used by \cocal{} can be found in \cite{Uryu2012} for black hole binaries
or \cite{Tsokaros2015} for neutron-star binaries. Here, we review the most
salient features that will be necessary for the \coctocac{} driver.

\subsection{The \coctocac{} driver}
\label{sec:coctocac}

As customary in a $3+1$ decomposition, the spacetime manifold
$\mathcal{M}=\mathbb{R} \times \Sigma_t$, is foliated by a family of
spacelike hypersurface $\Sigma_t$, parametrized by $t\in {\mathbb
  R}$. These hypersurfaces may represent data that is stationary (in
equilibrium), or quasi-stationary (in quasi-equilibrium) and they are
covered by overlapping multiple spherical coordinate patches. In
Fig.~\ref{fig:coctocac}, three such coordinate systems are used to cover
the hypersurface. One can think of Fig.~\ref{fig:coctocac} as the
equatorial plane of a neutron-star binary system. Two spherical
coordinate patches are used to cover the area around each neutron
star. They are called COCP-1 (from compact object coordinate patch) and
COCP-2 and are plotted with red and blue colors, respectively. COCP-1
(COCP-2) include all points inside the outer red (blue) sphere $S_b$ of
radius $r_b$\footnote{Note that the outer radii $r_b$, of COCP-1 and
  COCP-2 need not be equal, but in most cases we make such a choice.},
but outside the red (blue) excised sphere $S_e$. Note that these two
systems have opposite $(x,y)$ coordinates, but the same $z$
orientation. The reason for introducing the excised sphere $S_e$,
\cite{Tsokaros2007}, is to be able to resolve the second compact object
with reasonable resources. Without it, the size of the companion neutron
star has to be resolved by angular grids, while by using this concept, it
is enough to resolve the size of $S_e$, which is $\sim d_s/2$. This
implies that the angle to be resolved is $\sim 2 \arcsin 1/2 = \pi/3$. As
a rule of thumb, the angular resolution of a COCP is determined from the
degree of accuracy to resolve the deformation of the neutron stars
centered at the patch, and to resolve the size of their excised
sphere. The third patch, called the asymptotic-region coordinate patch or
ARCP, is denoted by green lines and includes all points outside the
sphere $S_a$ and infinity, typically a sphere $S_b$ not shown here at
very large distance from the center of mass $O$.

\begin{table}
\begin{tabular}{rll}
\hline
\hline
$r_a$: & Radial coordinate where the radial grids start. For      \\
\phantom{:}& the COCP patch it is $r_a=0$. \\
$r_b$: & Radial coordinate where the radial grids end. \\
$r_c$: & Center of mass point. Excised sphere is located   \\
\phantom{:}& at $2r_c$ in the COCP patch. \\
$r_e$: & Radius of the excised sphere. Only in the COCP patch. \\
$r_s$: & Radius of the sphere bounding the star's surface. \\
\phantom{:}& It is $r_s\leq 1$. Only in COCP. \\
$N_{r}$: & Number of intervals $\Dl r_i$ in $r \in[r_a,r_{b}]$. \\
$N_{r}^{1}$: & Number of intervals $\Dl r_i$ in $r \in[0,1]$. Only \\
\phantom{:}& in the COCP patch. \\
$\Nrf$: & Number of intervals $\Dl r_i$ in $r \in[0,r_s]$ in the COCP patch \\
\phantom{:}& or $r \in[r_a,r_a+1]$ in the ARCP patch. \\
$\Nrm$: & Number of intervals $\Dl r_i$ in $r \in[r_a,r_{c}]$. \\
$N_{\theta}$: & Number of intervals $\Dl \theta_j$ in $\theta\in[0,\pi]$. \\
$N_{\phi}$: & Number of intervals $\Dl \phi_k$ in $\phi\in[0,2\pi]$. \\
$d$: & Coordinate distance between the center of $S_a$ ($r=0$) \\
\phantom{:}& and the center of mass. \\
$d_s$: & Coordinate distance between the center of $S_a$ ($r=0$) \\
\phantom{:}& and the center of $S_e$. \\
$L$: & Order of included multipoles. \\
\hline
\hline
\end{tabular}  
\caption{Summary of grid the parameters used for the binary systems
  computed here.}
\label{tab:grid_param}
\end{table}

\begin{table*}
\begin{tabular}{cl|ccccccccccccc}
\hline
\hline
Type & Patch  & $\ r_a\ $ & $\ r_s\ $ & $\ r_b\ $ & $\ r_c\ $ & $\ r_e\ $ & 
$\ \Nrf\ $ & $\ N_r^1\ $ & $\ \Nrm\ $ & $\ N_r\ $ & $\ N_\theta\ $ & $\ N_\phi\ $ & $\ L\ $  \\
\hline
$\texttt{Hs2.0d}$ & ${\rm COCP-1}$ & $0.0$ & $0.7597667$ & $10^2$ & $1.25$ & $1.125$ 
               & $50$  & $64$ & $80$ & $192$ & $48$  & $48$ & $12$  \\
               & ${\rm COCP-2}$ & $0.0$ & $0.7597667$ & $10^2$ & $1.25$ & $1.125$ 
							 & $50$  & $64$ & $80$ & $192$ & $48$  & $48$ & $12$ \\
               & ${\rm ARCP}$   & $5.0$ & $-$         & $10^6$ & $6.25$ & $-$     
							 & $16$  & $-$  & $20$ & $192$ & $48$  & $48$ & $12$ \\
\hline
$\texttt{Hs2.5d}$ & ${\rm COCP-1}$ & $0.0$ & $0.7597667$ & $10^2$ & $1.25$ & $1.125$ 
               & $76$  & $96$ & $120$ & $288$ & $72$  & $72$ & $12$  \\ 
               & ${\rm COCP-2}$ & $0.0$ & $0.7597667$ & $10^2$ & $1.25$ & $1.125$ 
							 & $76$  & $96$ & $120$ & $288$ & $72$  & $72$ & $12$  \\
               & ${\rm ARCP}$   & $5.0$ & $-$         & $10^6$ & $6.25$ & $-$     
							 & $24$  & $-$  & $30$ & $192$ & $72$  & $72$ & $12$  \\
\hline
$\texttt{Hs3.0d}$ & ${\rm COCP-1}$ & $0.0$ & $0.7597667$ & $10^2$ & $1.25$ & $1.125$ 
               & $100$  & $128$ & $160$ & $384$ & $96$  & $96$ & $12$ \\ 
               & ${\rm COCP-2}$ & $0.0$ & $0.7597667$ & $10^2$ & $1.25$ & $1.125$ 
							 & $100$  & $128$ & $160$ & $384$ & $96$  & $96$ & $12$  \\
               & ${\rm ARCP}$   & $5.0$ & $-$         & $10^6$ & $6.25$ & $-$     
							 & $32$  & $-$  & $40$ & $384$ & $96$  & $96$ & $12$  \\
\hline
$\texttt{Hs3.5d}$ & ${\rm COCP-1}$ & $0.0$ & $0.7597667$ & $10^2$ & $1.25$ & $1.125$ 
               & $150$  & $192$ & $240$ & $576$ & $144$  & $144$ & $12$  \\
               & ${\rm COCP-2}$ & $0.0$ & $0.7597667$ & $10^2$ & $1.25$ & $1.125$ 
							 & $150$  & $192$ & $240$ & $576$ & $144$  & $144$ & $12$  \\
               & ${\rm ARCP}$   & $5.0$ & $-$         & $10^6$ & $6.25$ & $-$     
							 & $48$  & $-$  & $60$ & $384$ & $144$  & $144$ & $12$ \\
\hline
\hline
\end{tabular}
\caption{Four different grid structures parameters used for the circular
  binary computation in \cocal{}. All variables are explained in
  Table~\ref{tab:grid_param} and the distances are in normalized
  quantities. The \coctocac{} driver interpolates from COCP-1,2 when the
  normalized distance of the point under consideration from the center of
  mass is less than $R_i=69.125$, while from ARCP for larger
  values.}
\label{tab:cocgrids45}
\end{table*}

\begin{table*}
\begin{tabular}{l|cccccccc}
\hline
\hline
code  & $M_0$ & $M_{\rm ADM}$ & $\MK$ & $\GR_{c}\times 10^{-4}$ &
$J_{\rm ADM}$ & $\ \Omega\,[{\rm rad/sec}]\ $ & $\ d_s\,[{\rm km}]\ $ & $\ R_{\rm eq}\,[{\rm km}]\ $ \\
\hline
\cocal{} \texttt{Hs2.0d} & $\ 1.62504\ $ & $\ 2.99737\ $ & $\ 2.99716\ $ & $\ 9.563899\ $ & $\ 8.79553\ $ & 
$\ 1856.75\ $ & $\ 44.735\ $ & $\ 13.595$ \\
\hline
\cocal{} \texttt{Hs2.5d} & $\ 1.62505\ $ & $\ 2.99733\ $ & $\ 2.99718\ $ & $\ 9.577718\ $ &
$\ 8.81018\ $ & $1857.29$ & $44.722$ & $13.591$ \\
\hline
\cocal{} \texttt{Hs3.0d} & $\ 1.62505\ $ & $\ 2.99817\ $ & $\ 2.99804\ $ & $\ 9.582239\ $ &
$\ 8.82099\ $ & $1857.42$ & $44.718$ & $13.590$  \\
\hline
\cocal{} \texttt{Hs3.5d} & $\ 1.62505\ $ & $\ 2.99822\ $ & $\ 2.99811\ $ & $\ 9.585707\ $ &
$\ 8.82549\ $  & $1857.48$ & $44.715$ & $13.589$  \\
\hline
\lorene{}  & $\ 1.62504\ $ & $\ 2.99834\ $ & - & $\ 9.569626\ $ & $\ 8.81879\ $ & 
$\ 1867.49\ $ & $\ 44.707\ $ & $\ 13.605\ $  \\
\hline
\hline
\end{tabular}
\caption{Physical parameters of the irrotational binaries at the various
  resolutions of Table~\ref{tab:grid_param}. The columns denote the rest
  mass of each star, the ADM mass of the binary, the Komar mass, the
  central rest mass density, the ADM angular momentum in units of
  $G=c=M_\odot=1$, while the angular velocity, the separation and the
  equatorial radius are in physical units. The separation changes
  slightly with resolution as a result of iteration procedure followed by
  \cocal{}. Similar quantities are reported for the solution computed by
  \lorene{}. The ADM mass of a spherical solution that corresponds to a
  rest mass $M_0=1.62505$ is $M_{\rm ADM}=1.51481$ and the compactness is
  $\mathcal{C}:=M/R=0.1401$. }
\label{tab:loco45}
\end{table*}

The values of the radii $r_a$, $r_b$, and $r_e$ that correspond to
spheres $S_a$, $S_b$, $S_e$ for each of the coordinate patches used are
set as follows. For the case of ARCP, the radius $r_a$ of the inner
boundary $S_a$ is taken large enough to be placed outside of the excised
spheres $S_e$ for each COCP, but small compared to the radius $r_b$ of
the outer boundary $S_b$ for each COCP. Typically, for a neutron star
with a mass $M$, $r_b =\mathcal{O}(100 M)$, and $r_e = \mathcal{O}(M)$
for COCP, while $r_a =\mathcal{O}(10M)$, and $r_b =\mathcal{O}(10^6 M)$
or larger for ARCP. At present, although no compactification of the ARCP
is done, no obvious problem related to our results has been detected.

Another important feature used in \cocal{}, which is relevant for
importing correctly the initial data to an evolution code, is the
normalization of all its quantities. This is discussed in detail in
Section IIIB of \cite{Tsokaros2015}, but let us mention the most import
facts. In particular, we rescale the spatial coordinates $x^i$ as
\begin{equation}
\hat{x}^i:=\frac{x^i}{R_0} \,.
\end{equation}
We do this in order to stabilize the root-finding method for the
eigenvalues $C,\ \Omega$, the constant of the Euler integral, and the
angular velocity of the compact object, as well as for controlling the
star surface. For single rotating neutron stars \cite{Huang08,Uryu2016a},
the rescaling factor $R_0$ is chosen so that the coordinate equatorial
radius of the star is unity (stated differently, the radius of the star
along the positive $x$-axis is $R_0$). For neutron-star binaries
\cite{Tsokaros2015}, the scaling factor is chosen in such a way that the
coordinate equatorial radius of the star has a fixed value $r_s\leq 1$
(stated differently, the radius of the star along the positive $x$-axis
is $r_s R_0$). In typical evolution codes, such as the one employed here,
the units are also $G=c=M_\odot=1$, so that for an arbitrary point
$(x,y,z)_{\rm cac}$, the correspondent \cocal{} point is
\begin{equation}
(x,y,z)_{\rm cac} \longrightarrow (x,y,z)_{\rm coc}=
\left(\frac{x_{\rm cac}}{R_0},\frac{y_{\rm cac}}{R_0},\frac{z_{\rm cac}}{R_0}\right)\,,
\label{eq:xyz_coctocac}
\end{equation}
and similar care has to be paid when one taking derivatives as, for
example, in the extrinsic curvature, \ie
\begin{equation}
(K_{ij})_{\rm cac} = \frac{(K_{ij})_{\rm coc}}{R_0}  \,.
\label{eq:kij_coctocac}
\end{equation} 
For simplicity, hereafter we will assume that one has taken into account
the normalizing factor $R_0$ when translating points and variables from
an evolution code to \cocal{}, and we will describe only the choice that
has to be made regarding the coordinate systems.

Figure~\ref{fig:coctocac} shows a with light gray color the $z=0$ plane
of a Cartesian grid used by an evolution code, as well as the three
spherical coordinate systems that are typically used by \cocal{}. The
hypersurface $\Sigma_t$ where a solution is provided by \cocal{} has the
same $z=0$ plane with the evolutionary Cartesian grid whose origin is
also identified by the ``center of mass'' $O$ of \cocal{}. In other
words, the asymptotic patch, ARCP, of \cocal{} has the same origin as the
evolutionary Cartesian grid, and the $z=0$ plane is the same for all
grids. The problem is then to interpolate for each Cartesian gridpoint,
$P(x_p,y_p,z_p)$, from the nearby \cocal{} spherical points. We note that
$(x_p,y_p,z_p)$ are also the coordinates of $P$ with respect to ARCP. To
perform such an interpolation, a choice has to be made regarding the
position of $P$ relative to the \cocal{} coordinate systems. Since all
distances are measured with respect to $O$, the general rule of thumb is
that if the distance $r_p=\sqrt{x_p^2+y_p^2+z_p^2}$ is large enough, then
the interpolation will be performed in the ARCP. Otherwise for points
close to $O$ the interpolation will be done either from COCP-1 or
COCP-2. Inside the COCPs (spheres $S_b$ in Fig.~\ref{fig:coctocac})
points are not uniformly distributed and, in addition, there are
``holes'', \ie regions devoid of coordinate points, which are the regions
inside the spheres labeled as $S_e$. One simple solution is to consider
the $x_p$ coordinate of $P$. If $x_p\leq 0$ then we perform a fourth
order Lagrange interpolation from nearby COCP-1 points, otherwise from
COCP-2.

As a more concrete example of the procedure followed in the driver, we
can adopt the same notation as in \cite{Uryu2012, Tsokaros2015} and
denote by $d_s$ the distance between the two stars (\ie between the
geometric centers of the two stars). We also denote by $d$ the distance
from the center of mass of the system to the geometric center of the star
on the negative $x$-axis of ARCP. Without loss of generality, we then
assume that the heavier star is on the negative $x$-axis, so that
$d_s\geq 2d$, and that the radii $r_b$ of COCP-1 and COCP-2 are the same
(we can always make such a choice). As a result, the outermost point of
COCP-2 along the negative ARCP $x$-axis is at a distance $r_b-d_s+d$ from
$O$, while the outermost point of COCP-1 along the positive ARCP $x$-axis
is at a distance $r_b-d$. Let therefore
\begin{equation}
R_s:=\min\{r_b-d_s+d,r_b-d\} = r_b-d_s+d \,,
\label{eq:Rc}
\end{equation}
and consider the cube centered at $O$ with each face having a length
$2R_i$, $R_i:=R_s/\sqrt{2}$. In practice we take $R_i=0.7 R_s$. 
Then, for each Cartesian
point $P$, if $r_p\geq R_i$, we interpolate from ARCP, otherwise we
examine the sign of $x_c$. For $x_c\leq 0$ and $r_p<R_i$, we interpolate
from COCP-1, while we interpolate from COCP-2 otherwise. Notice also that
in a region with $x_c\leq 0$, COCP-1 is denser than COCP-2, so that the
interpolations will be more accurate. The contrary is true for
$x_p>0$. Typical values for the relevant quantities are $r_b=100$,
$d_s=2.5=2d$, which means that $R_s=98.75$ while $R_i=69.125$. As a
concluding remark, we note that Fig.~\ref{fig:coctocac} is not
in scale. For example, the inner boundary of ARCP (green sphere $S_a$)
has radius $r_a=5.0$, so that, in reality, there is quite a large space
between that and the sphere of radius $R_i$, while in the figure they
appear quite close.

\subsection{Local and global comparison of initial data 
from \lorene{} and \cocal{}}
\label{sec:comp_irrot}

In this Section we carefully compare the initial data produced by two
different codes, namely, \cocal{} and \lorene{}, which use completely
different numerical methods for the solution of the constraint
equations. In order to do so, we compute the solutions for the physically
same irrotational binary having the same gravitational (rest) mass and
where the two stars are at a distance of approximately $44.7\,{\rm
  km}$. The reason we use the adverb ``approximately'' is because the two
codes obtain the final solutions in rather different ways. On the one
hand, \lorene{} allows one to set up explicitly the masses of the binary
and the distance between the two stars, and an iteration is then carried
out until a circular solution is obtained at the desired accuracy. In
\cocal{}, on the other hand, distances are expressed in terms of the
normalizing factor $R_0$, which is only found at the end of the
computation.

Details of the logical flow followed by \cocal{} can be found in
\cite{Tsokaros2015} Section III-B, with the relevant radii summarised in
Table~\ref{tab:grid_param}. Note that $r_s$ is the radius that
corresponds to the inner point of the neutron-star's surface closer to
the center of mass, and $d_s$ the coordinate distance between the two
stars. The physical lengths, though, are $r_s R_0$ and $d_s R_0$, so that
as one sets the coordinate distance $d_s$ and the star radius $r_s$,
\cocal{} computes binaries whose separation is expressed in terms of the
star's radius. When a converged solution is obtained, the code finds the
value of $R_0$ (as well as of $\Omega$ and the constant of the Euler
integral $C$) and can then compute the physical separation in ${\rm km}$
of the binary. As the resolution changes, $R_0$ also changes slightly,
with the consequence that the distance $d_s$ between the two stars
changes too. Of course, this change is only very small and we can safely
assume that the binary systems are at the same separation. In the future
we plan to address this issue by changing $r_s$ and employing a
root-finding method to arrive exactly at the requested distance between
the two stars.

\begin{figure}
\begin{center}
\includegraphics[width=0.45\columnwidth]{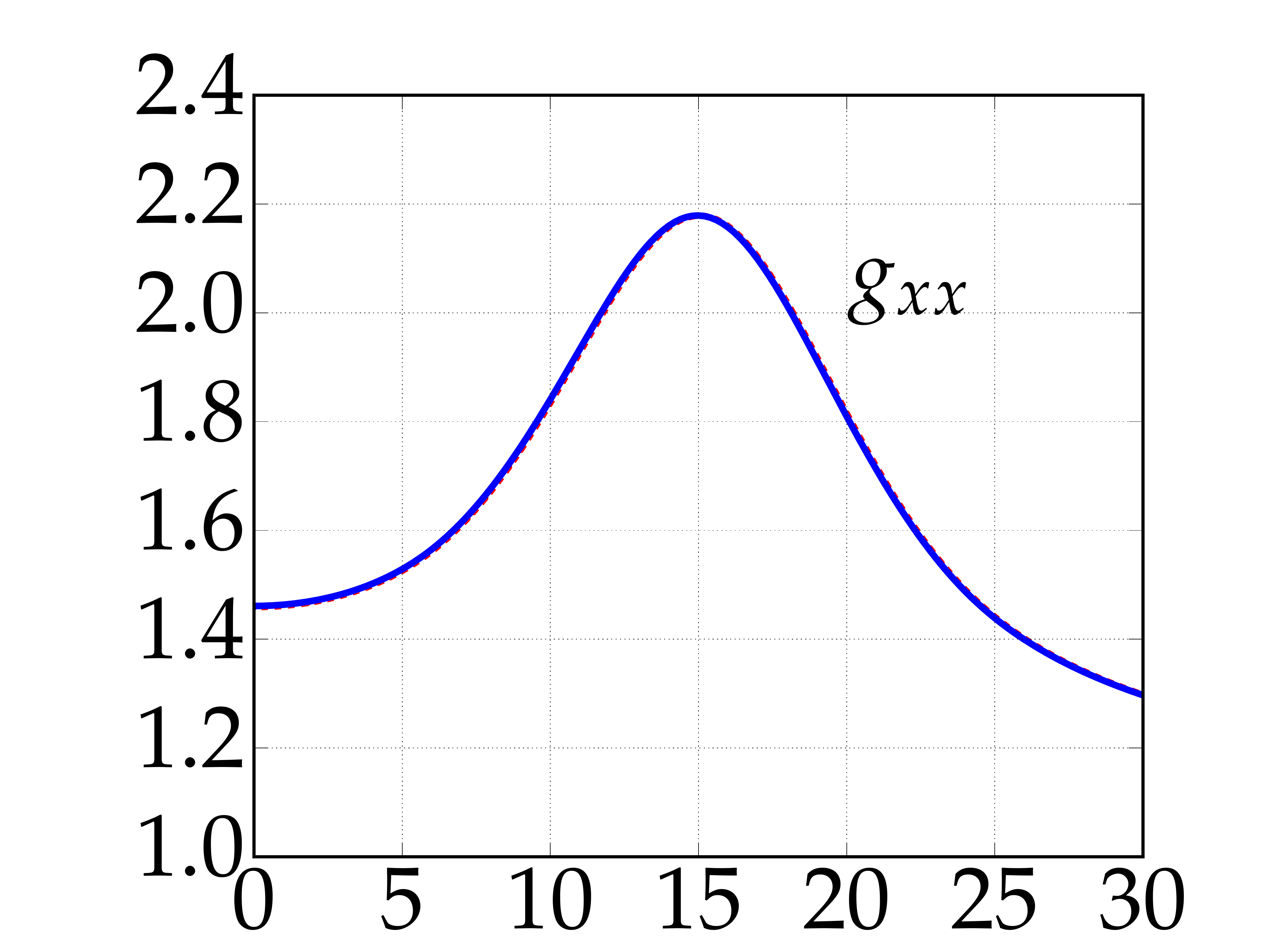}
\includegraphics[width=0.45\columnwidth]{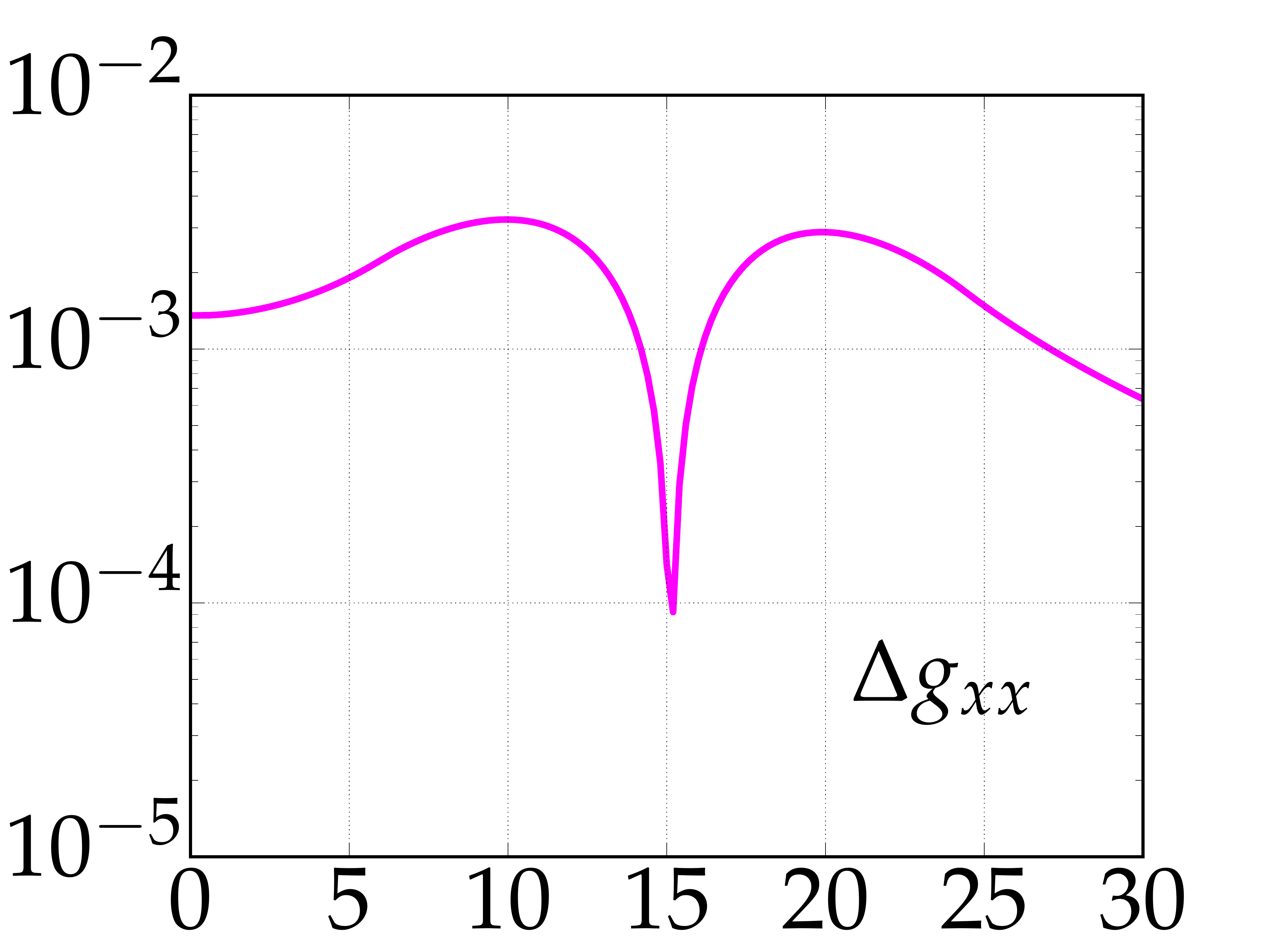}
\end{center}
\begin{center}
\includegraphics[width=0.45\columnwidth]{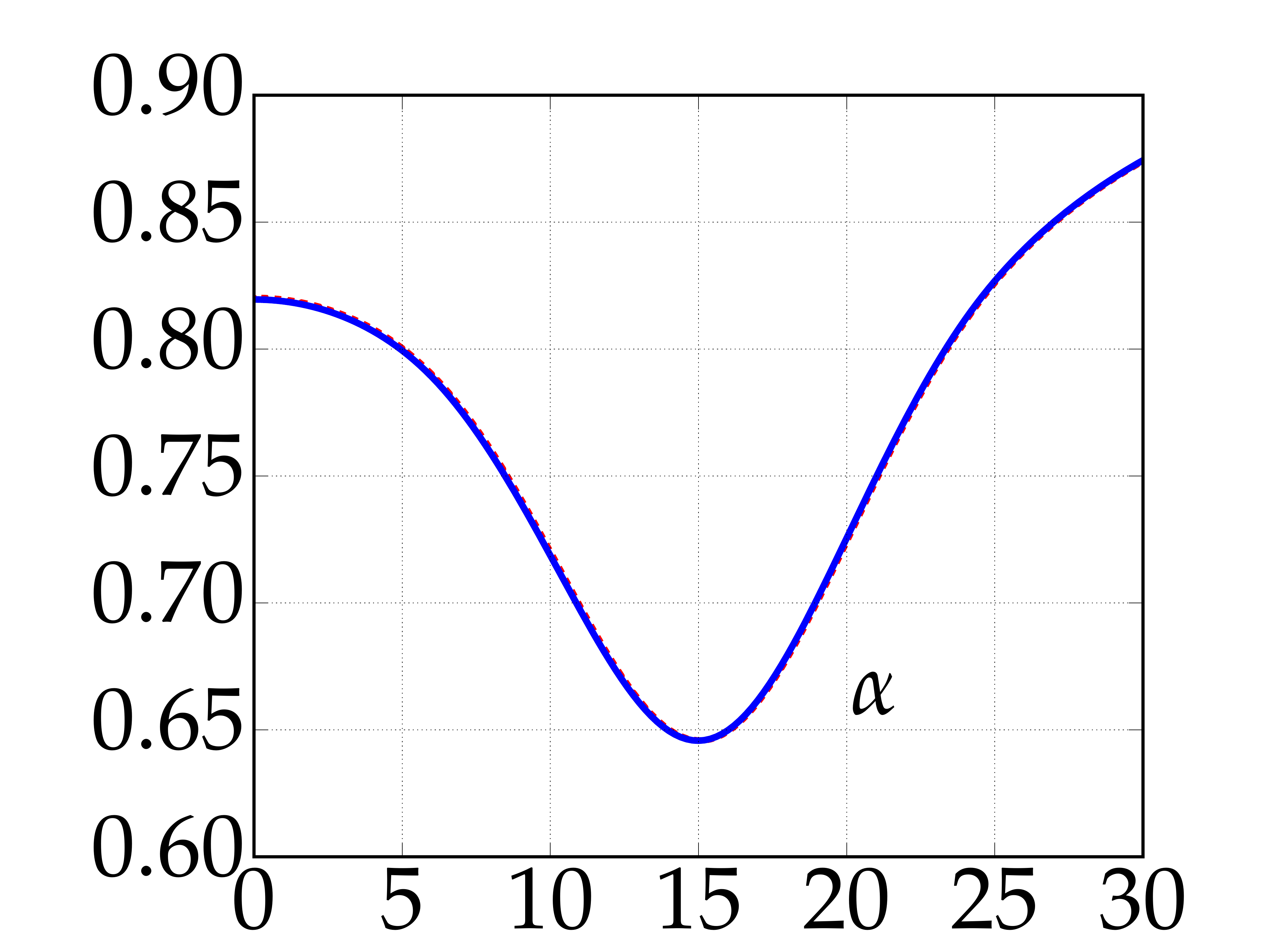}
\includegraphics[width=0.45\columnwidth]{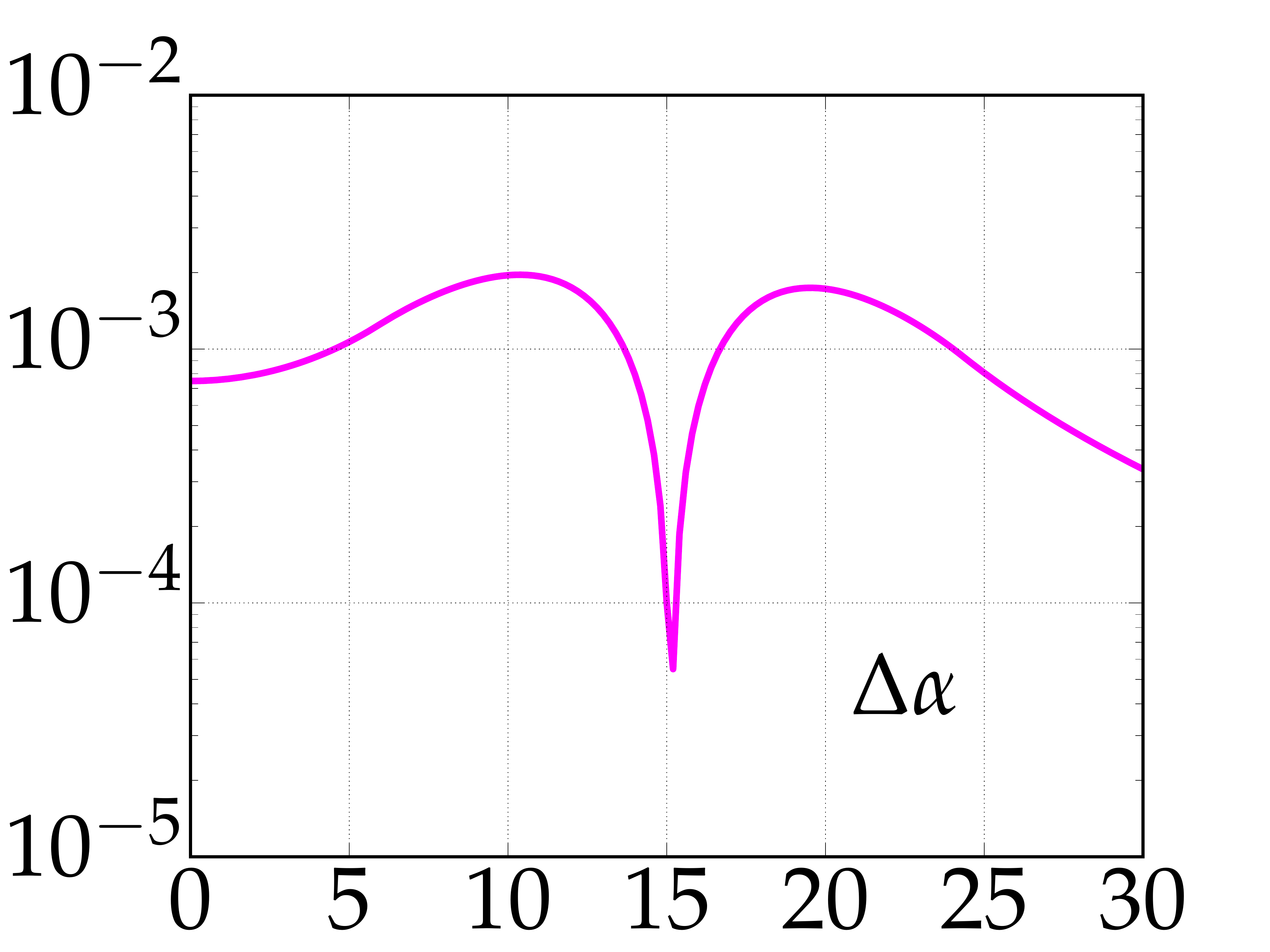}
\end{center}
\begin{center}
\includegraphics[width=0.45\columnwidth]{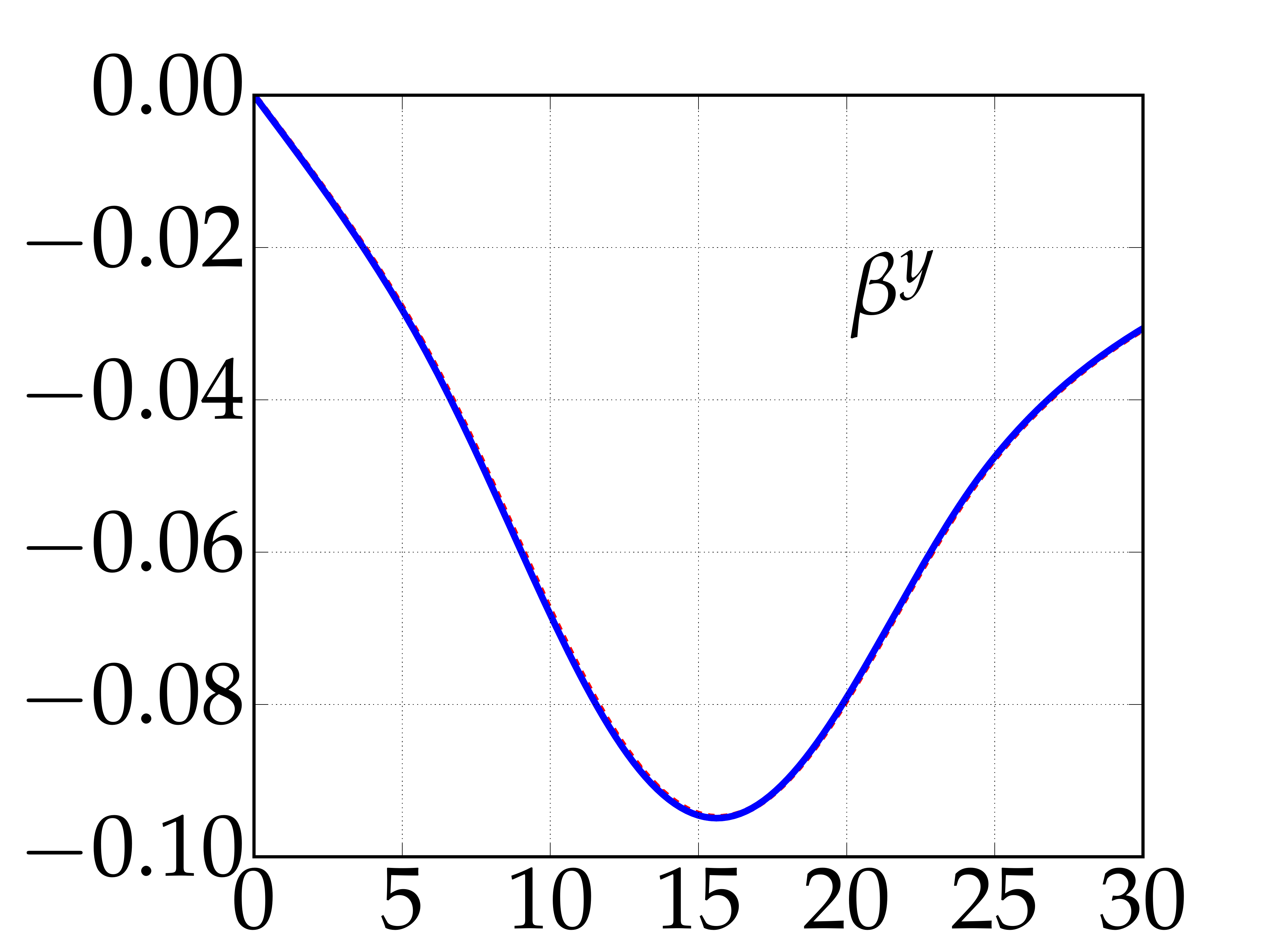}
\includegraphics[width=0.45\columnwidth]{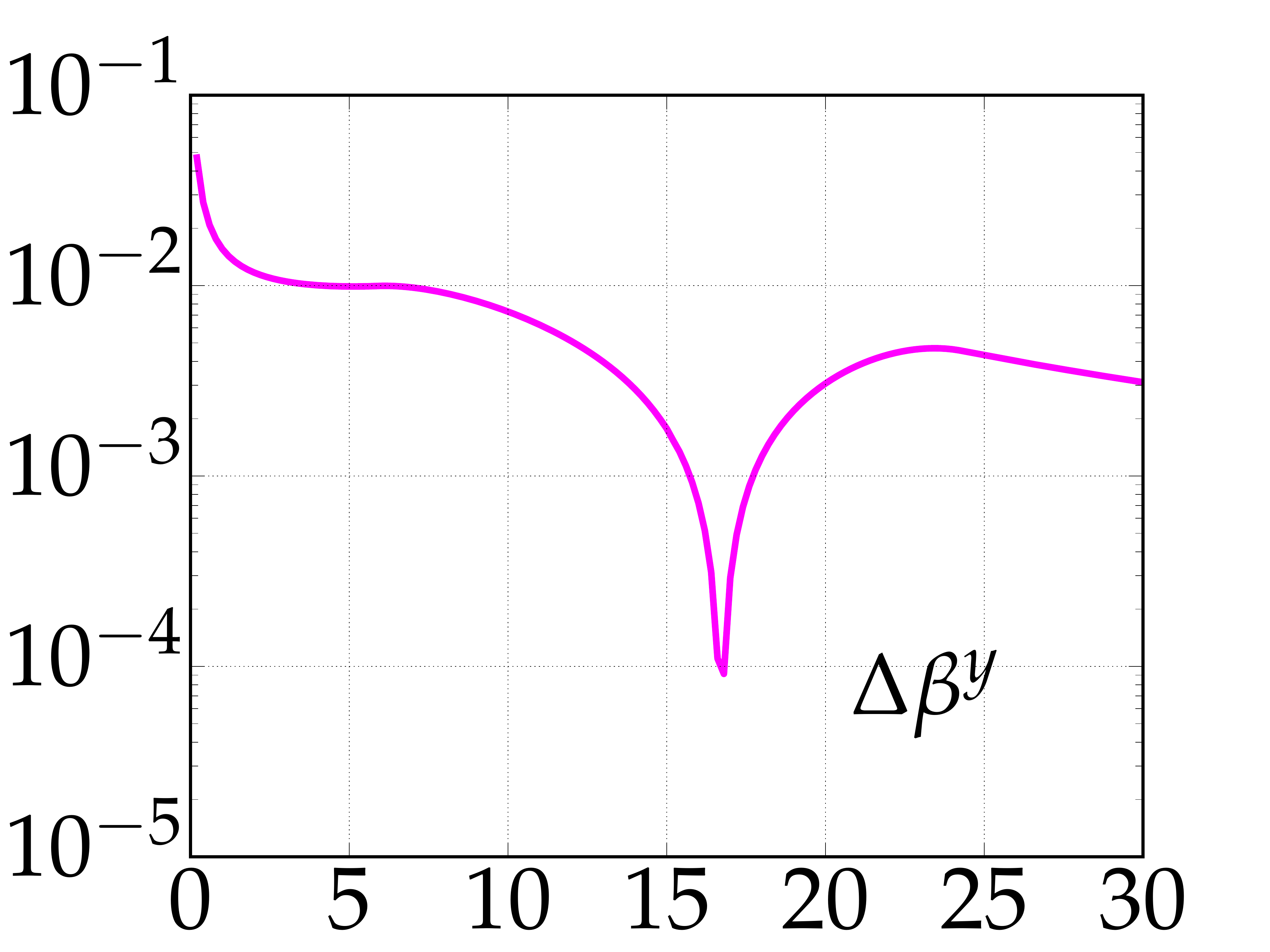}
\end{center}
\begin{center}
\includegraphics[width=0.45\columnwidth]{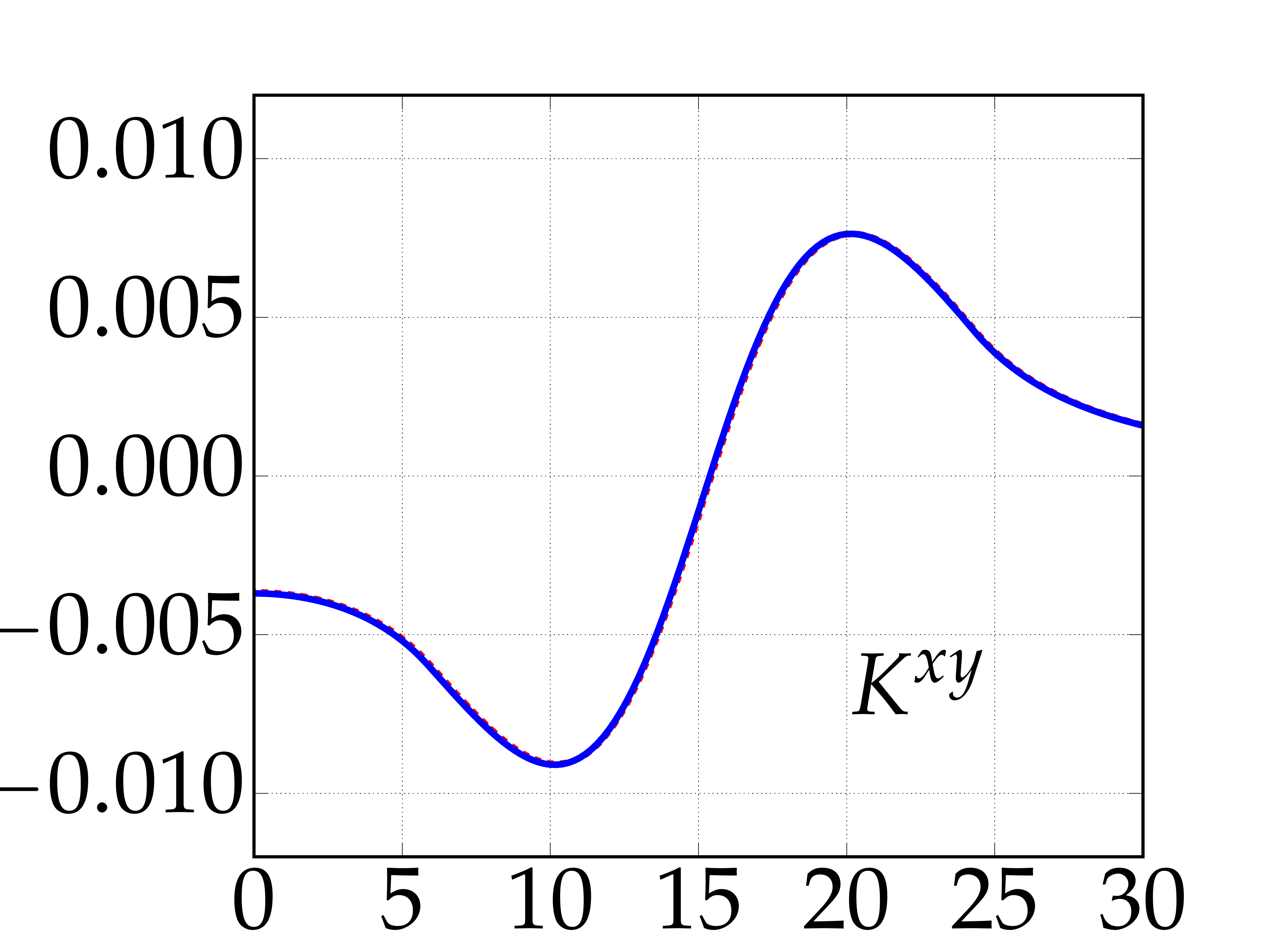}
\includegraphics[width=0.45\columnwidth]{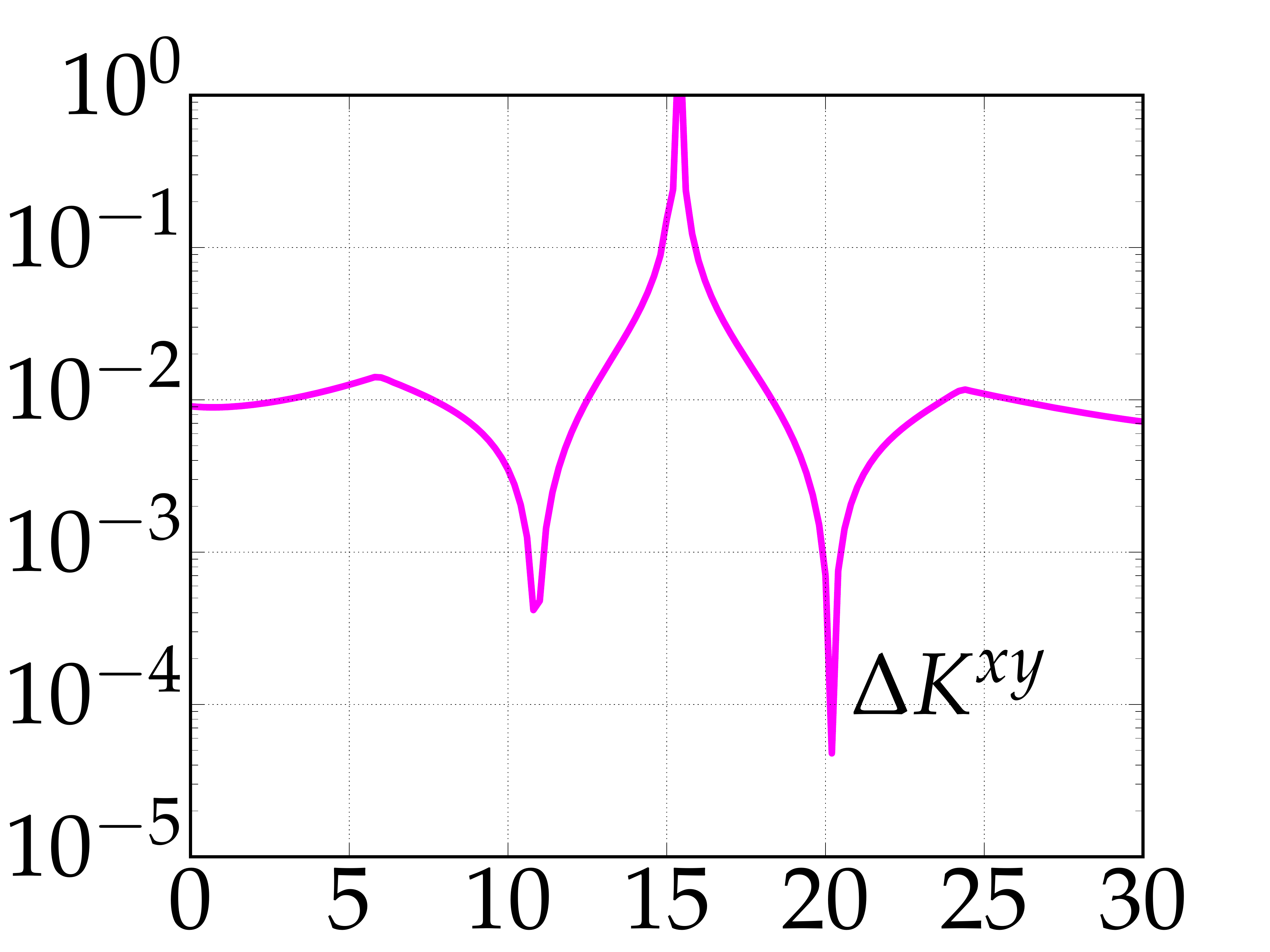}
\end{center}
\begin{center}
\includegraphics[width=0.45\columnwidth]{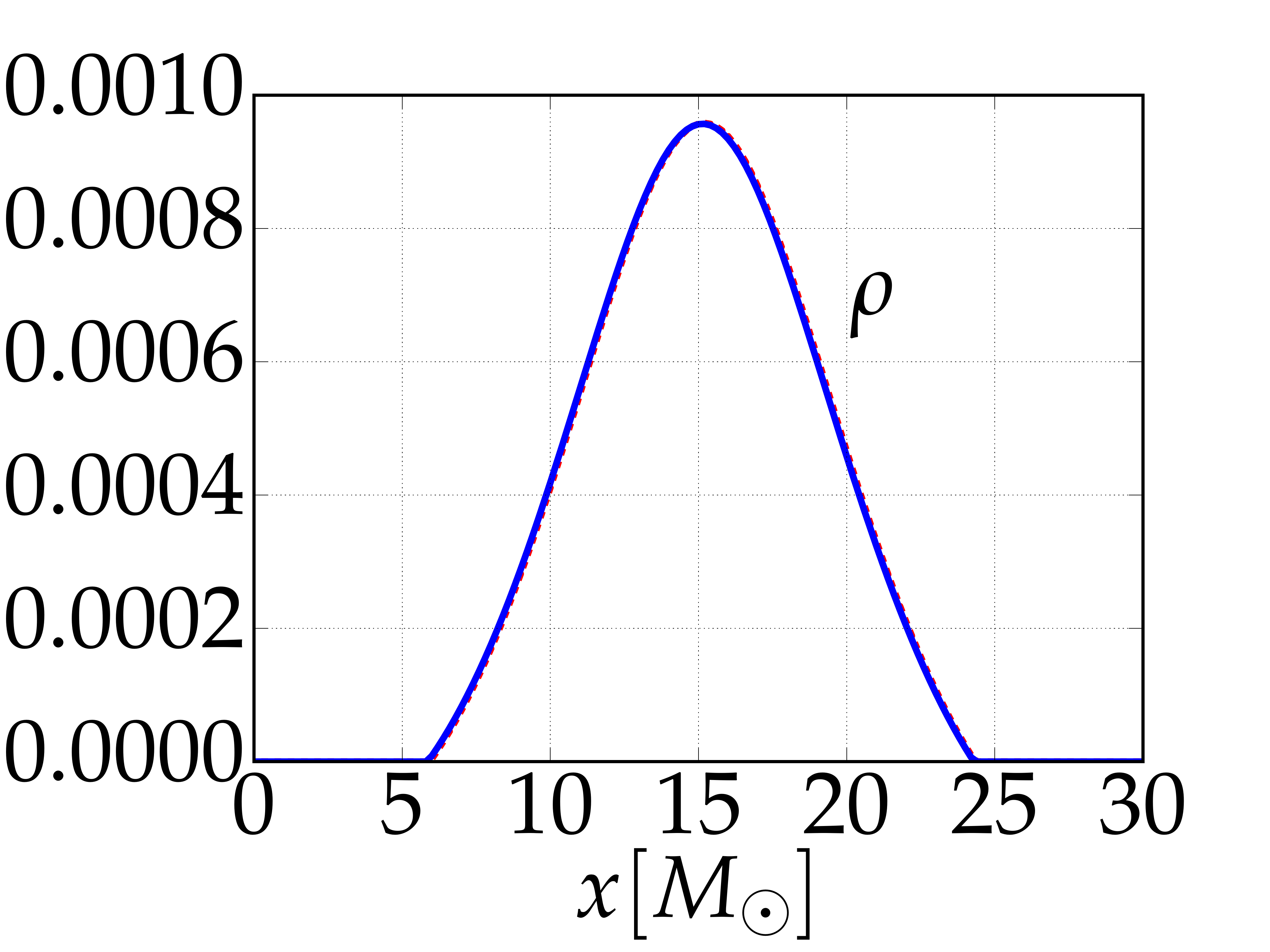}
\includegraphics[width=0.45\columnwidth]{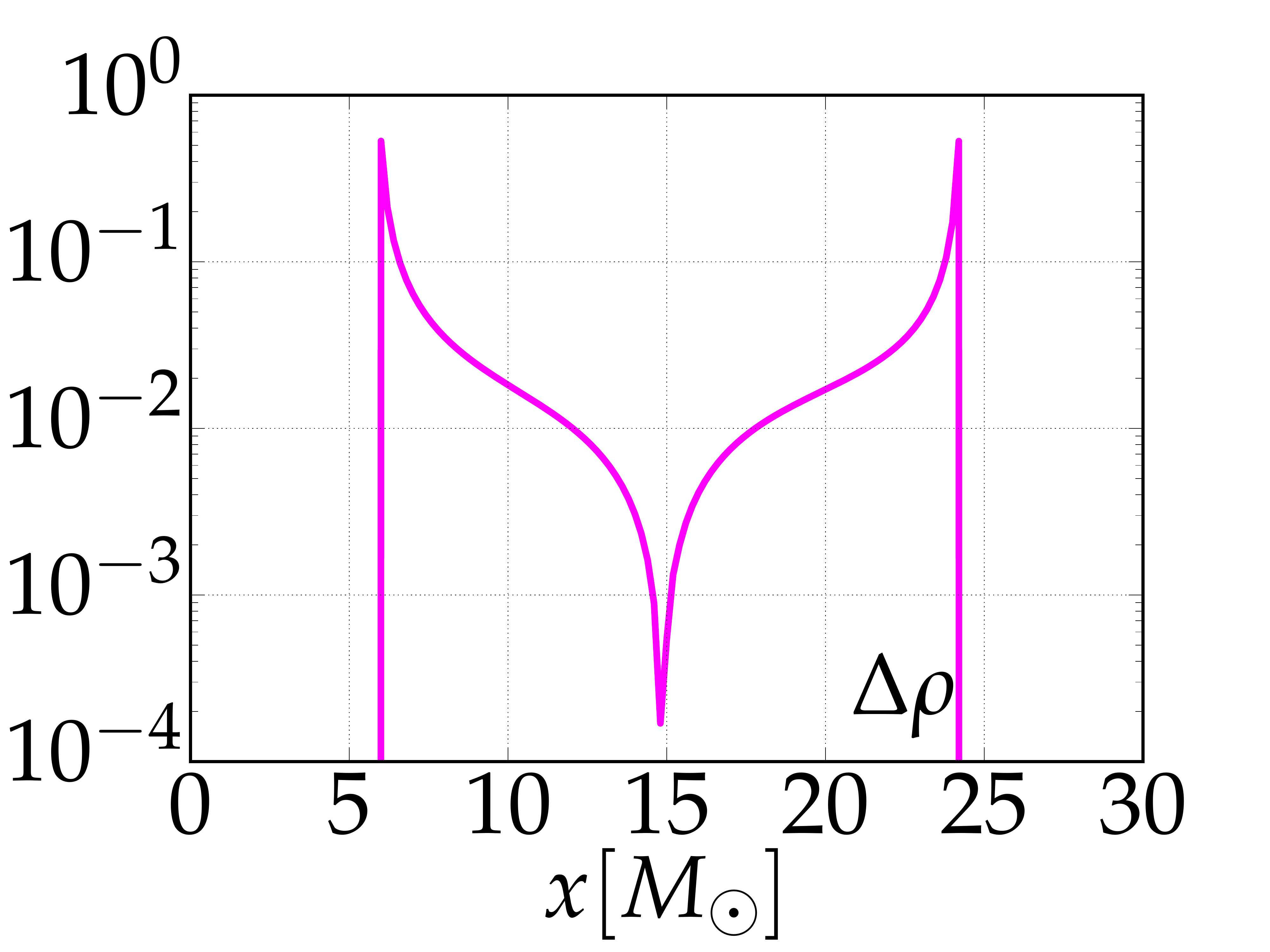}
\end{center}
\caption{Left column: from top to bottom, initial data quantities
  relative to the metric function $g_{xx}=\GC^4$, the lapse function
  $\alpha$, the $y$-component of the shift, the $xy$-component of the
  extrinsic curvature, and the rest mass density $\rho$, as computed by
  \cocal{} (red lines) and \lorene{} (blue lines). The $x$-axis is the 
  positive $x$-axis of the Cartesian grid  with $x=0$
  corresponding to the center of mass of the binary. Right column:
  relative difference between \cocal{} and \lorene{} as computed from
  Eq.~\eqref{eq:reldif}.}
\label{fig:irrot_var1}
\end{figure}

\begin{figure}
\begin{center}
\includegraphics[width=0.48\columnwidth]{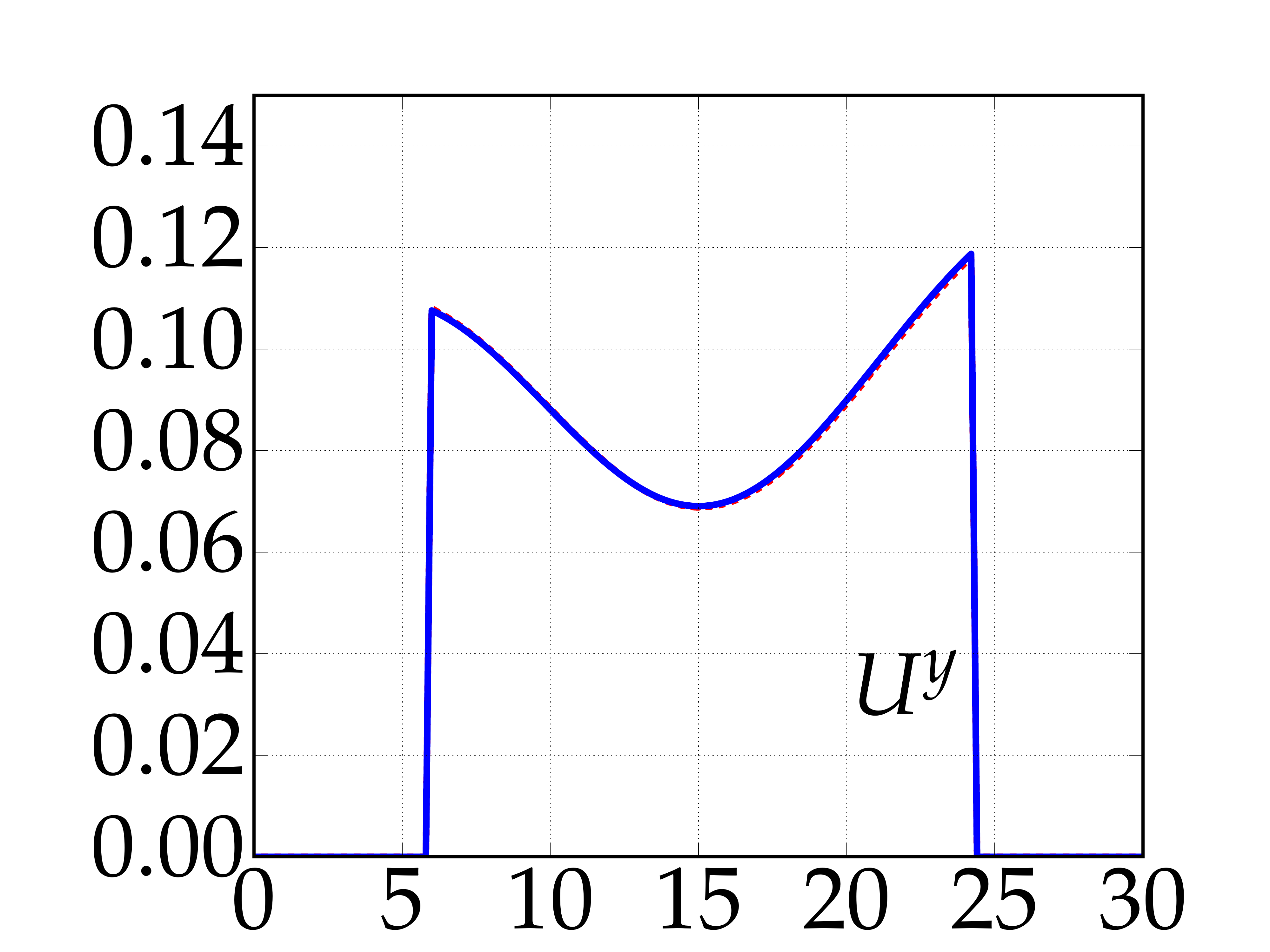}
\includegraphics[width=0.48\columnwidth]{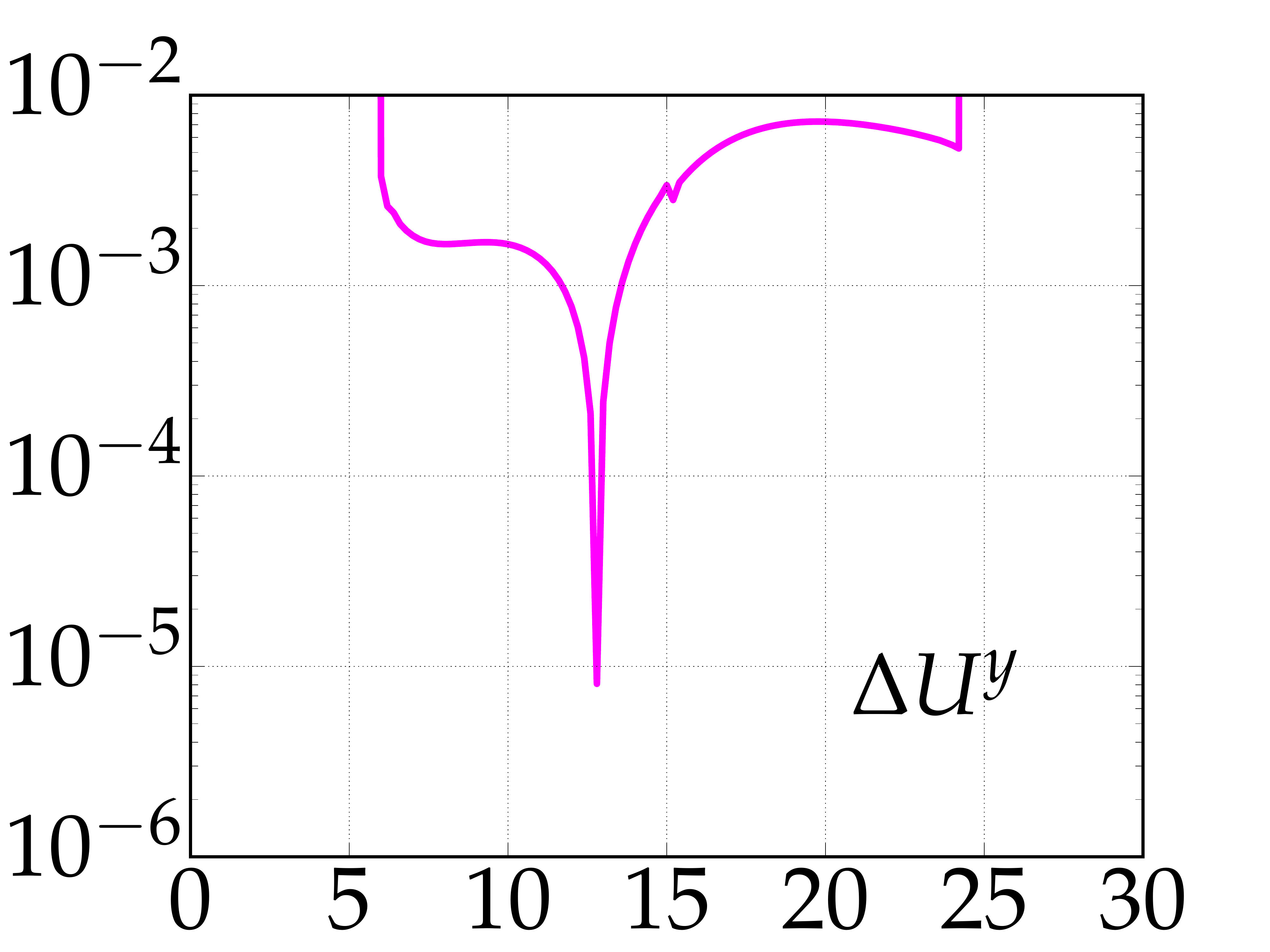}
\end{center}
\begin{center}
\includegraphics[width=0.48\columnwidth]{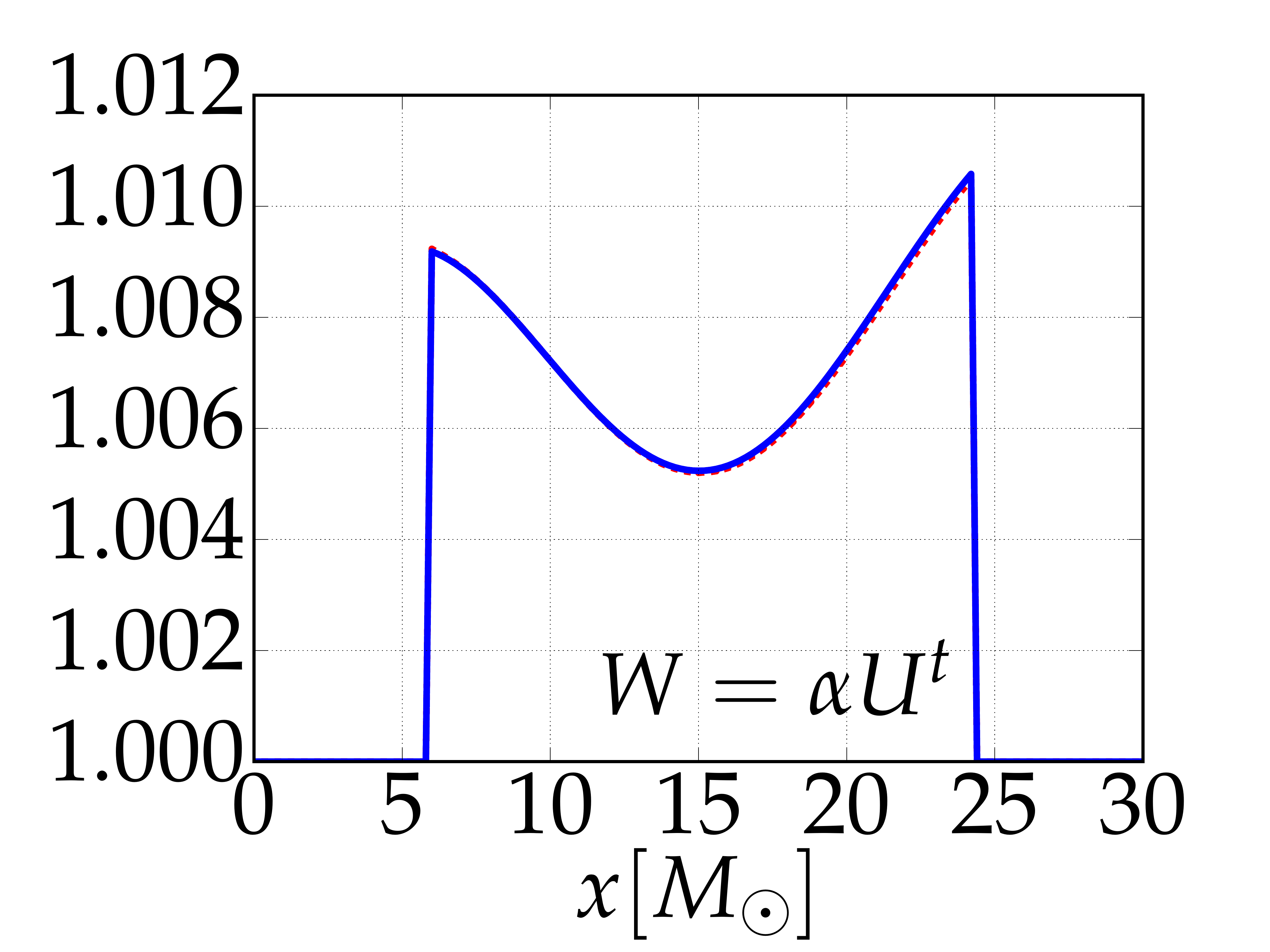}
\includegraphics[width=0.48\columnwidth]{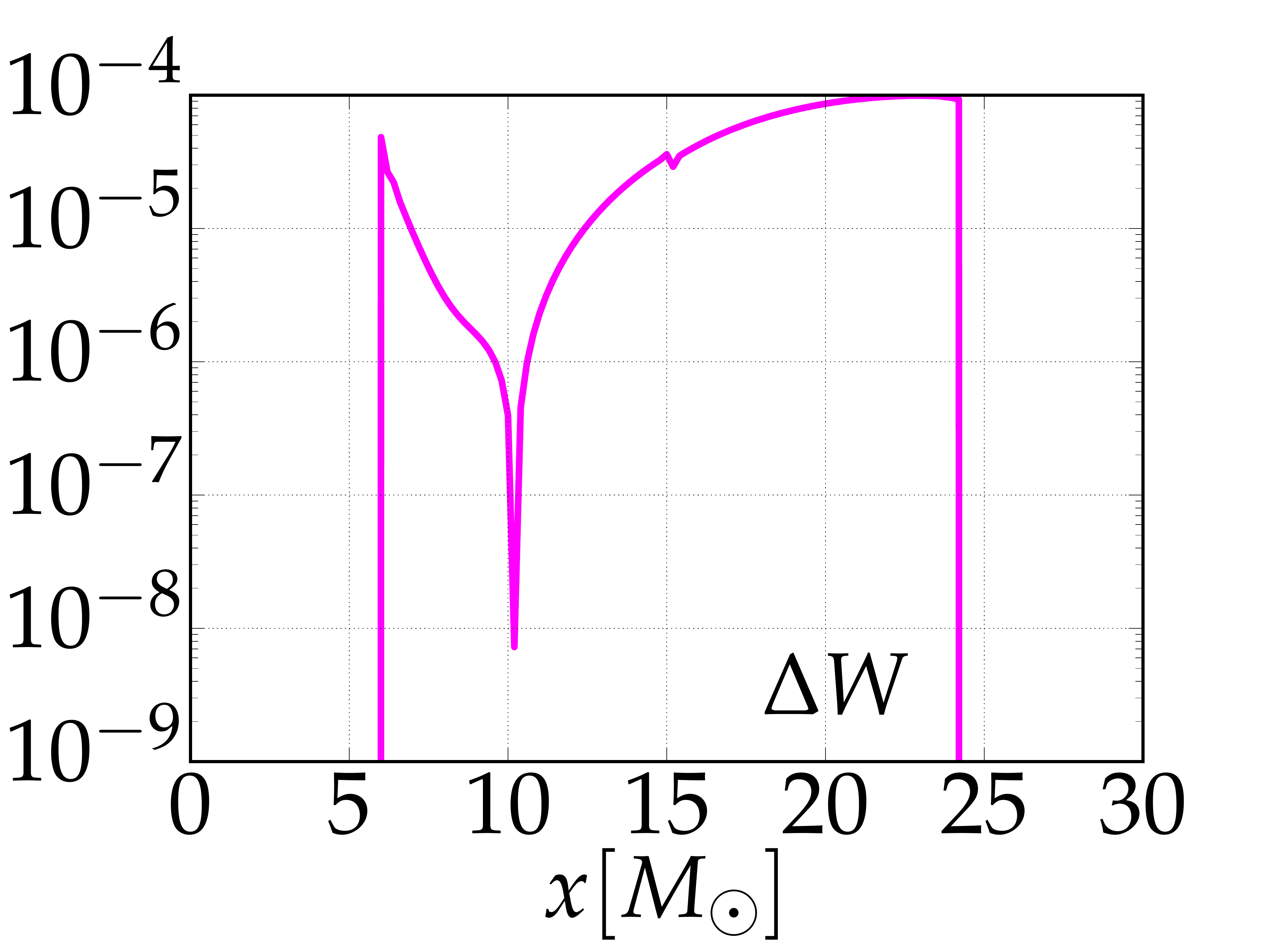}
\end{center}
\caption{The same as Fig.~\ref{fig:irrot_var2} for the $y$-component of
  the three velocity relative to the Eulerian observers and the
  corresponding Lorentz factor.}
\label{fig:irrot_var2}
\end{figure}

\begin{figure}
\begin{center}
\includegraphics[width=\columnwidth]{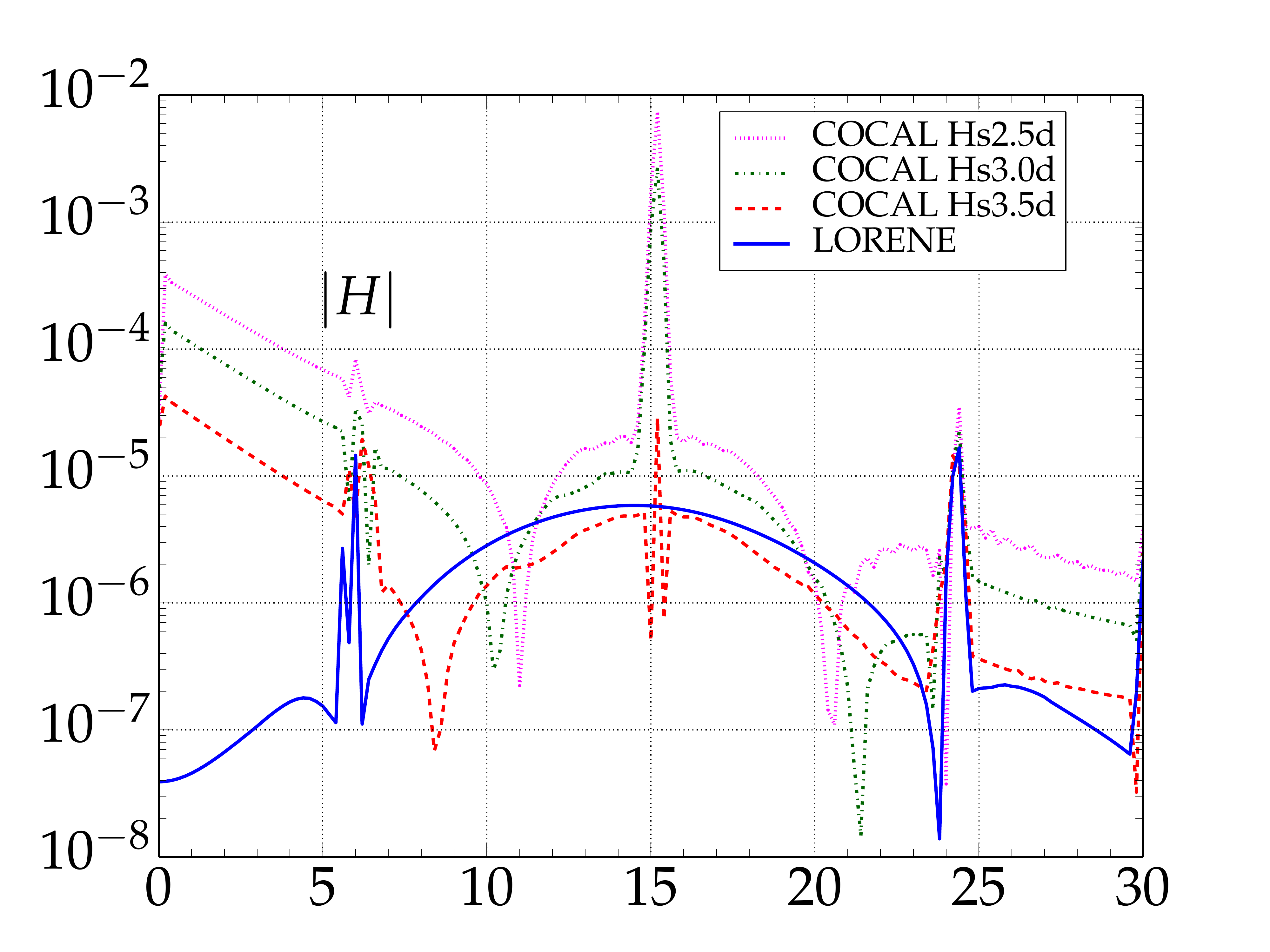}
\includegraphics[width=\columnwidth]{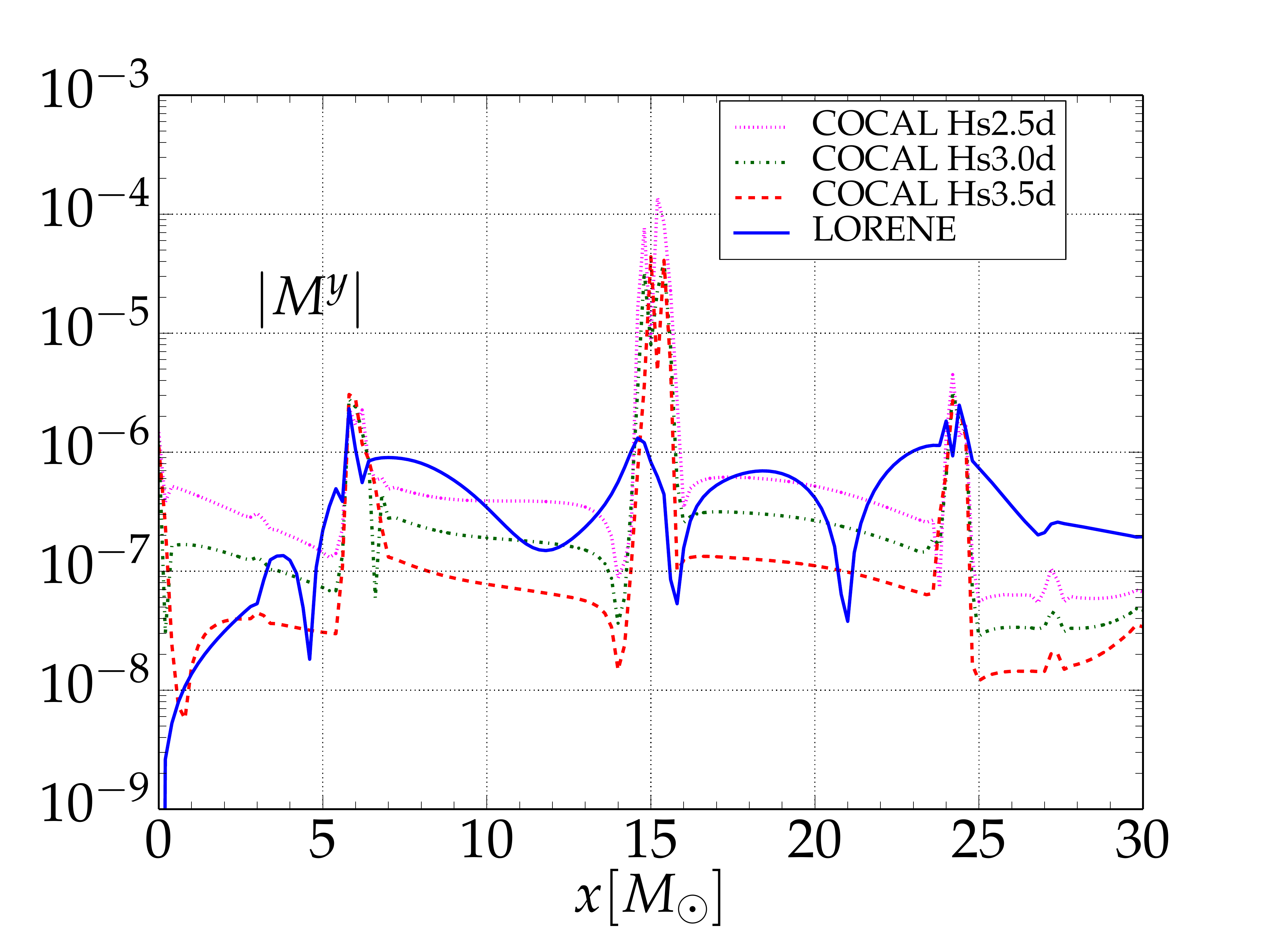}
\caption{Hamiltonian (top) and momentum-constraint violations (bottom)
  for the $y$-component of the shift ($\GB^y$) along the $x$-axis for the
  irrotational binary system at the initial time. The origin $x=0$
  corresponds to the center of mass of the binary, with the surface of
  the star to be located at $x\approx 6\,M_{\odot}$ and at 
  $x\approx 24\,M_{\odot}$.}
\label{fig:xaxis_cv}
\end{center}
\end{figure}

At present, however, we compute the initial data for an irrotational
binary at separation of $\simeq 44.7\,{\rm km}$ by fixing $r_s=0.7597667$
and $d_s=2r_c=2.5$, and report in Table~\ref{tab:cocgrids45} the four
different resolutions used by \cocal{} to obtain the solutions presented
here. Each symbol is explained in Table~\ref{tab:grid_param} and in more
detail in \cite{Tsokaros2015}. For simplicity, and because we are not
interested in microphysical effects here, the equation of state is set to
be a simple polytrope with polytropic index $\Gamma=2$ and polytropic
constant $K=123.6$.

The initial data computed by \lorene{} employs six different domains to
cover the computational region around each star, with a number of
collocation points for the spectral expansion given by $N_r \times
N_\theta \times N_\phi = 33 \times 25 \times 24$, where $N_r$,
$N_\theta$, and $N_\theta$ denote the number of points for the radial,
polar, and azimuthal directions, respectively. In our model the ratio
between the star radius and the separation is roughly three, so that,
according to Ref. \cite{Taniguchi2010}, the resolution that we employ is
sufficient to achieve a fractional error of $10^{-5}$ in the ADM mass
comparable to the one obtained by \cocal{}.

The physical parameters of the binary are presented in
Table~\ref{tab:loco45}. Each star of the binary is constructed to
correspond to a spherical solution of rest mass $M_0=1.62505$ or $M_{\rm
  ADM}=1.51481$, with a relative accuracy of $\mathcal{O}(10^{-6})$ in
the rest mass, which is computed as
\begin{equation}
M_0 =  \int_{\Sigma_t} \GR u^\GA dS_\GA \,,
\label{eq:restmass}
\end{equation}
while the ADM and Komar mass are computed as  
\begin{eqnarray}
\Madm &=& -\frac{1}{2\pi}\int_{S_\infty} \pd^i\GC\, dS_i \,,  \label{eq:adm} \\
\MK &=& \frac{1}{4\pi}\int_{S_\infty} \pd^i \GA\, dS_i \,. \label{eq:komar}
\end{eqnarray}
The surface integrals are calculated at a certain finite radius,
typically around $r\sim 10^4\,M$, and the relative differences found
between the Komar and ADM mass is of the order of $10^{-5}$ even for the
\cocal{} initial data with the coarsest resolution $\texttt{Hs2.0d}$,
thus providing a simple measure of the overall error of the code. The
ADM angular momentum is instead computed as
\begin{equation}
J = \frac{1}{8\pi}\int_{S_\infty} \Kab\, \phi^b\, dS_a\,.
\label{eq:angmom}
\end{equation}

In Fig.~\ref{fig:irrot_var1} we report various quantities of the
irrotational solution along the positive $x$-axis of the Cartesian grid,
so that $x=0$ is the center of mass of the binary. 
The star of radius $R_{\rm eq}\approx 9\,M_{\odot}$ is positioned approximately 
at $x\approx 15\,M_{\odot}$. In both
., on the left column we plot the quantity as computed with
\cocal{} (red lines) and \lorene{} (blue lines), relative to the
$\texttt{Hs3.0d}$ resolution, while on the right column we plot the
relative difference
\begin{equation}
\Delta f := \left|1 - \frac{f_{_{\rm COCAL}}}{f_{_{\rm LORENE}}}\right|
\,.
\label{eq:reldif}
\end{equation}
Going from top to bottom in Fig.~\ref{fig:irrot_var1}, the quantities
plotted are the metric $g_{xx}=\GC^4$ (note that $g_{ij}=\GC^4
\GD_{ij}$), the lapse function $\alpha$, the $y$-component of the shift,
the x$y$-component of the extrinsic curvature, and the rest mass density,
while in Fig.~\ref{fig:irrot_var2}, the $y$-component of the fluid
velocity with respect to the Eulerian observer and the corresponding
Lorentz factor. 

The 4-velocity can also be written as $u^\mu=\GA u^t (n^\mu + U^\mu)$
with $n^\mu$, the unit normal to the hypersurface (Eulerian 4-velocity)
and
\begin{equation}
 U^y = \frac{1}{\GA}\left(\frac{u^y}{u^t}+\GB^y\right) = 
\frac{\GG^y_{\ \mu} u^\mu}{\GA u^t}= \GA \frac{\GC^{-4} \pd^y \Phi}{\GL} \,,
\label{eq:irrot_vy}
\end{equation}
where, we recall, $\GL:=C+\GO^i D_i\Phi$. As it can be seen in
Fig. \ref{fig:irrot_var2}, the difference in the computed variables
between the two codes is of the order of $1\%$ or less, except for points
at or near zero crossings, where the relative error, Eq.~\ref{eq:reldif}, 
produces large values.

Comparing the right columns of Figs.~\ref{fig:irrot_var1} with Fig. 6 of
Ref. \cite{Uryu:2012b}, where a similar comparison was made between
\cocal{} and \kadath{} for black-hole binary initial data, we note that
the difference between the two codes is approximately one order of
magnitude larger than in \cite{Uryu:2012b}. There are two main reasons
behind this. 

First, in Ref. \cite{Uryu:2012b} the comparison was direct in the sense
that the \kadath{} code evaluates the solution at exactly the same
gridpoints used by \cocal{}, so that no interpolation needs to be done;
here, on the other hand, comparison is done after the solutions of both
\lorene{} and \cocal{} are interpolated on the Cartesian grids. Second,
and more importantly, the black-hole binary problem is scale free, thus
allowing Ref. \cite{Uryu:2012b} to compare \textit{exactly} the same
physical system. This is no longer true for the neutron-star binaries that
we explore here, since the two binaries have slightly different central
rest-mass densities and also different separations, radii, etc. (see
Table~\ref{tab:loco45}). This is also manifested by the fact that
Figs.~\ref{fig:irrot_var1}, \ref{fig:irrot_var2} do not change
considerably if we increase or decrease the \cocal{} resolution, implying
that the observed differences in the metric functions are already
dominated by the intrinsic differences in the physical models considered.

Having examined some of the representative variables of the initial
dataset, we next move into an analysis of the constraint equations on the
initial spacelike hypersurface.
%
%
In Fig.~\ref{fig:xaxis_cv} we show the residuals for both the Hamiltonian
constraint equation and the $y$-component of the momentum constraint
equation along the $x$-axis. Here too,
$x=0$ corresponds to the center of mass of the binary with the star
surface located at $x\approx 6\,M_{\odot}$, and at $x\approx 24\,M_{\odot}$. 
For the initial data computed with \cocal{} we show the three
highest resolutions $\texttt{Hs2.5d,\ Hs3.0d,\ Hs3.5d}$ of Table
\ref{tab:grid_param} and note that since the star radius is $13.59\,{\rm
  km}$ and the number of points across the star are $\Nrf=76,\ 100\,{\rm
  and}\ 150$ at these three resolutions, the spatial resolution along the
$x$-axis is $179,\ 136$, and $91\,{\rm m}$, respectively.

A first reading of these plots reveals that inside the star both codes
produce errors of approximately the same magnitude. For \cocal{},
however, the Hamiltonian violations have a spike at the center of the
star, \ie at $x\approx 15\,M_{\odot}$, which converges away with
resolution (\cf initial dataset \texttt{Hs3.5d}). This spike, which
involves $\sim 4-5$ points around the center, is not a reason of major
concern and for two distinct reasons. First, the localized violation is
rapidly removed when the initial data is actually evolved leaving no
apparent influence on the evolution (see also discussion in Section
\ref{sec:comp_evol}). 

Second, as we can see from Fig.~\ref{fig:irrot_var1}, the conformal
factor $\GC$ is computed very accurately in the region around the stellar
center; indeed, a closer inspection of the terms that produce this
violation reveals that it is the result of the location of the origin of
the spherical COCP, which induces local inaccuracies in the second
spatial derivatives of the conformal factor, $\pd_i^2\GC$, near the
stellar center. Similarly, the violations of the momentum constraint
inside the star are of the same order (or even smaller) than those
produced by \lorene{}. Around the stellar surface, both codes exhibit a
jump in the violations due to the existing discontinuity in the first
derivatives of the matter fields. Outside the star and towards the center
of mass, the \cocal{} code produces violations that three orders of
magnitude larger than \lorene{} in the Hamiltonian constraint, but of the
same order for the momentum constraint. The reason for this behaviour is
probably to be found in the resolution of the radial grid, since in that
region we have an increasing step of $\GD x$. We plan to study the source
of this error in the future, by modifying the grid structure there. From
the opposite side of the star and moving towards spatial infinity, again
we have a reasonable agreement between the three sets of initial data. It
is also important to notice that the \cocal{} violations converge away
with the expected second-order accuracy of the finite-difference scheme.

\section{Impact on the evolutions of different initial-data solvers} 
\label{sec:comp_evol}

In order to evolve the initial datasets introduced in the previous
section, we have used the high-order evolution code \whiskythc{}
\cite{Radice2013b, Radice2013c, Radice2015}, which solves the equations
of general-relativistic hydrodynamics in the \emph{Valencia
  formulation}~\cite{Banyuls97} using a finite-difference scheme that
reconstructs the fluxes in local-characteristic variables using a
high-order reconstruction scheme (MP5~\cite{suresh_1997_amp}). In these
simulations, we also employed a positivity preserving limiter, which is
crucial to treat properly the low-density regions of the flow
\cite{Radice2013b}. The evolution of the spacetime is provided by the
\mclachlan{} code \cite{Brown:2008sb}, which solves a conformal-traceless
``$3+1$'' formulation of the Einstein equations either in the
BSSNOK~\cite{Nakamura87, Shibata95, Baumgarte99} or in the CCZ4 form
\cite{Alic:2011a}; we have here employed the BSSNOK formulation, leaving
to future work the investigation with the CCZ4 formulation. The
\mclachlan{} code is part of the open source software framework
\etoolkit{}~\cite{loeffler_2011_et, EinsteinToolkit:web}, which is based
on the \cactus{}~\cite{cactus_web} computational toolkit. We use a
fourth-order finite differencing and the very robust Gamma-driver shift
condition together with the '$1 + \log$' slicing, which have been shown
to be numerically well-behaved for spacetimes describing both isolated
and neutron-star binaries~\cite{Baiotti04, Baiotti06, Baiotti08}.

\begin{figure*}
\begin{center}
\includegraphics[width=0.50\columnwidth]{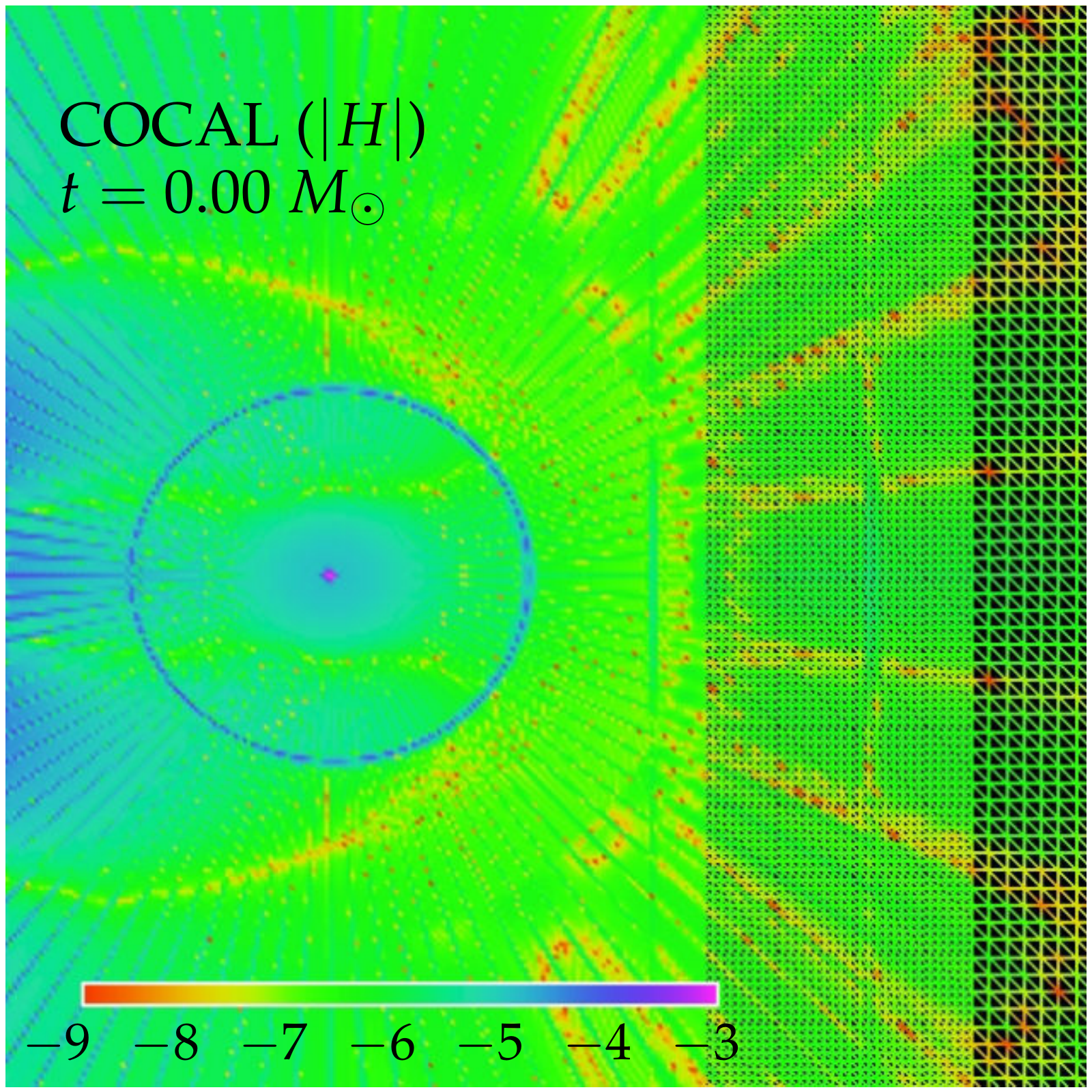}
\includegraphics[width=0.50\columnwidth]{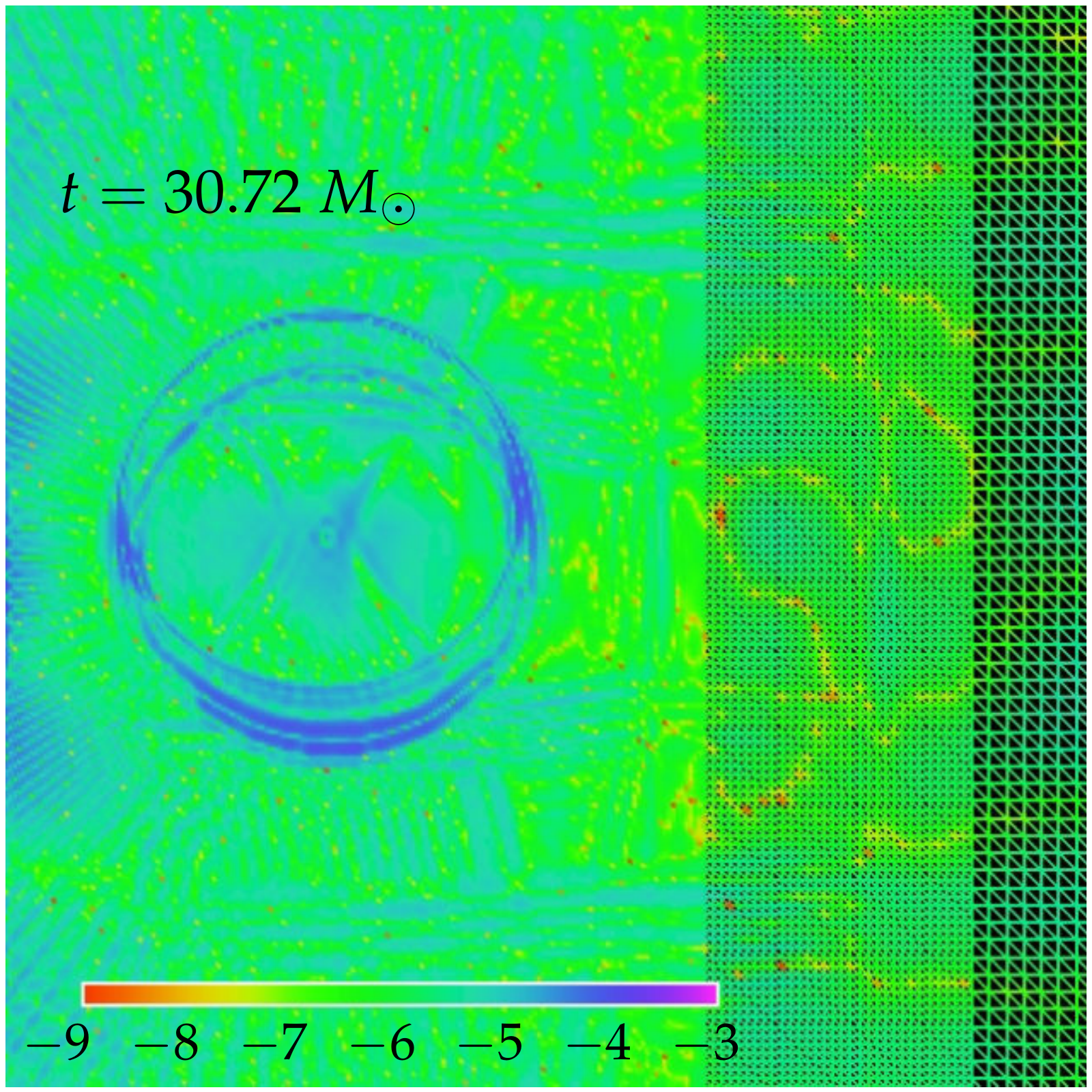}
\includegraphics[width=0.50\columnwidth]{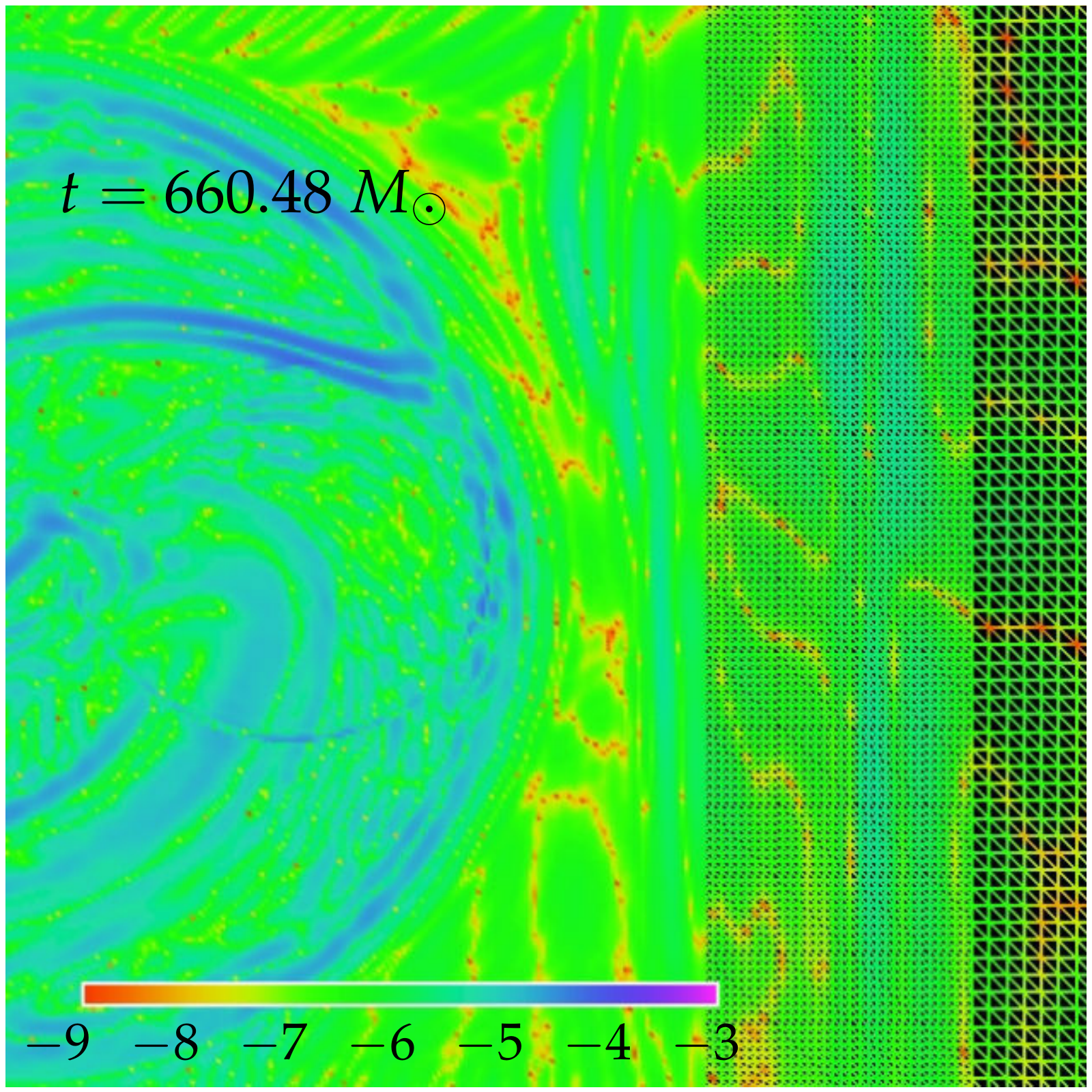}
\end{center}
\begin{center}
\includegraphics[width=0.50\columnwidth]{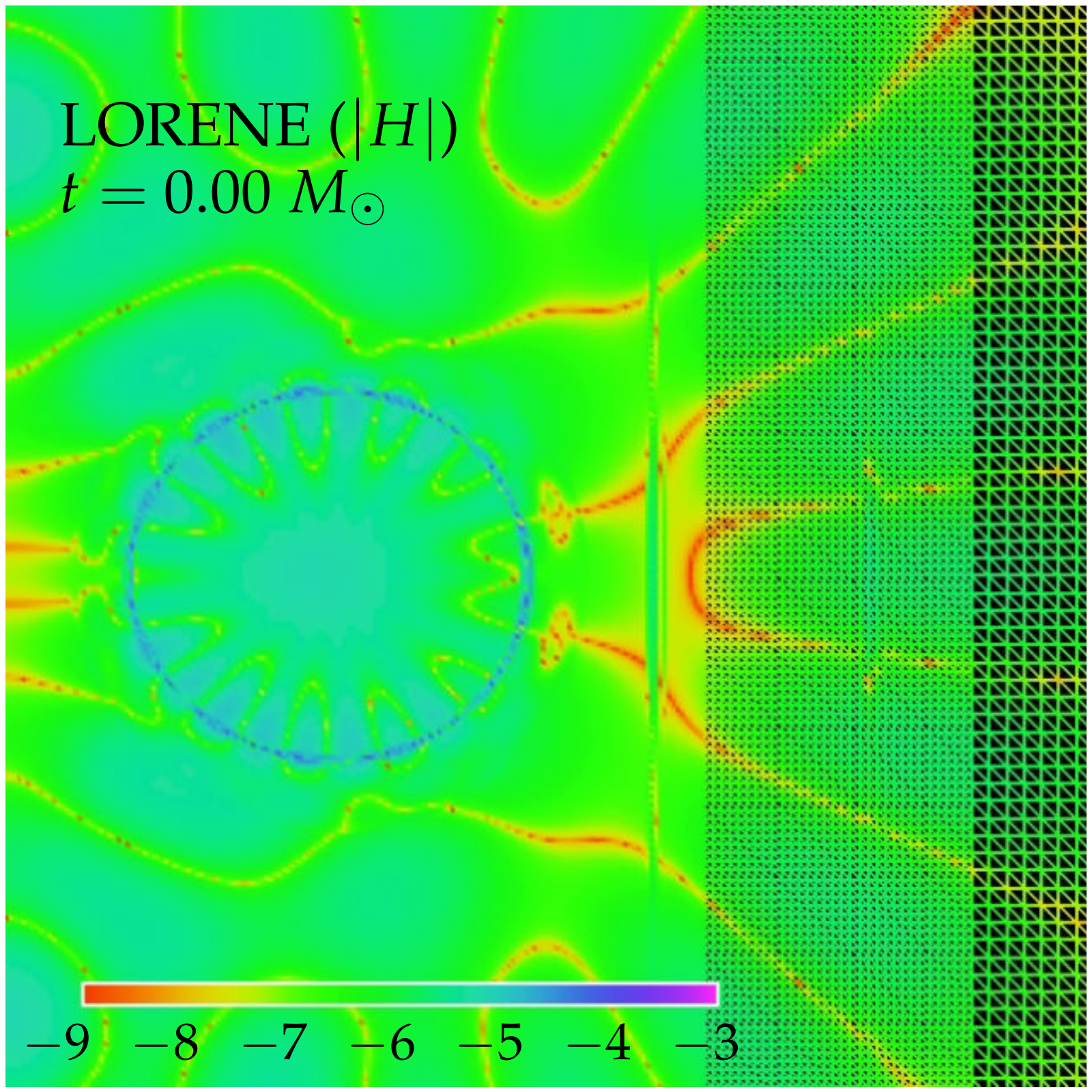}
\includegraphics[width=0.50\columnwidth]{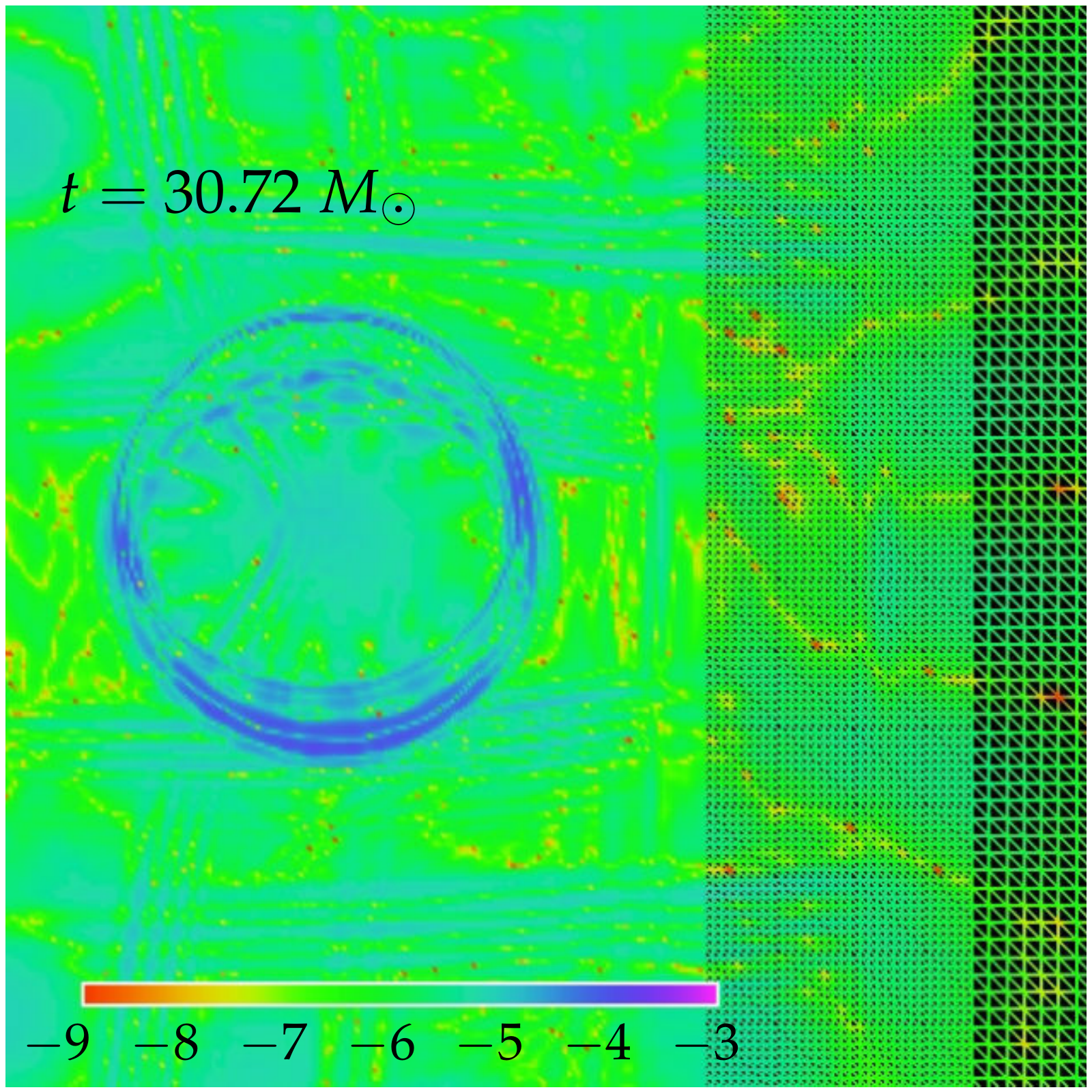}
\includegraphics[width=0.50\columnwidth]{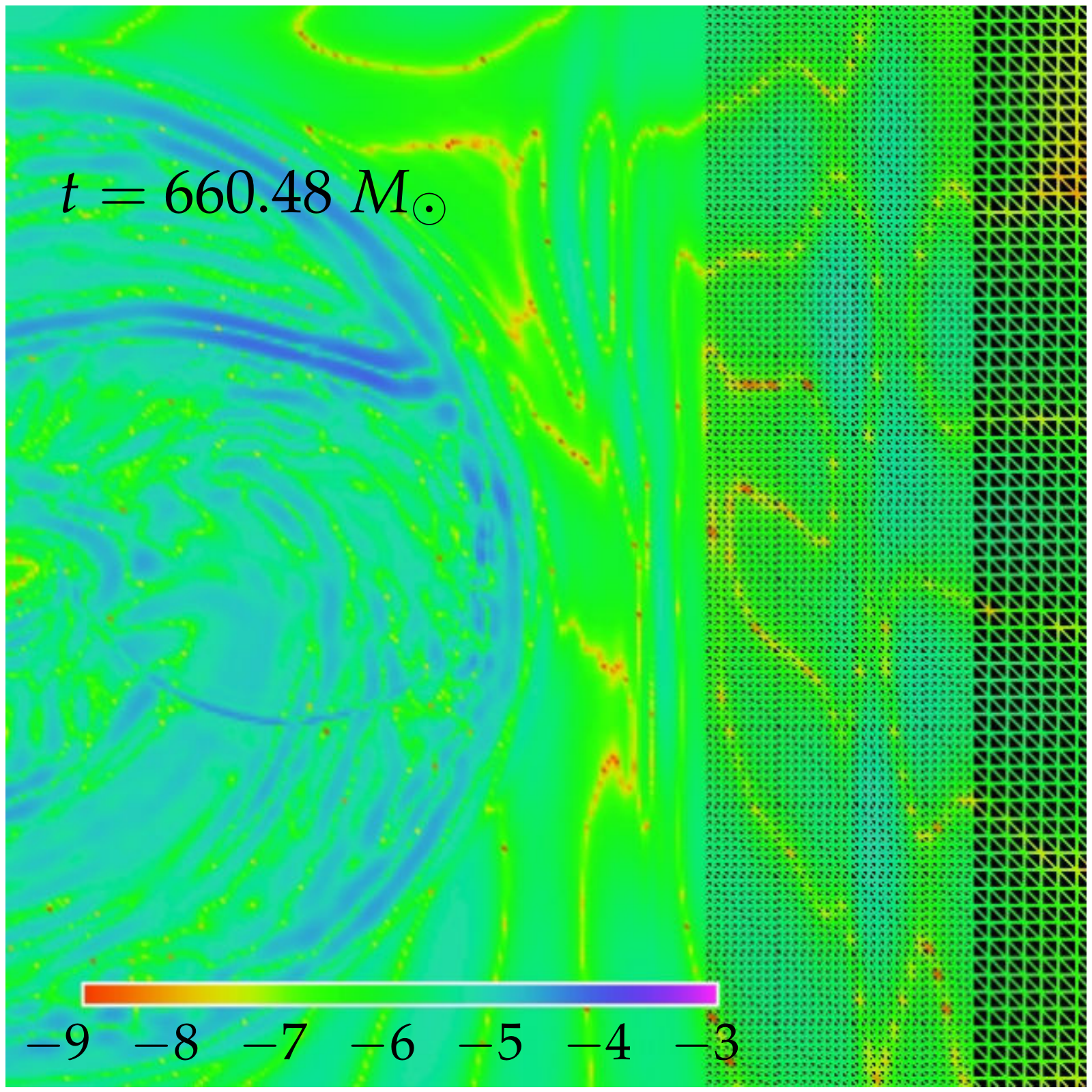}
\end{center}
\begin{center}
\includegraphics[width=0.50\columnwidth]{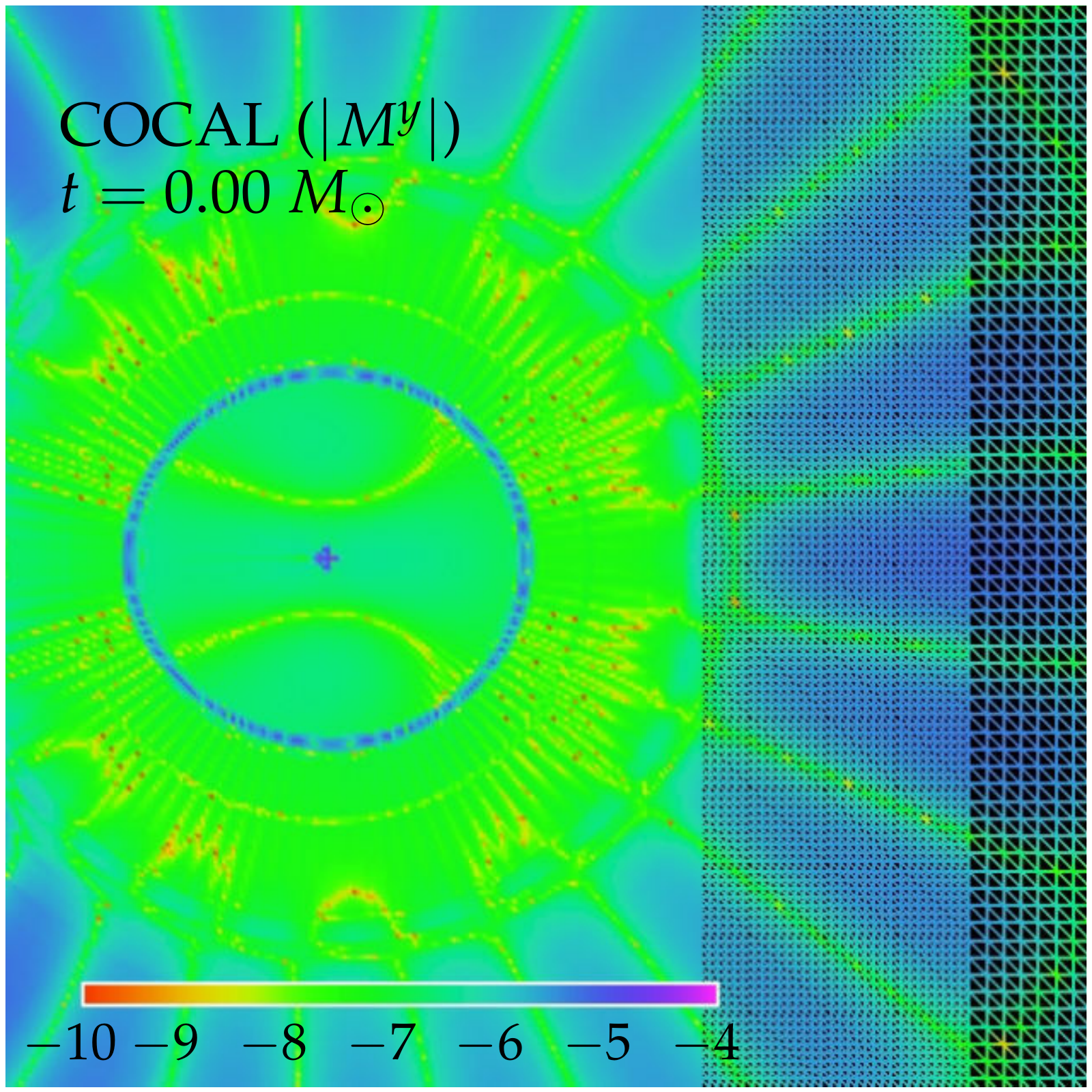}
\includegraphics[width=0.50\columnwidth]{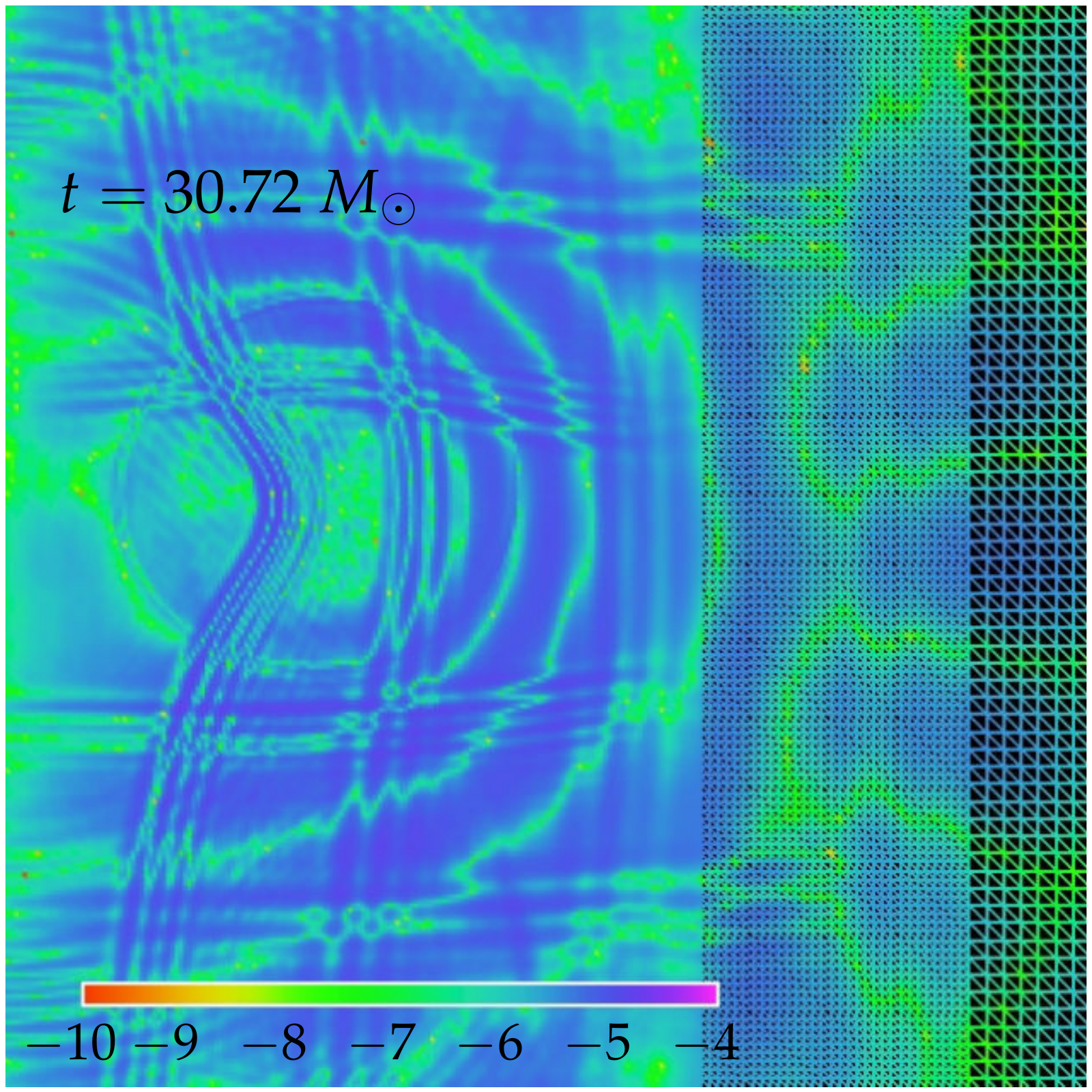}
\includegraphics[width=0.50\columnwidth]{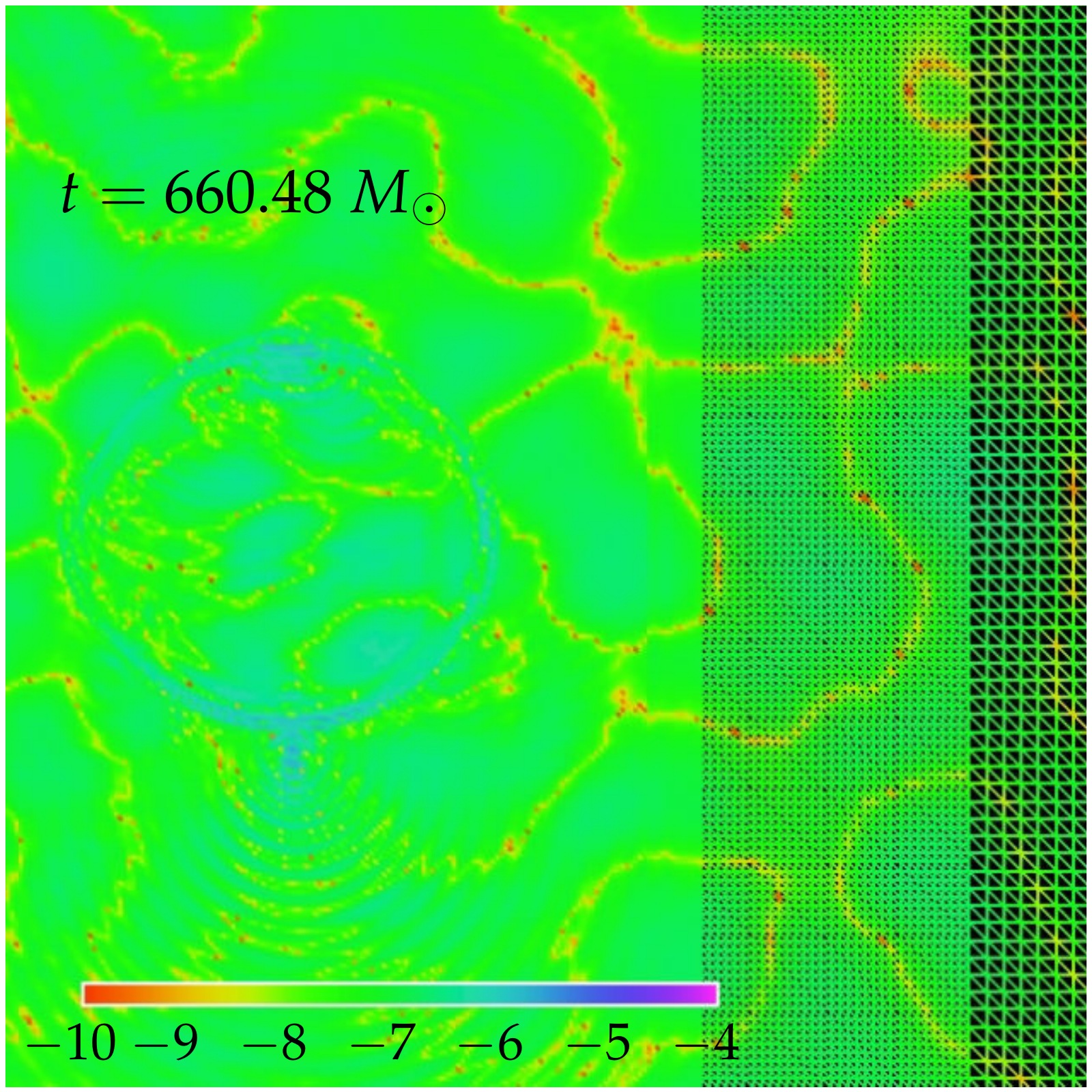}
\end{center}
\begin{center}
\includegraphics[width=0.50\columnwidth]{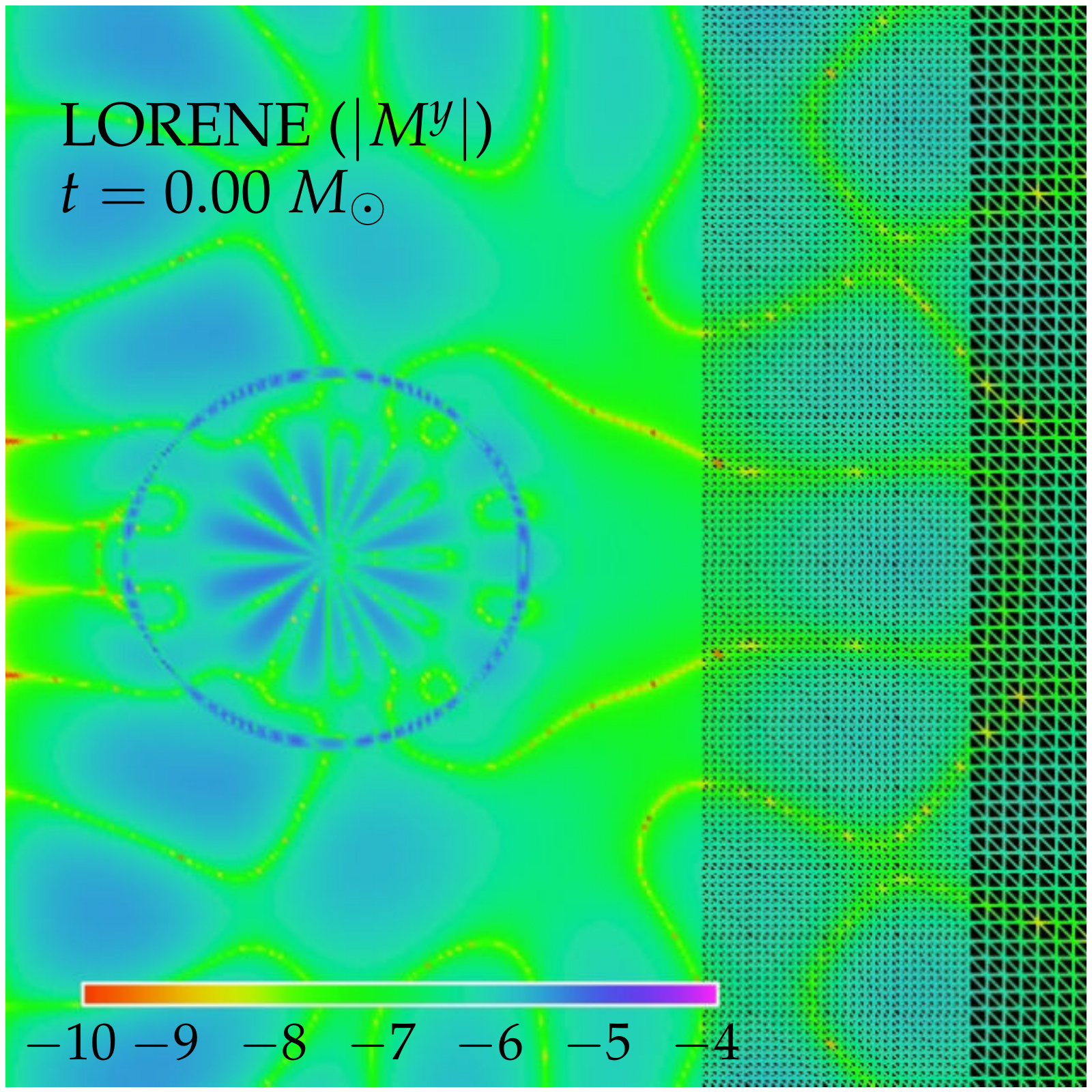}
\includegraphics[width=0.50\columnwidth]{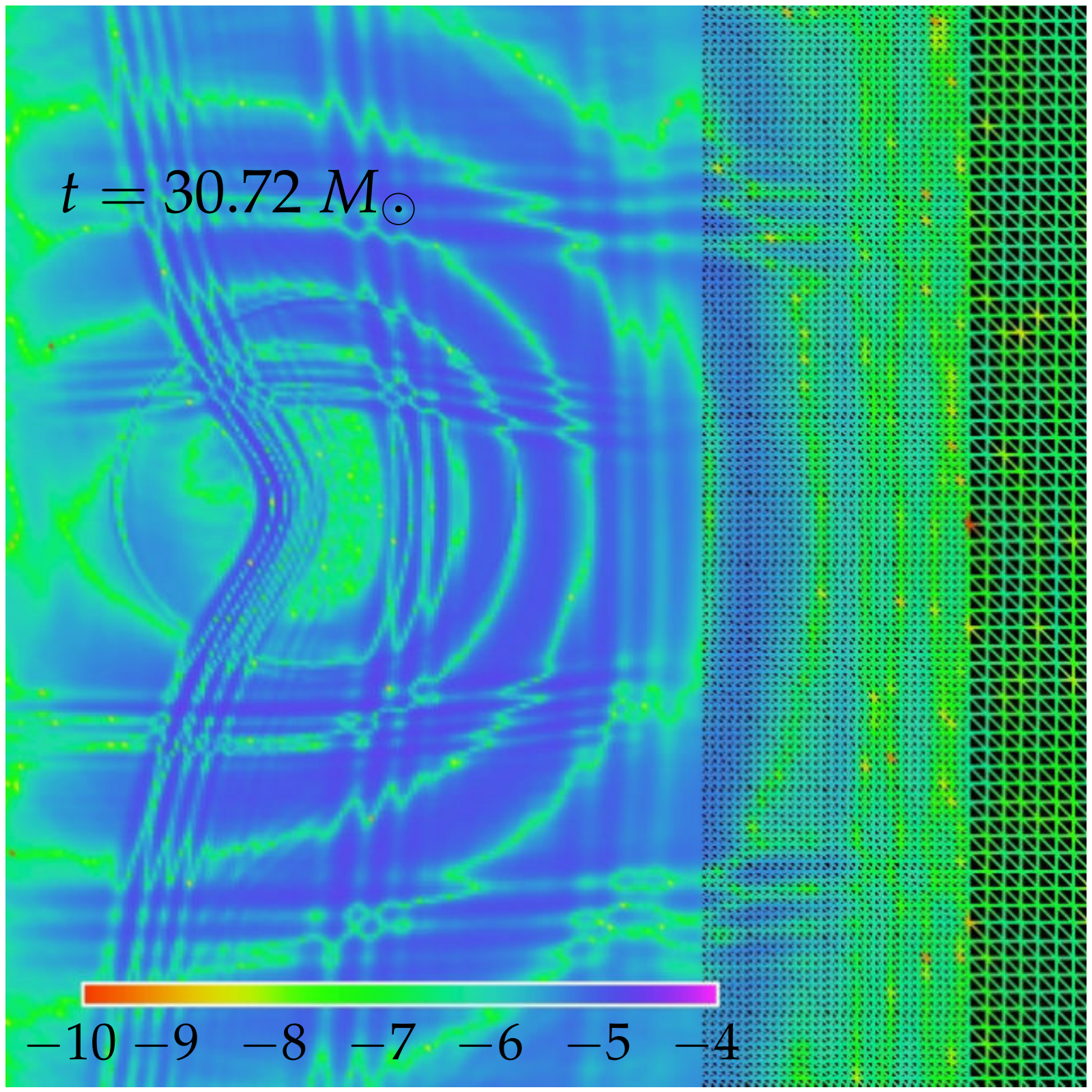}
\includegraphics[width=0.50\columnwidth]{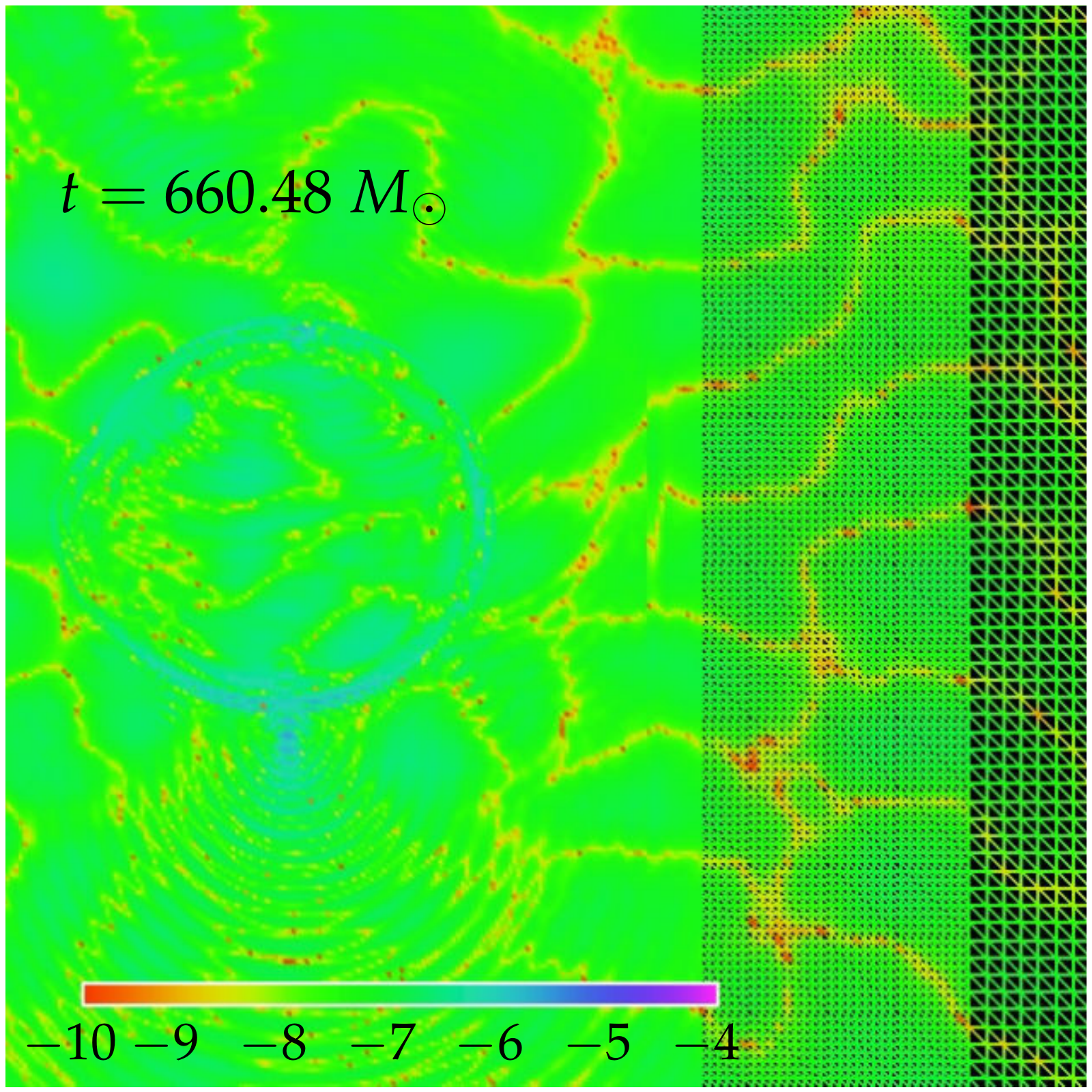}
\end{center}
\caption{Logarithmic violations of the constraint equations shown at
  three different times in the different columns: initial time ($t=0$),
  just after the beginning of the evolution ($t=30.72\,M_{\odot}$), and
  one orbit later ($t=660.48\,M_{\odot}$). From top to bottom, the first
  row shows the violations of the Hamiltonian constraint on the $(x,y)$
  plane from \cocal{}, while the second row the corresponding violations
  from \lorene{}. The third and fourth rows show the violations of the
  $y$-component of the momentum constraint from \cocal{} and \lorene{},
  respectively. Note that all panels show data on three finest levels of
  refinement, with two borders clearly visible. The bounding box in the
  $(x,y)$ plane encompassing each of the panels spans roughly the range
  $[0,50] \times [-25,25]\,M_{\odot}$. The oval shape indicates the neutron
  star surface at every moment.}
\label{fig:logCOir}
\end{figure*}

In particular, we use for these simulations a computational domain in
which $0<x,z\leq 1024\,M_{\odot}$ and $-1024\,M_{\odot}\leq y\leq 1024
\,M_{\odot}$, \ie~we assume $\pi$ symmetry along the $(x,z)$ plane and
reflection symmetry on the $(x,y)$ plane. It is important to remark that
placing the outer boundary at a sufficiently large radius is crucial to
avoid that spurious and constraint-violating reflections from the outer
boundaries spoil the convergence order; for example, we have experienced
that having a computational domain with outer boundary at $512\,M_{\odot}
\simeq 755\,{\rm km}$, which is quite common for neutron-star binary 
simulations \cite{Rezzolla2016}, would not yield convergence waveforms.

An adaptive mesh-refinement grid (AMR) hierarchy is provided by the
\carpet{} driver~\cite{Schnetter-etal-03b,carpet_web}, and we use six
levels of refinement, the finest of which has three different
resolutions: low (L), medium (M), and high (H). These three resolutions
correspond respectively to spatial mesh spacings of $h
=0.2,\ 0.133,\ 0.1\,M_\odot \simeq 295,\ 197,\ 148\,{\rm m}$, or,
equivalently, to $80,\ 120,\,{\rm and}\ 160$ cells along the $x$-axis for
the coarsest grid. See Table~\ref{tab:grid_hierarchy} for more details
on this grid hierarchy.

\begin{table}
\begin{tabular}{ccccc}
\hline
\hline
Level & \multicolumn{3}{c}{AMR Box Extent} & Mesh Spacing \\
& $x$ & $y $& $z$ & $h$ \\
\hline
$0$ & $[0,1024]$ & $[-1024,1024]$ & $[0,1024]$ & $3.2$ \\
$1$ & $[0,240]$  & $[-240,240]$   & $[0,240]$  & $1.6$ \\
$2$ & $[0,120]$  & $[-120,120]$   & $[0,120]$  & $0.8$ \\
$3$ & $[0,64]$   & $[-64,64]$     & $[0,48]$   & $0.4$ \\
$4$ & $[0,40]$   & $[-40,40]$     & $[0,22]$   & $0.2$ \\
$5$ & $[0,30]$   & $[-30,30]$     & $[0,11]$   & $0.1$ \\
\hline
\hline
\end{tabular}
\caption{AMR grid hierarchy: reported are the boxes' extents along the
  $x$, $y$ and $z$ directions, which reflect whether the symmetry
  conditions imposed on them, \ie $\pi$-symmetry along the $x$ coordinate
  at $x=0$ (on the $(y,z)$ plane); reflection symmetry along the $z$
  coordinate at $z=0$, \ie on the $(x,y)$ equatorial plane. The grid
  hierarchy was kept fixed throughout the simulation for each one of the
  different simulations. The mesh spacings listed on the last column are
  the ones used for the highest-resolution simulation.}
\label{tab:grid_hierarchy}
\end{table}

The initial data, computed either with \lorene{} and \cocal{} (for the
latter we use the $\texttt{Hs3.5d}$ dataset) is then evolved with a
Courant factor set to to $0.3$. We note that we reset the shift vector to
zero at the start of each evolution, \ie we do not use the shift as
provided be the initial data codes. The two stars inspiral for about
three orbits (\ie approximately seven gravitational-wave cycles) and then
merge. Because the initial masses have been chosen to be sufficiently
large, the merger leads to a prompt collapse to a black hole surrounded
by an accretion torus \cite{Baiotti08}.

\begin{figure*}
\begin{center}
\includegraphics[width=\columnwidth]{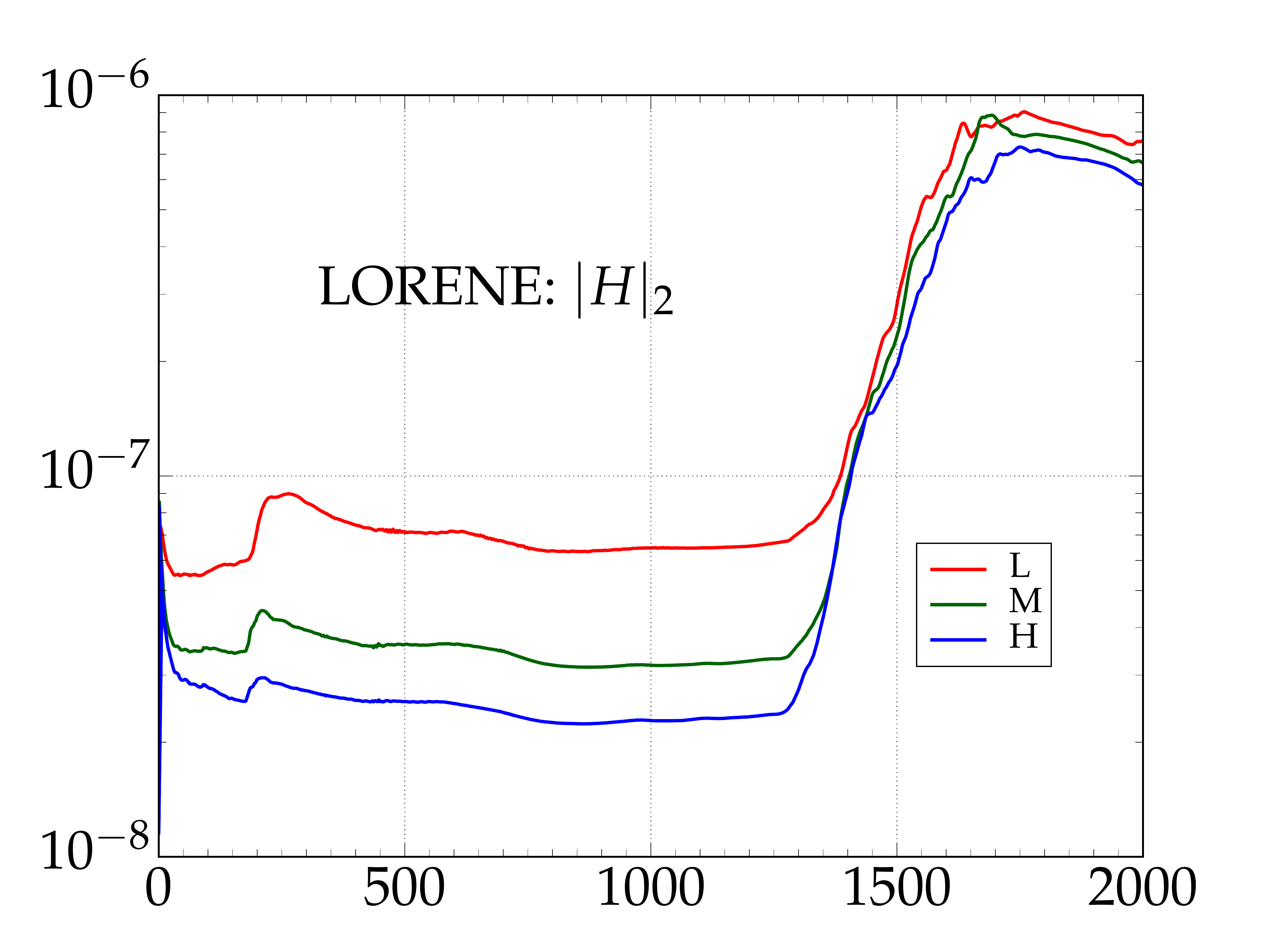}
\includegraphics[width=\columnwidth]{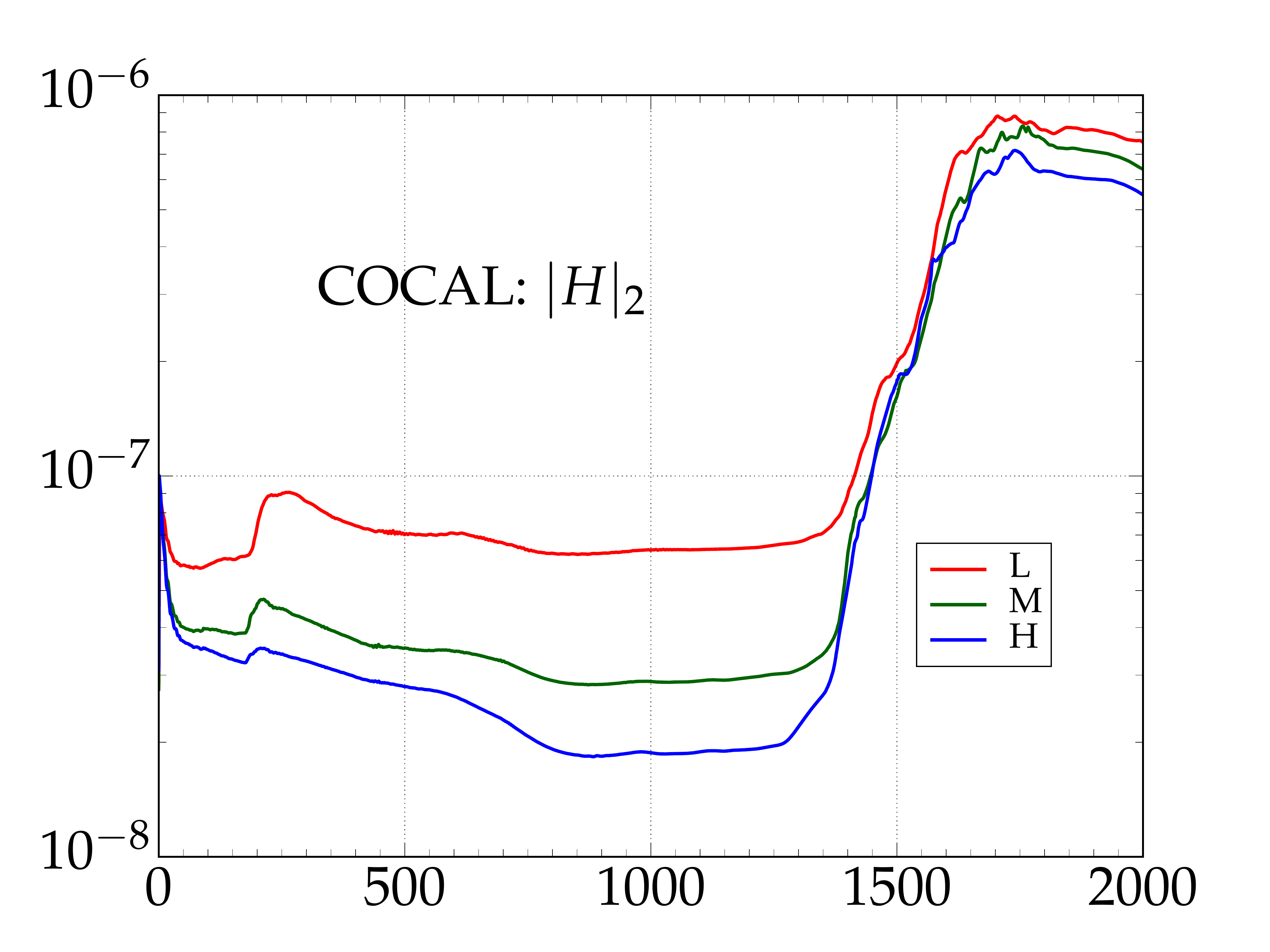}
\end{center}
\begin{center}
\includegraphics[width=\columnwidth]{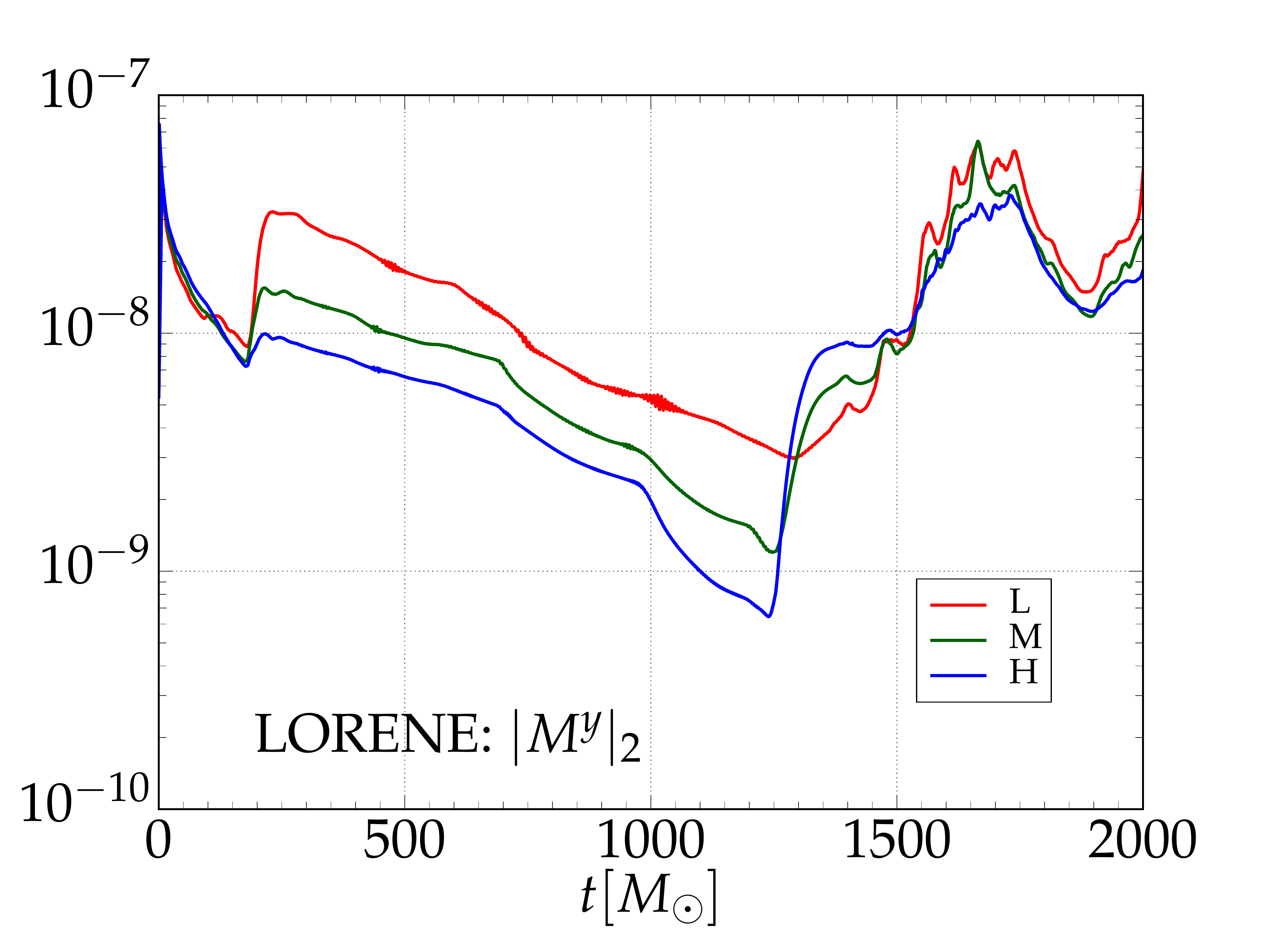}
\includegraphics[width=\columnwidth]{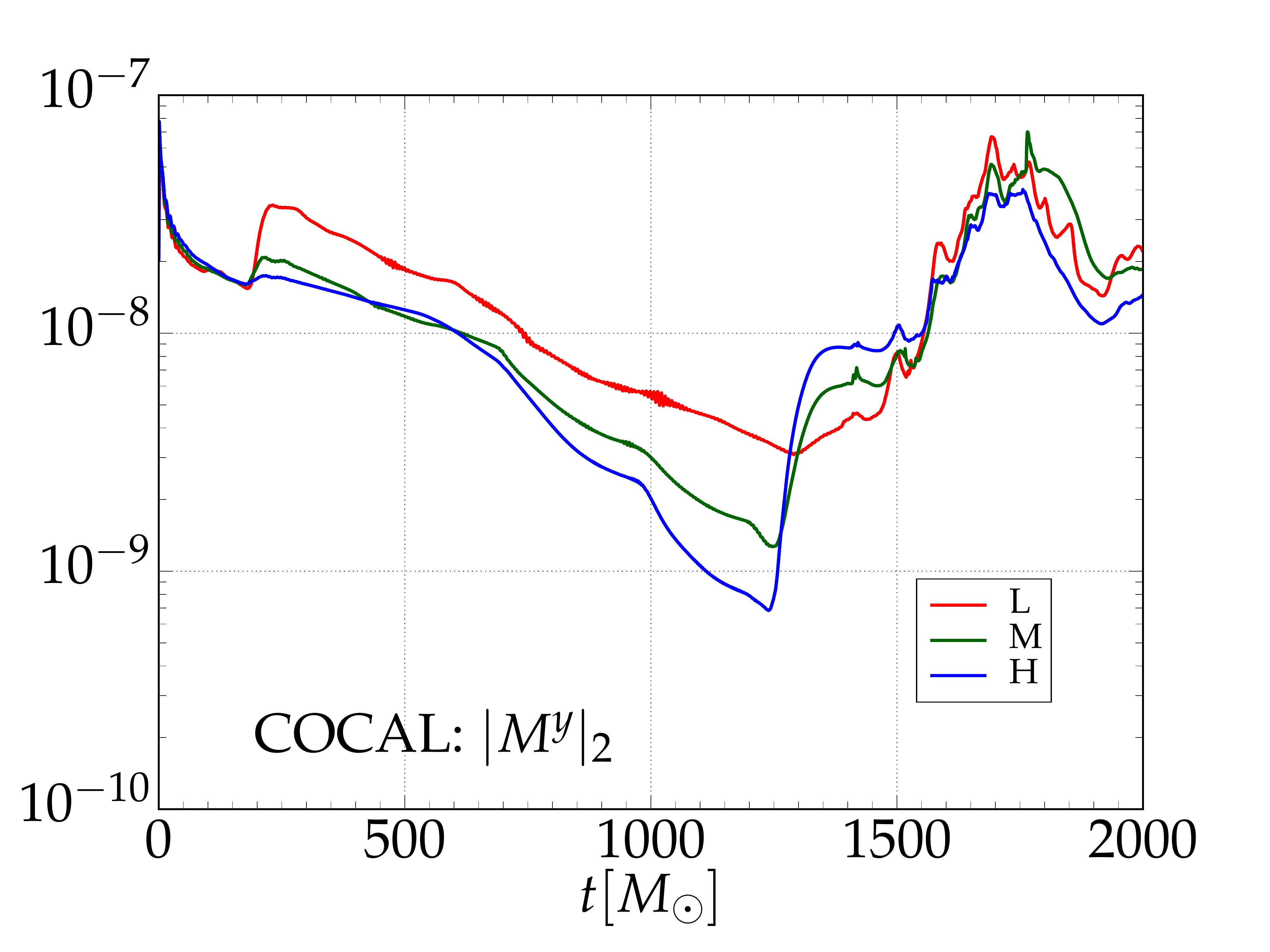}
\end{center}
\caption{Constraint violation $L_2$ norms for \cocal{} and \lorene{} as a
  function of time. The first row shows the Hamiltonian constraint, while
  the second row represents the $y$-component of the momentum
  constraint. Each color refers to a given resolution for the evolution
  grid: low (red), medium (green), and high (blue).}
\label{fig:cv_evol}
\end{figure*}

A more complete picture of the constraint violations as a function of
time is shown in Fig.~\ref{fig:logCOir}, where each panel shows the
constraint violations in the equatorial plane, or $(x,y)$ plane, of the
binary, focusing on the region from the center of mass (middle of left
side on each panel) to approximately six neutron-star radii. From top to
bottom, the first row represents \cocal{} $\texttt{Hs3.0d}$ initial data,
while the second row shows the \lorene{} Hamiltonian violations at three
different times: at $t=0$, which corresponds to the initial data, just
after the simulation is launched, at $t=30.72\,M_{\odot}$, and after one
orbit, at $t=660.48\,M_{\odot}$. When considering the properties of the
initial data, it is possible to note the characteristic
spherical-coordinates pattern of \cocal{}, while in case of \lorene{}
data, one has a wavy kind of structure which reflects the spectral
methods used. The surface of the neutron star is easily noticeable as
violations of the constraints tend to create a discontinuity there. Also
visible is the increase of \cocal{}'s violations towards the center of
mass as was seen in Fig.~\ref{fig:xaxis_cv}.  Apparently these violations
exist in the region around the $(y,z)$ plane close to the center of
mass. The small spike of violations at the center of the neutron star is
also visible. 

Soon after the beginning of the evolution, at $t=30.72\,M_{\odot}$
(middle column), the stars have rotated of about one degree and the
violations of both codes become very similar both inside the star as well
as near the center of mass. This tendency continues one orbit afterwards
(third column) at $t=660.48\,M_{\odot}$ up until the merger. In the third
and fourth rows, we show the momentum violations for \cocal{} and
\lorene{}, respectively. Again the characteristic patterns of both codes
are visible in the initial data first column, with \cocal{} having less
violations inside and around the star. As the binary evolves differences
are washed out and both codes produce similar behaviours.

Up until now all convergence analysis has been done with respect to the
resolution of the initial data. In what follows we fix the initial data
(\texttt{Hs3.5d} for \cocal)
and perform a convergence analysis with respect to the resolution of the
evolution code. In Fig.~\ref{fig:cv_evol}, we monitor the $L_2$ norm
indicator for the Hamiltonian (first row) and $y$-component of the
momentum constraint (second row). It is defined as
\begin{equation} 
|f|_2 := \sqrt{\frac{1}{N} \sum_{i=1}^{N}|f_i|^2 }\,,
\end{equation} 
where $N=N_r\times N_\GU \times N_\GP$ is the total number of
points. Merger happens at approximately $1500\,M_\odot$ or $8\,{\rm
  ms}$.

Every plot has three solid lines that correspond to the three different
evolution resolutions: red is for low, green is for medium, and blue is
for high, with outer boundaries at $1024\,M_\odot$ as stated earlier.
%
%
A first feature to be noticed in these plots is the presence of a local
maximum around $180\,M_\odot$, and the behavior of the violations until
that time. It is possible to see in Fig.~\ref{fig:cv_evol} that this
maximum is reduced as the resolution of the evolution increases and that
its position in time changes as the position of the second AMR refinement
boundary is varied. Together, these considerations clearly indicate that
the first local maximum in the constraint violations is simply due to the
position of the second AMR box and, albeit, annoying has a clear origin
and is not particularly harmful for the subsequent evolution.

A second feature to notice when considering the constraint violations in
the time interval $0<t<180\,M_\odot$ is that although the ones
coming from the Hamiltonian equation scale according to the resolution
(except for an initial interval $0<t<50\,M_\odot$), this is not happening
for the momentum-constraint equation. There, the violations monotonically
decrease until the starting of the "bump" at $t=180\,M_\odot$, and
increasing the resolution does not affect them. Since the initial data
computed with \cocal{} and \lorene{} are already at high resolution, and
since for $t=0$ the violations are approximately more than five times the
ones at $t=180\,M_\odot$, we believe that this behavior is caused by
inaccuracies inherent in the initial data formalism, like the omission of
certain equations or terms in the Euler and the gravitational field
equations. After a certain time ($180\,M_\odot$ in our case) these
violations are washed out and then evolution errors scale accordingly.

In addition to the $L_2$ norm shown here, we have also computed and
studied the behaviour of the $L_1$ norm (\ie $|f|_1 :=
\sum_{i=1}^{N}|f_i|/N$) and of the $L_\infty$ norm (\ie $|f|_\infty :=
\max_i \{|f_i|\}$). More specifically, the $L_1$ norm is of the order
$\lesssim 10^{-8}$ for all the resolutions considered, both for the
\cocal{} and for the \lorene{} initial data, while the $L_\infty$ norm is
the largest of all, with values of the order of $\lesssim 10^{-6}$. Also
this quantity, however, shows a clear convergence scaling in the
Hamiltonian violations. Overall, it is evident that the behavior of the
evolution of the constraint violations is extremely similar both for
\cocal{} as well as for \lorene{} initial data.

\begin{figure*}
\begin{center}
\includegraphics[width=0.9\columnwidth]{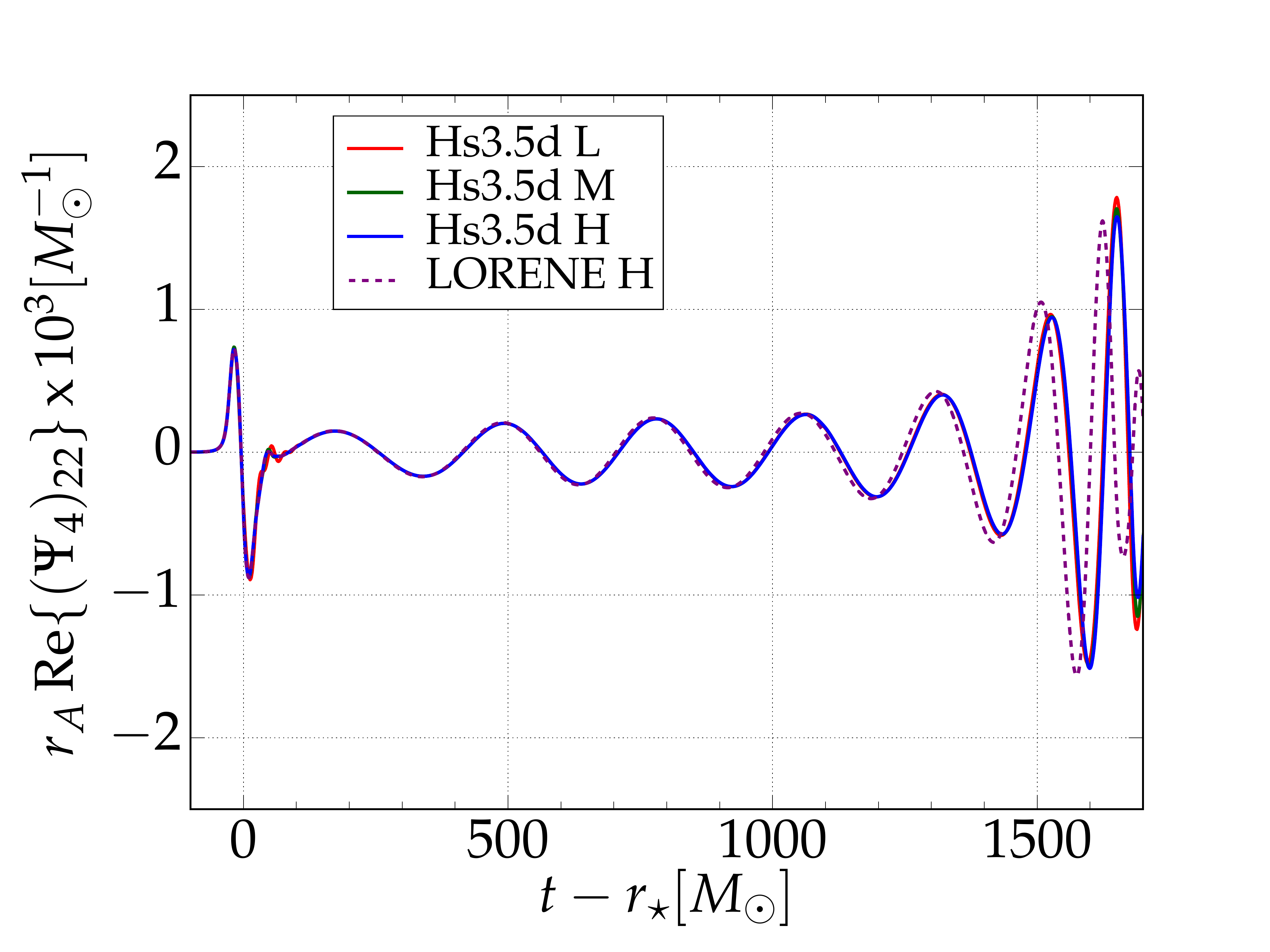}
\includegraphics[width=0.9\columnwidth]{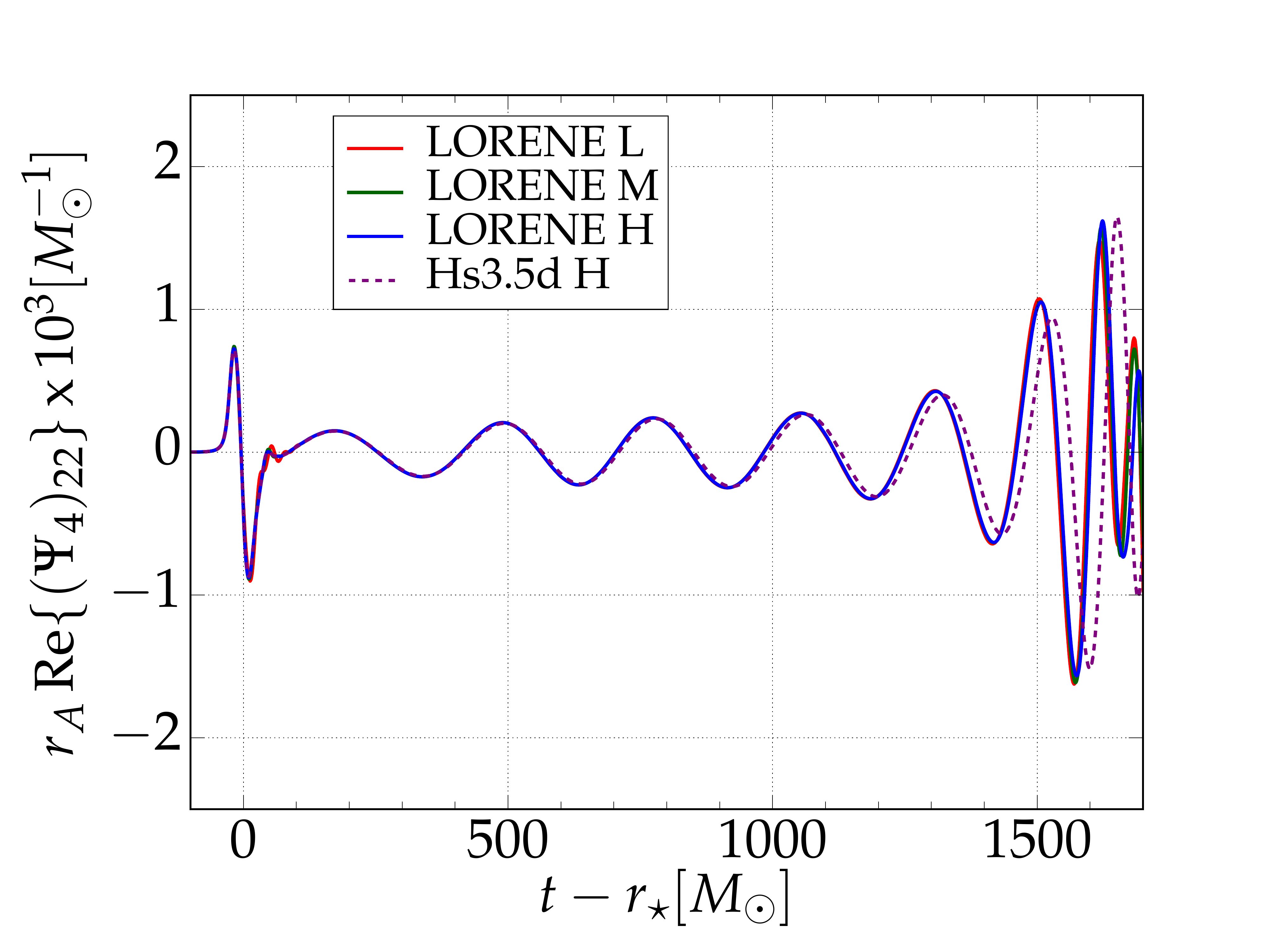}
\end{center}
\begin{center}
\includegraphics[width=0.9\columnwidth]{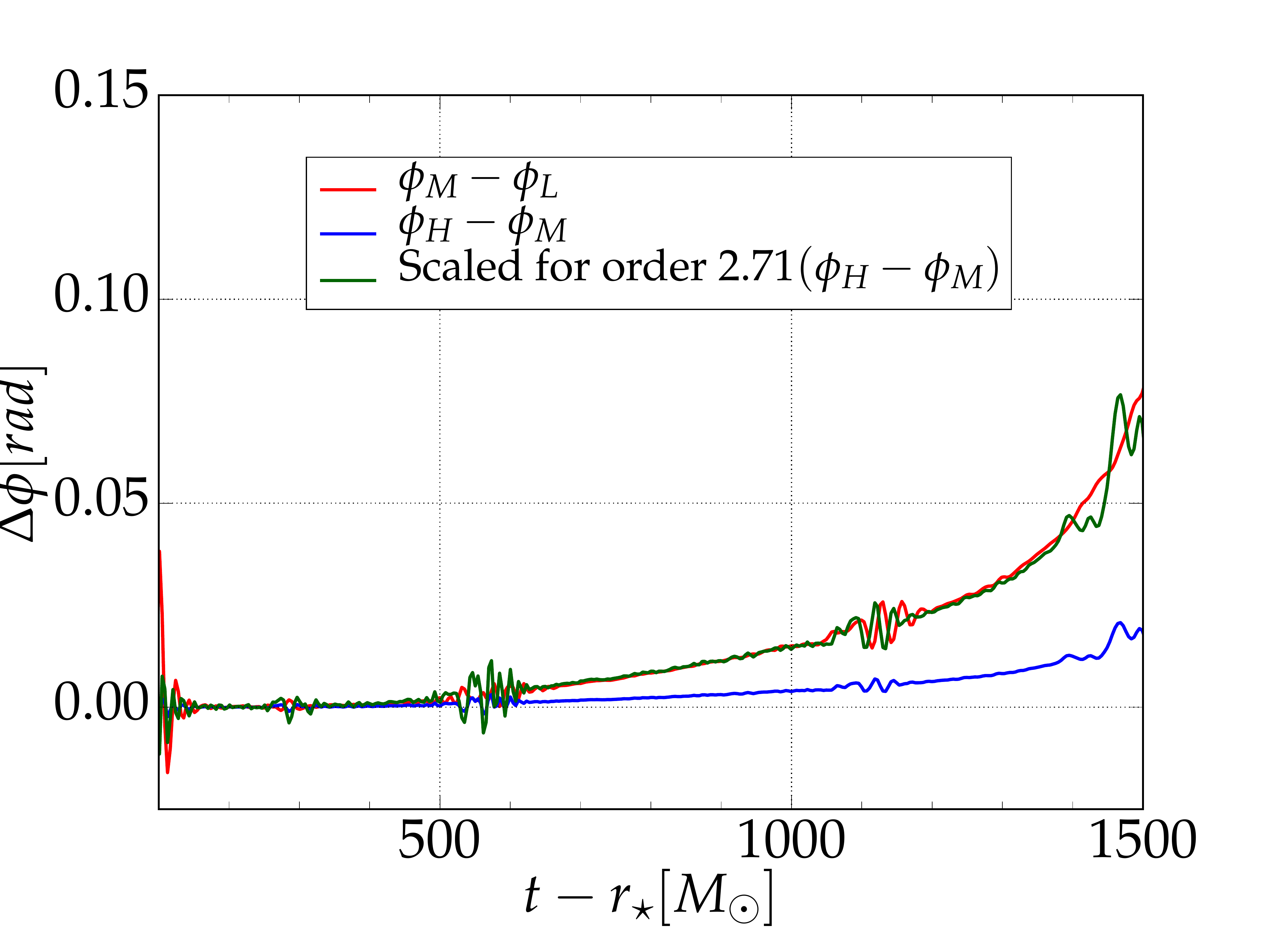}
\includegraphics[width=0.9\columnwidth]{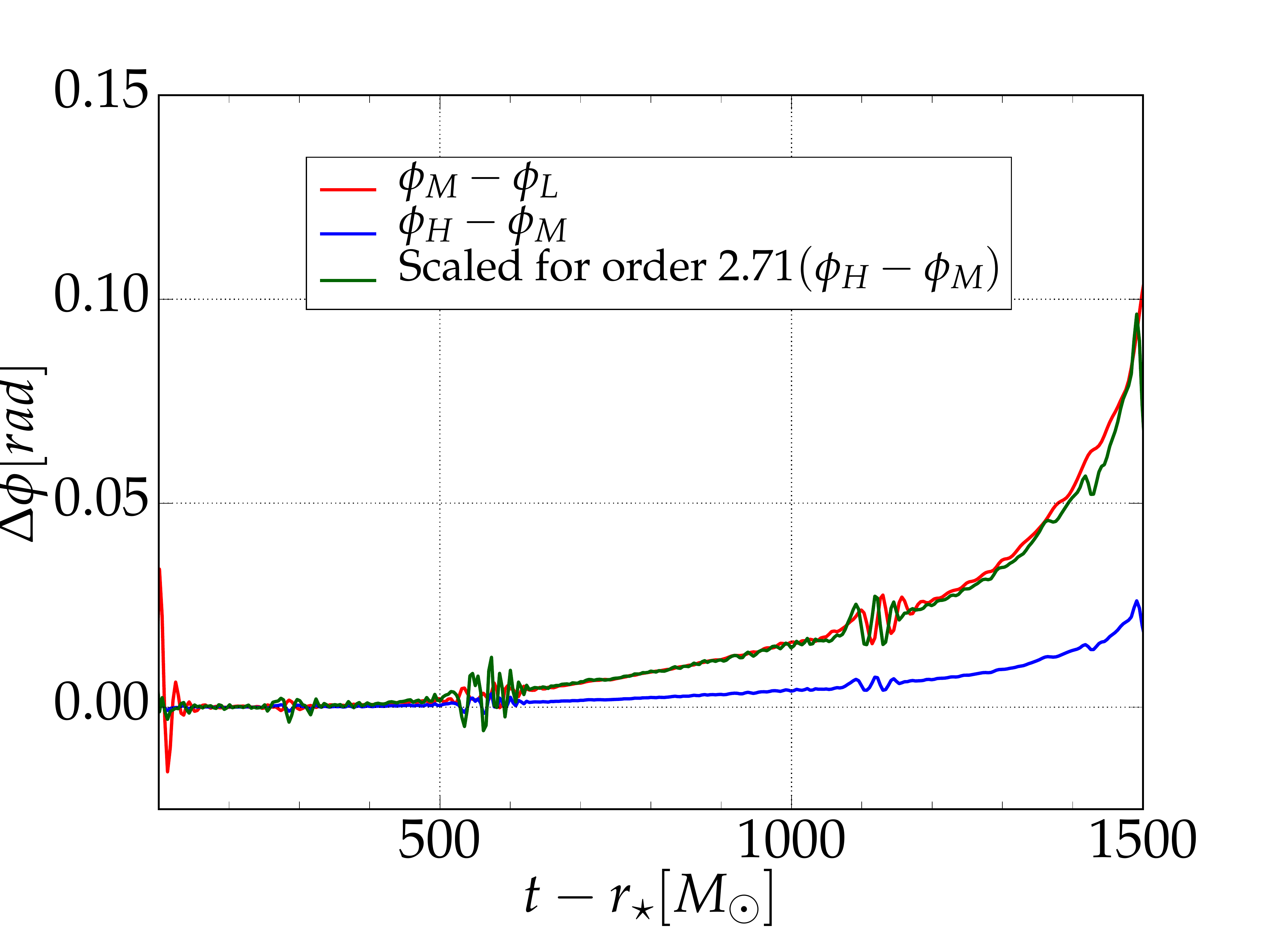}
\end{center}
\begin{center}
\includegraphics[width=0.9\columnwidth]{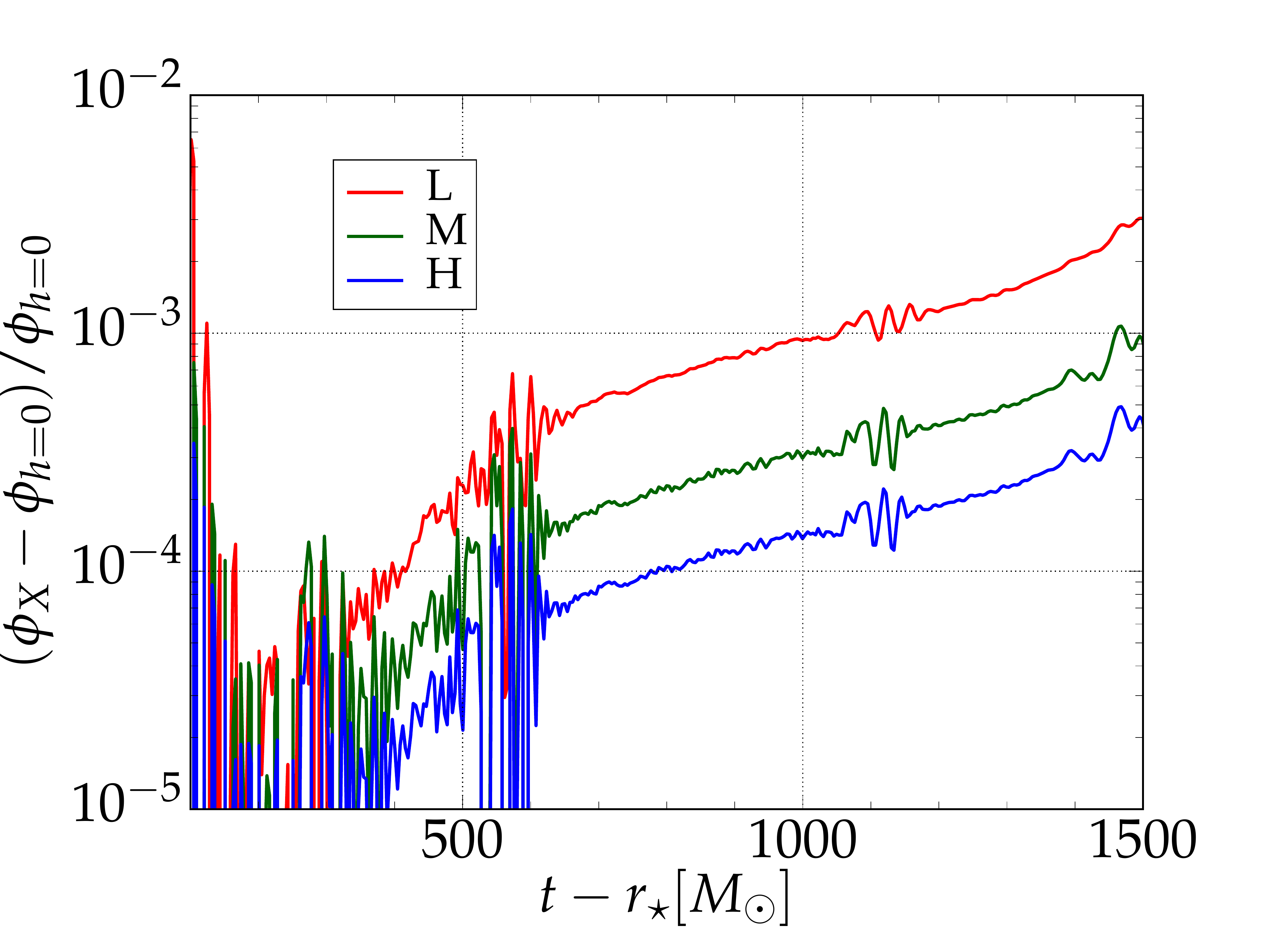}
\includegraphics[width=0.9\columnwidth]{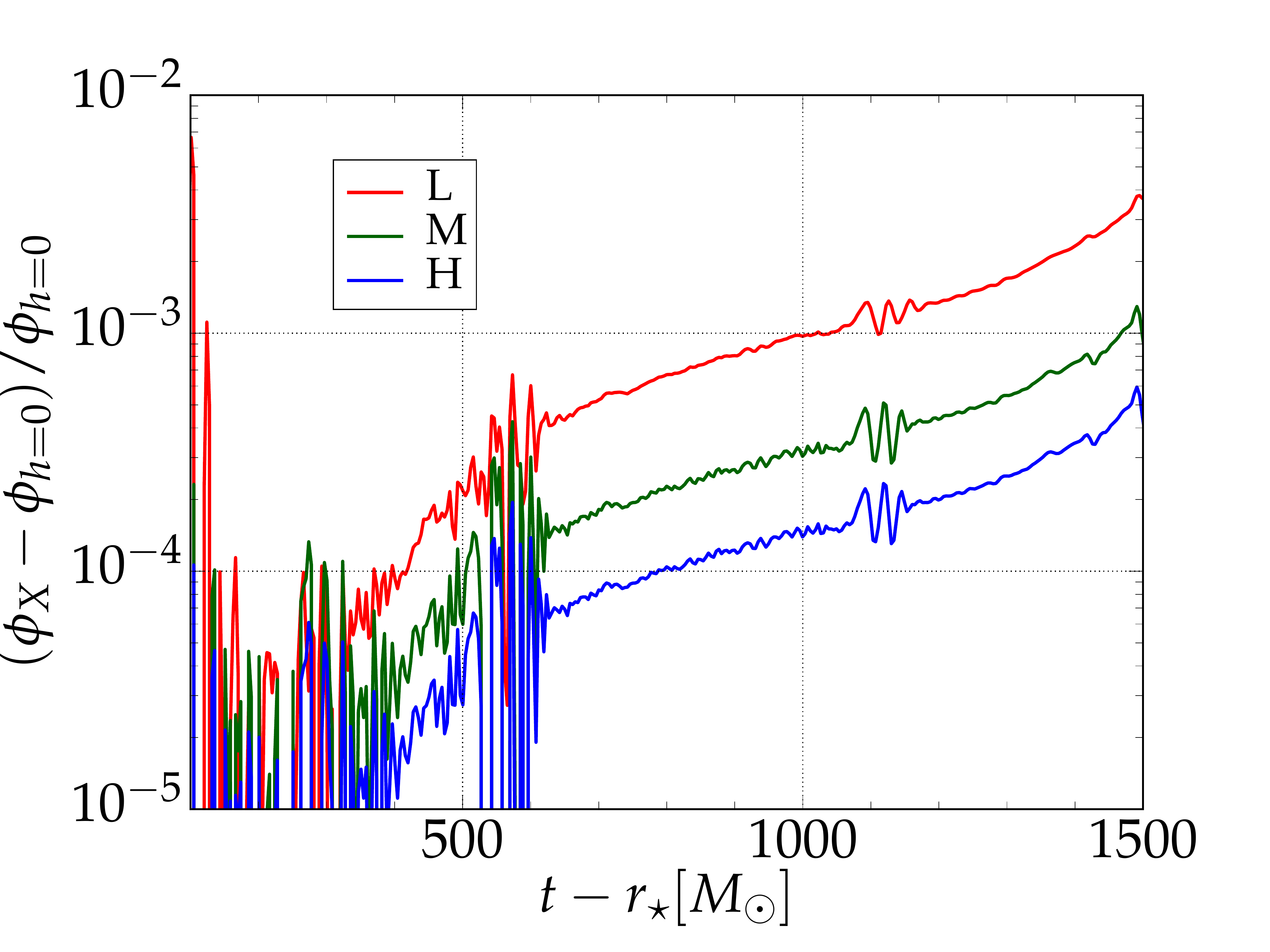}
\end{center}
\caption{First row: real part of $(\Psi_4)_{22}$ extracted at
  $\bar{r}=450\,M_\odot$ as a function of the retarded time for both
  \cocal{} \texttt{Hs3.5d} (left panel) and \lorene{} (right panel)
  initial data and for the three evolution resolutions (L, M, H). On each
  plot with dashed line we denote the evolution with the highest
  resolution of the other code initial dataset so that the dephasing
  between the two datasets to become apparent. Second row: dephasing
  between different resolutions and the rescaled dephasing between the
  high and medium resolution assuming a convergence order $p=2.71$. The
  left panel is for \cocal{}, while right one is for \lorene{}. Third
  row: relative phase difference for the $\ell=m=2$ mode of $\Psi_4$ with
  respect to the Richardson-extrapolated value (computed assuming a
  convergence order of $p=2.71$).}
\label{fig:psi4}
\end{figure*}

One of the main goals in this work is to estimate the impact that
slightly different initial data can have on the observed
gravitational-wave signal. It is well known that the Einstein equations
are highly nonlinear and is therefore possible that even minute
differences in the initial data can result into large and indeed
measurable differences in the radiated quantities. The ability of
measuring how large this impact is of course essential to weigh it in in
the overall budget of numerical-relativity calculations and hence to
measure how the extraction of physical parameters of the sources can be
affected. Hence, we next concentrate here on the gravitational-wave
emission on the $\ell=m=2$ mode of the Weyl scalar $\Psi_4$ which we
extract at $\bar{r}=450\,M_\odot$ 
\begin{equation}
(\Psi_4)_{22}=A(t) e^{i\GP(t)} \,.
\label{eq:psi4}
\end{equation}
The real part of $(\Psi_4)_{22}$ with respect to the retarded time
$t-r_\star$, is plotted in the top row of Fig.~\ref{fig:psi4}, where
\begin{equation}
r_\star:=r_A+2M_{\rm ADM}\ln\left({r_A}/{2M_{\rm ADM}}-1\right)\approx
478.8 \,M_\odot\,,
\end{equation}
is the tortoise radius and $r_A := \bar{r} (1+M_{\rm ADM}/2\bar{r})^2$ is
the approximated areal radius\footnote{We have compared this
  approximation against a numerical computation of the areal radius based
  on the proper area computation of the extraction surfaces. For a
  surface at $\bar{r}=450\,M_\odot$, the relative differences between the
  approximation and the numerically computed radius was $\sim 2\times
  10^{-6}$ during the inspiral and around $\sim 4\times 10^{-5}$ as it
  peaks during the merger.}.

\begin{figure}
\begin{center}
\includegraphics[width=0.75\columnwidth]{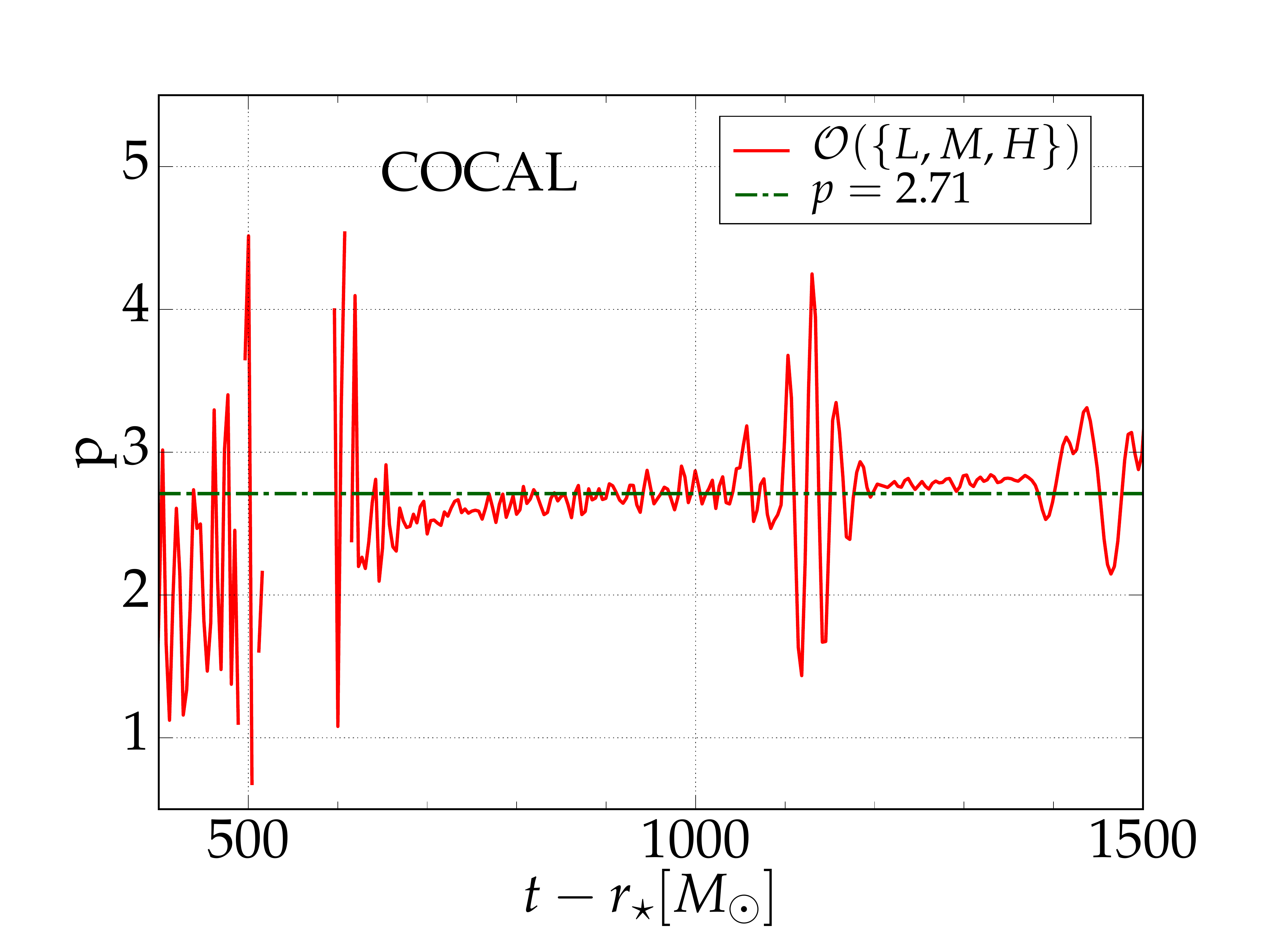}
\includegraphics[width=0.75\columnwidth]{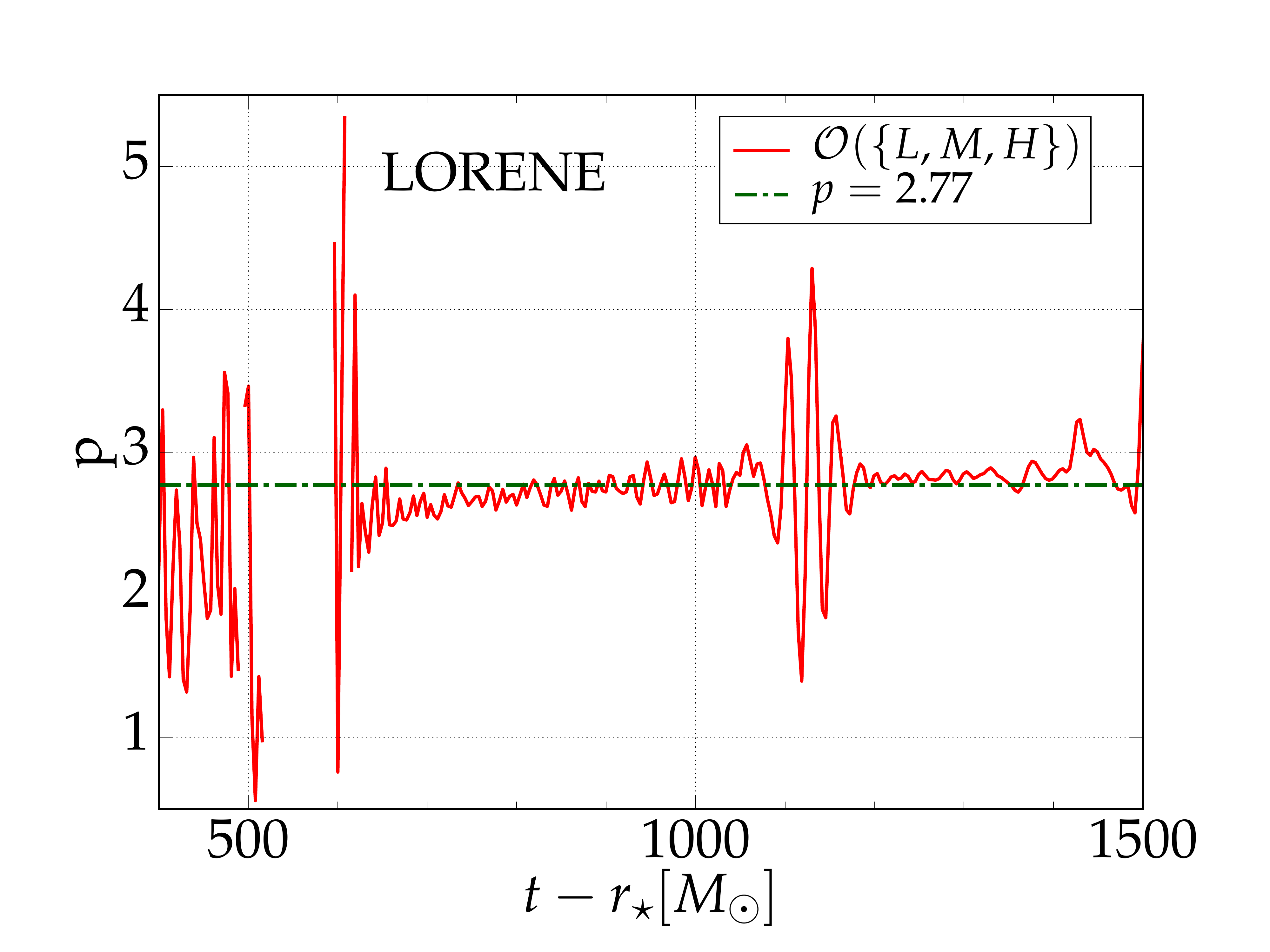}
\end{center}
\caption{Convergence order $p$ as a function of time as computed by
  Eq.~\ref{eq:convor} for \cocal{} (top panel) and \lorene{} (bottom
  panel) initial data. The average values for \cocal{} (\lorene{}), \ie
  $p=2.71\pm 0.27 (2.77\pm 0.24)$, are computed as arithmetic averages over
  the time interval $[650,1500]\,M_{\odot}$ where outlier data points,
  $p<1$ and $p>4$ are excluded from the average and represent the
  uncertainty range.}
\label{fig:conv_order}
\end{figure}

The left panel in the top row of Fig.~\ref{fig:psi4} refers to the
\cocal{} \texttt{Hs3.5d} initial data and we report the waveforms as
computed at the the three different resolutions L (red line), M (green
line) and H (blue line), which, we recall, are relative to spatial mesh
spacings of $0.2,\ 0.1333,\ 0.1\,M_\odot$ on the finest grid. Note that
at these resolutions the differences among the various waveforms are
extremely small, both in phase and in amplitude and one needs to zoom-in
in the figure to appreciate them. Similar waveforms are shown in the
right panel in the top row of Fig.~\ref{fig:psi4}, which instead refers
to the \lorene{} initial data. On each of these plots we also include a
dashed magenta line with the highest resolution run of the other initial
dataset in order to emphasize the dephasing that is instead observed when
comparing the two initial datasets.

\begin{figure}
\begin{center}
\includegraphics[width=\columnwidth]{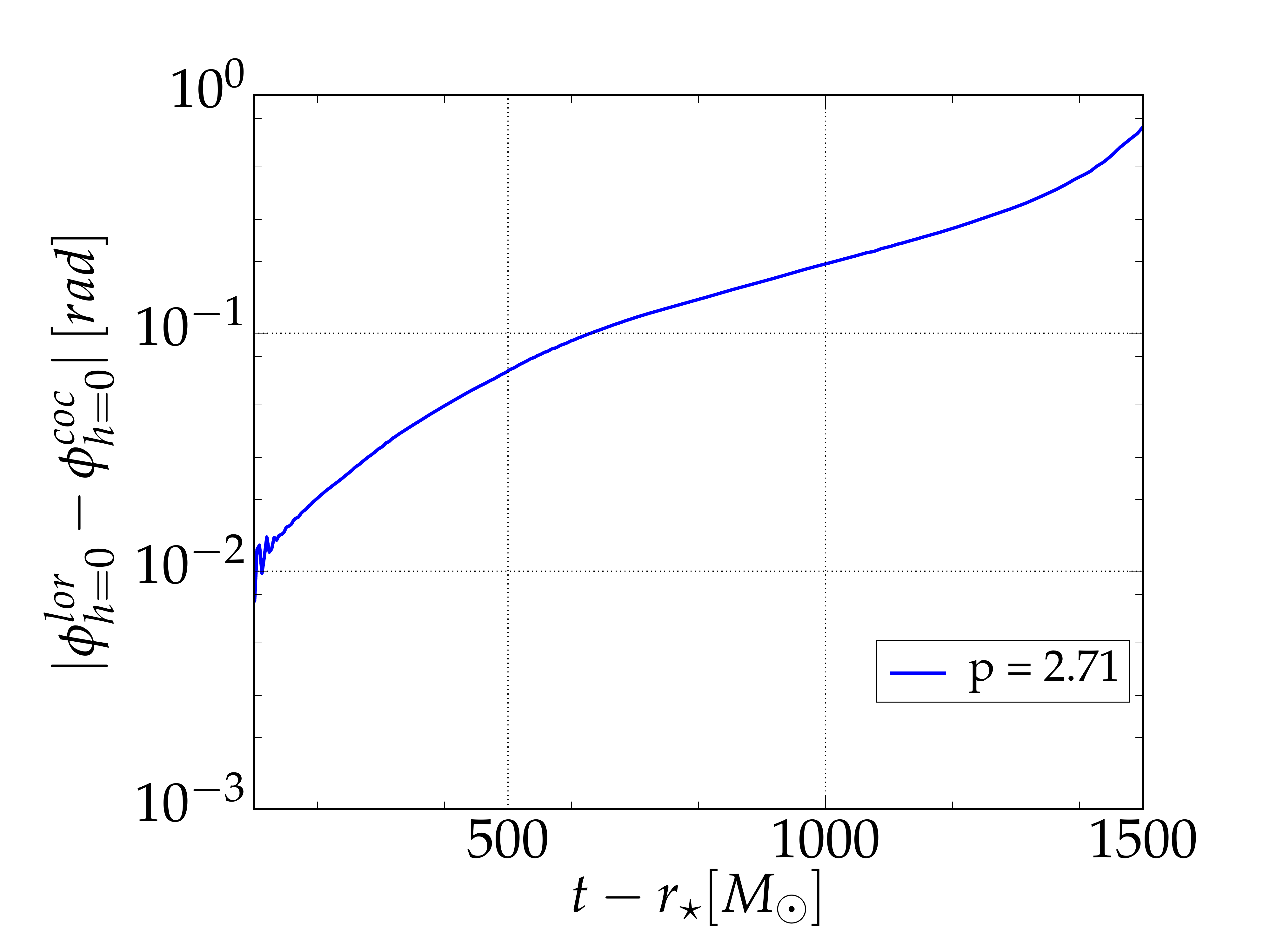}
\end{center}
\caption{Difference between the Richardson-extrapolated phases for
  \cocal{} and \lorene{} initial data using the three resolutions L, M,
  and H.}
\label{fig:dph}
\end{figure}

This dephasing observed in the top row of Fig.~\ref{fig:psi4} is
reminiscent of the behaviour observed in \cite{Radice2013c}, where a
comparison between two evolution codes of different convergence order,
\whisky{} \cite{Baiotti06, Baiotti08} and \whiskythc{}, has been made.
In that work, it was shown that given the exactly same initial data, a
second-order evolution code (\whisky{}) produces a significant phase
difference for the gravitational wave at different resolutions. This
phase difference was as large as $\sim 2$ radians between a low and a
high-resolution simulation. When the same experiment was repeated using
the higher-order \whiskythc{} code the dephasing between different
resolutions became as small as $\sim 0.6$ radians. Here, the evolution
runs have been done with \whiskythc{} only and the small differences in
phase are due uniquely to small differences in the initial datasets. In
other words, the evolution of the two slightly different initial datasets
resembles the dephasing measured when using evolution codes with
different orders of accuracy.
 
To gain a better understanding of the dephasing and to compare the
convergence properties for both sets of initial data, we report the
change $\Delta\GP$ between medium and low, as well as the high and medium
resolutions in the middle row of Fig.~\ref{fig:psi4}. The left plot
refers to the \cocal{} initial data, while the right plot to the
\lorene{} initial data. Also plotted is the rescaled $\Delta\GP$ for the
high-minus-medium resolution, and after employing a convergence order of
$p=2.71$ (see Fig.~\ref{fig:conv_order} and the discussion below). This
exponent $p$ is a genuine measure of the convergence order of our code
and we believe similar measurements should accompany any work reporting
high-quality gravitational waveforms. Here, $p$ has been computed by
solving the equation \cite{Rezzolla_book:2013, Radice2013b}
\begin{equation}
\frac{\GP_{h_1}-\GP_{h_2}}{\GP_{h_2}-\GP_{h_3}}\ =\ \frac{h_1^p-h_2^p}{h_2^p-h_3^p}\ ,
\label{eq:convor}
\end{equation}
where $(h_1,h_2,h_3)=(0.2,0.1333,0.1)$ are the intervals of the three
resolutions L, M, and H employed. Note that because $p$ is a function of
time (see Fig.~\ref{fig:conv_order}), the value reported refers to the
average over time of all convergence orders, after discarding an initial
noisy time interval. In this way, we obtain $p=2.71\pm 0.27$ for the
\cocal{} initial data and essentially the same value, \ie $p=2.77\pm
0.24$, for the \lorene{} initial data. A convergence order of this
magnitude is consistent with previous studies \cite{Radice2013c} of
binaries at close separations. At the last row of Fig.~\ref{fig:psi4}
(again left plot refers to \cocal{} while right plot to \lorene{} initial
data) we calculate the relative difference between the
Richardson-extrapolated phase for the three resolutions used. The value
at infinite resolution ($h=0$) is calculated from Eq. \eqref{eq:convor}
by setting, for example $h_1=0$, and solving for $\GP_{h_1}$, using the
previously calculated convergence order $p=2.71$, this is computed as
\begin{equation}
\GP_{h=0} = \GP_{h_2}+\frac{\GP_{h_2}-\GP_{h_3}}{({h_3}/{h_2})^p-1}\,.
\label{eq:richa}
\end{equation}
In all cases, although the overall behavior looks extremely similar the
significant dephasing can result to different observables.

In Fig. \ref{fig:dph} we plot the difference between the
Richardson-extrapolated ($h=0$) phases of the \cocal{} and \lorene{}
initial data using the L, M, H resolutions. As it is quite apparent, even
after approximately one orbit, the evolutions resulting from \cocal{} and
\lorene{} initial data differ by as much as $0.1$ radians and the
difference is approximately $0.5$ radians at merger time. Stated
differently, despite employing initial data referring to essentially the
same physical binary and computed by two highly accurate numerical codes
yielding global (local) differences that are $\lesssim 0.02\%\ (1\%)$,
the extrapolated gravitational-wave phases at the merger time can differ
by $\sim 0.5$ radians already after $\sim 3$ orbits. Considering that
these results have been obtained after using rather high spatial
resolutions, we believe that the use of a high-order numerical code such
as \whiskythc{} has been crucial in bringing out these differences.

\section{Conclusions}
\label{sec:conclusions}

We have presented the first evolutions of our newly constructed
initial-data code \cocal{} \cite{Tsokaros2015}, and performed an accurate
study on the role that slightly different initial data play on the
evolution of neutron-star binaries. The \coctocac{} driver, that enables
communication with existing evolution codes in \cactus{} toolkit, was
presented and a detailed converge analysis both with respect to the
initial data itself, as well as with respect to the \whiskythc{}
evolution code was performed for the case of irrotational neutron-star 
binaries separated at $45\,{\rm km}$. In addition, for benchmark purposes
regarding future spinning simulations, we have also examined a corotating
solution at $45\,{\rm km}$.

Our main goals in this work have been, on the one hand, to validate the
accuracy of the initial data constructed by this new initial data code
and, on the other hand, to estimate potential differences on the
gravitational-wave signal as it is produced by different initial data
codes. For this purpose, we have used the widely used, open-source code
\lorene{} and have carried out a close comparison for the initial data
computed with the codes when considering the same physical binary. For
the first time, we have also explored the impact that the minute
differences in the two initial-datasets have on the extrapolated
gravitational-wave signal. 

In this way, we have found that although the initial data between the the
two initial-data codes have global (local) differences that are $\lesssim
0.02\%\ (1\%)$, the extrapolated gravitational-wave signal at the merger
time and after about three orbits can have a dephasing of half a
radian. This is an alarming reminder of the care that needs to be paid
when comparisons are performed between results that start from slightly
different initial data or when the initial data errors are not properly
taken into account in the simulation error budget.

\acknowledgments 

We thank David Radice for help with \whiskythc{} and for the analysis of
the gravitational waves. A.\, T. is supported by Vibetech
Consultants. This work was supported by JSPS Grant-in-Aid for Scientific
Research(C) 15K05085 and 25400262, by ``NewCompStar'', COST Action
MP1304, from the LOEWE-Program in HIC for FAIR, and the European Union's
Horizon 2020 Research and Innovation Programme under grant agreement
No. 671698 (call FETHPC-1-2014, project ExaHyPE). The simulations were
performed on SuperMUC at LRZ-Munich and on LOEWE at CSC-Frankfurt.

\appendix
\section{Pointwise comparison of corotating solutions} 
\label{sec:comp_corot}

\begin{figure}
\begin{center}
\includegraphics[width=0.48\columnwidth]{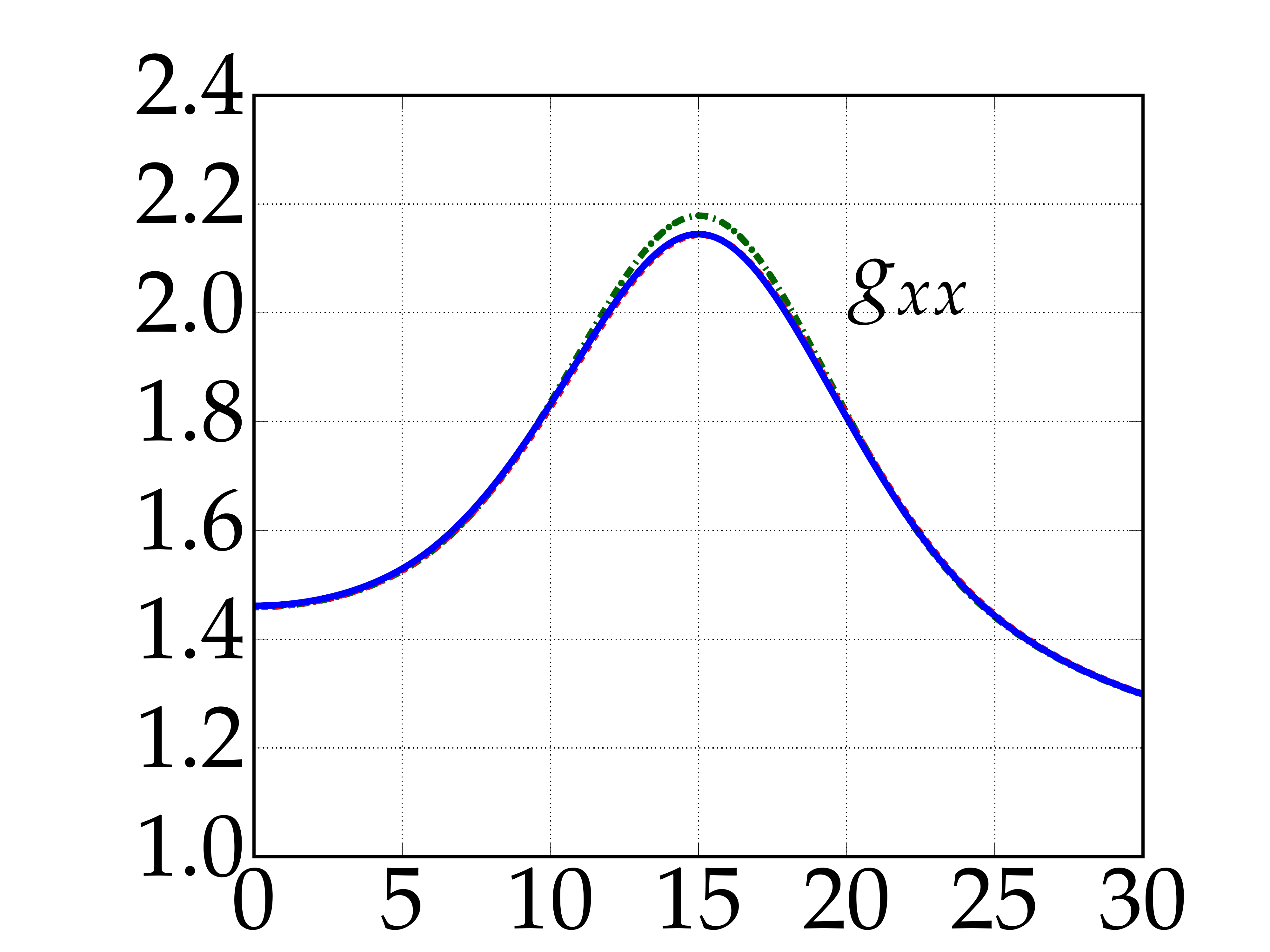}
\includegraphics[width=0.48\columnwidth]{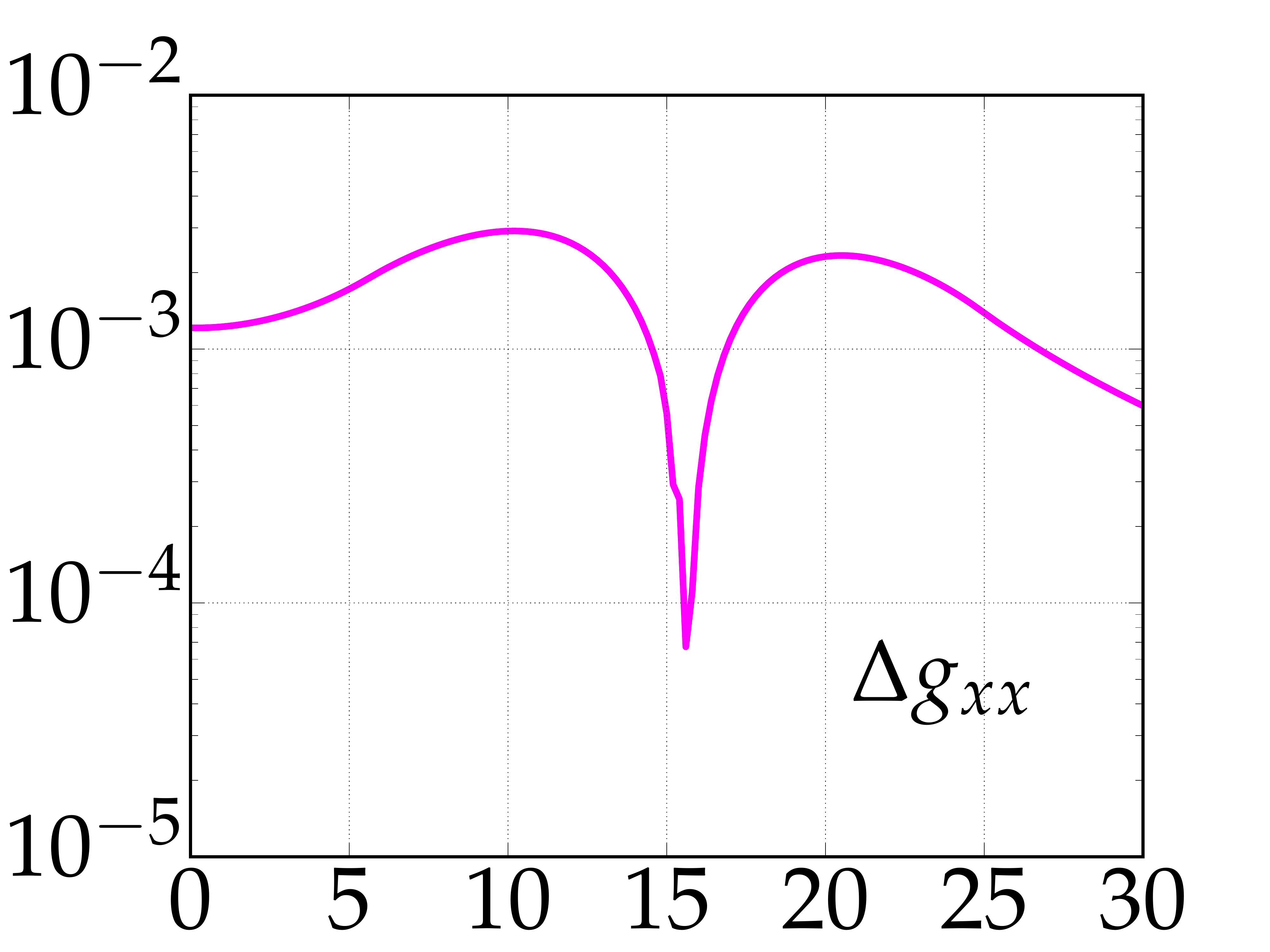}
\end{center}
\begin{center}
\includegraphics[width=0.48\columnwidth]{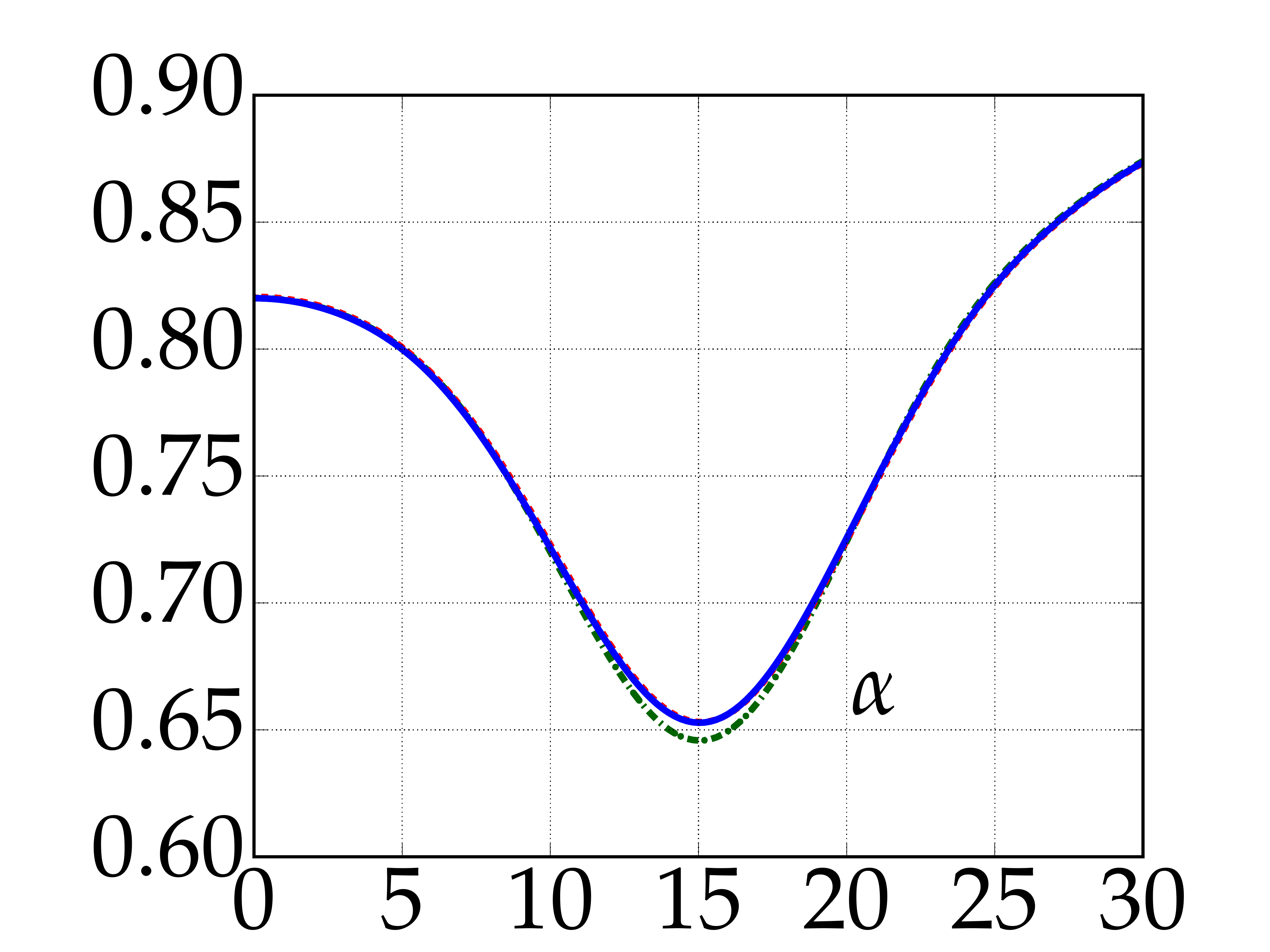}
\includegraphics[width=0.48\columnwidth]{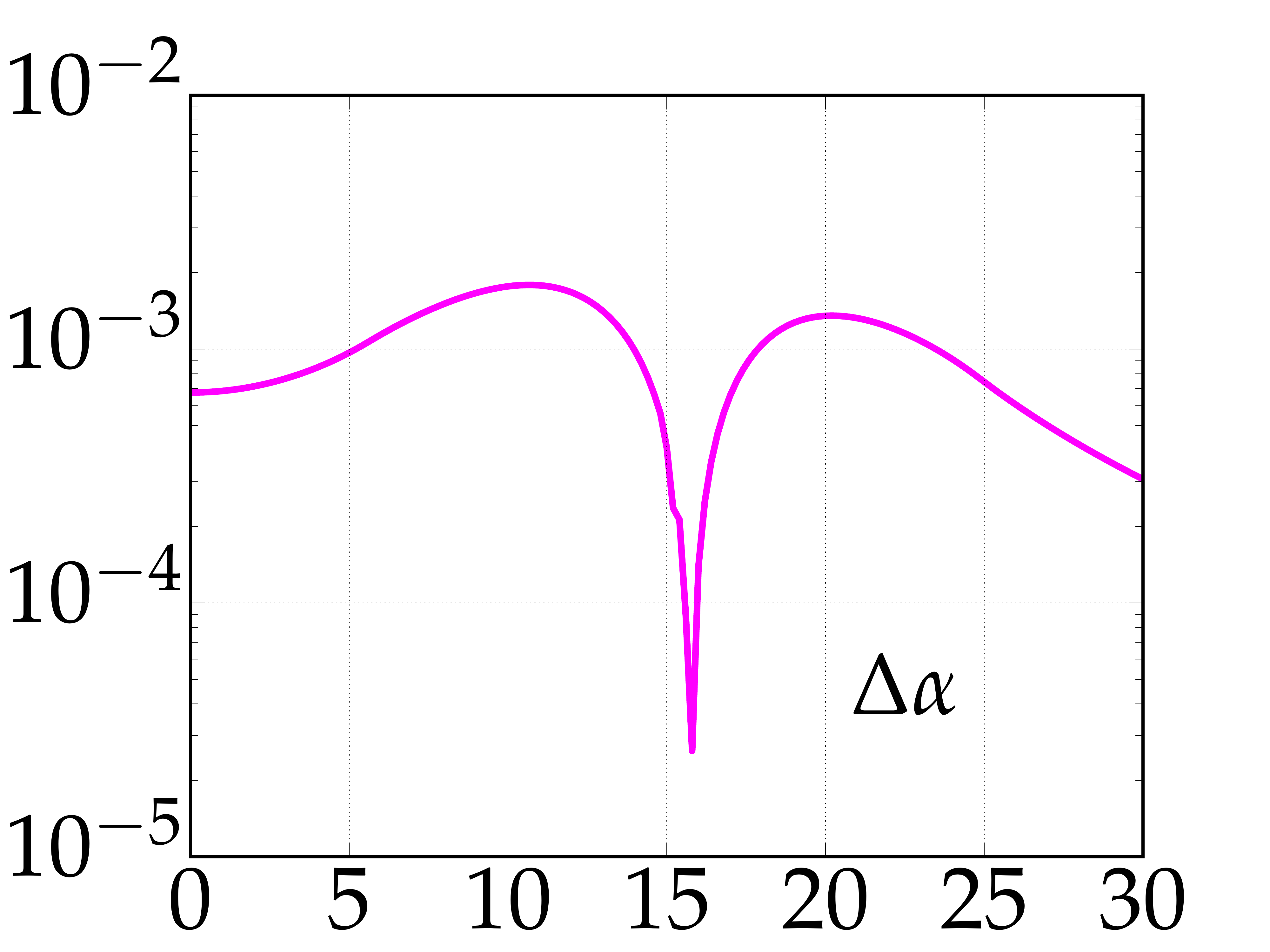}
\end{center}
\begin{center}
\includegraphics[width=0.48\columnwidth]{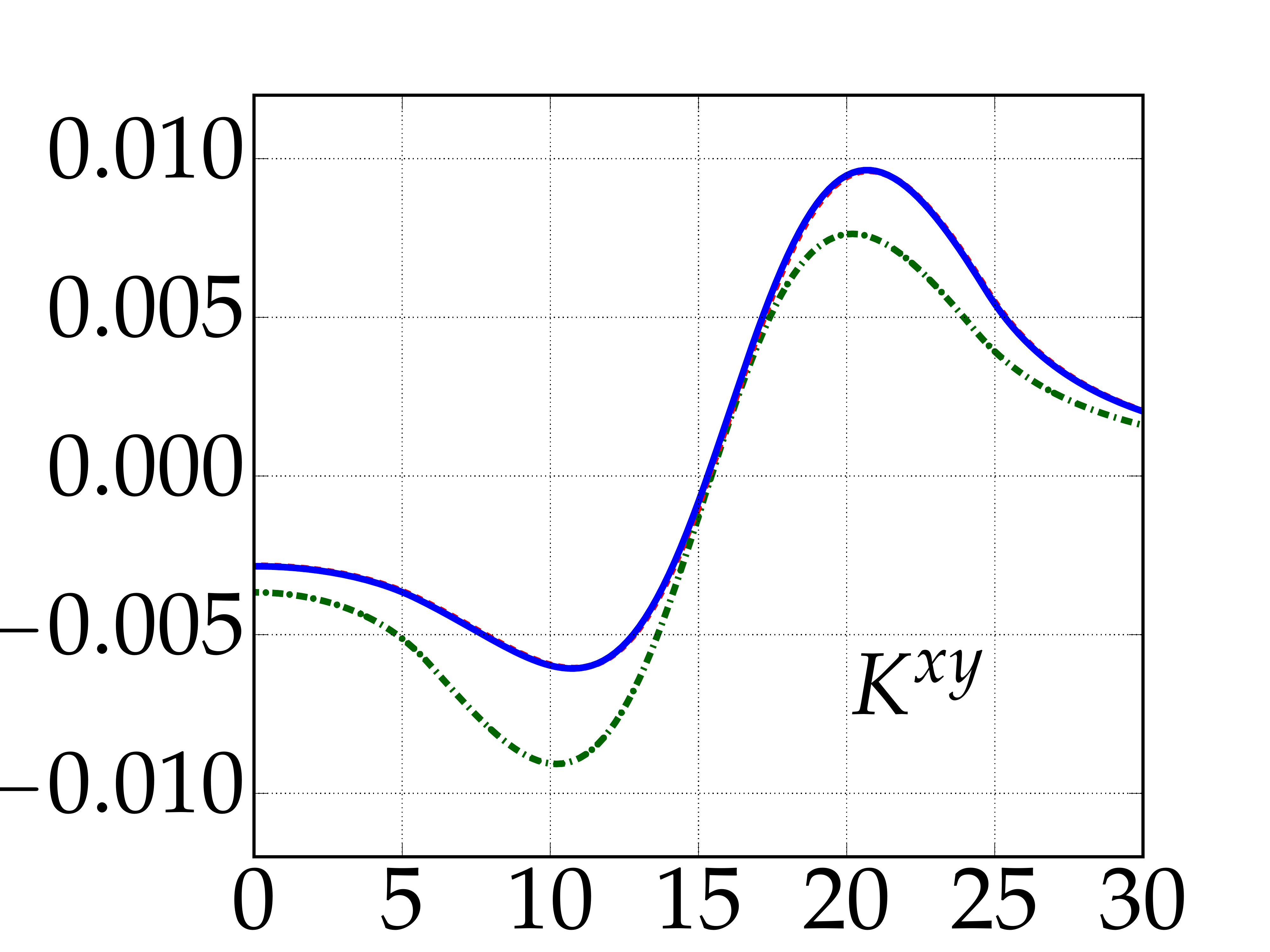}
\includegraphics[width=0.48\columnwidth]{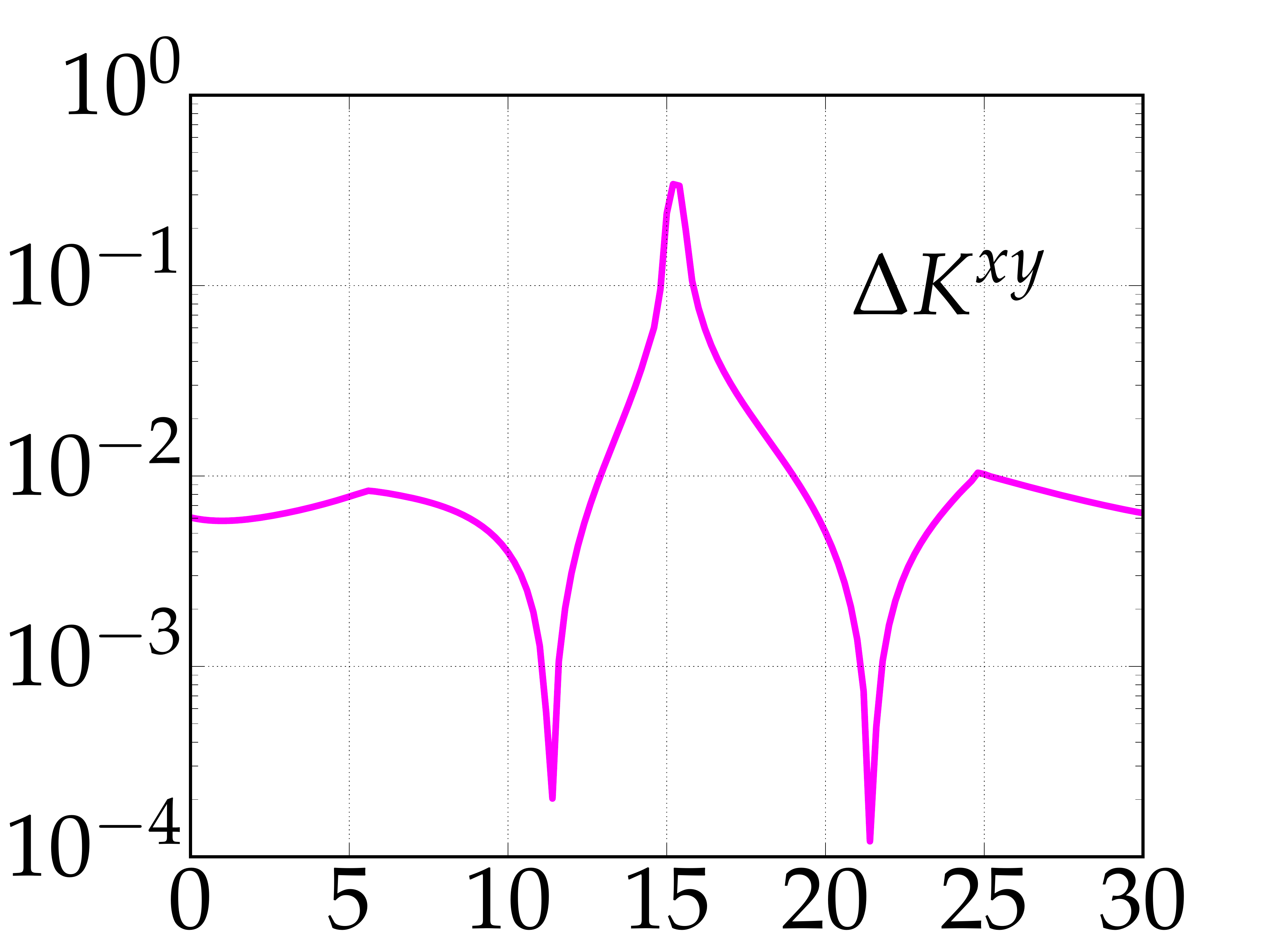}
\end{center}
\begin{center}
\includegraphics[width=0.48\columnwidth]{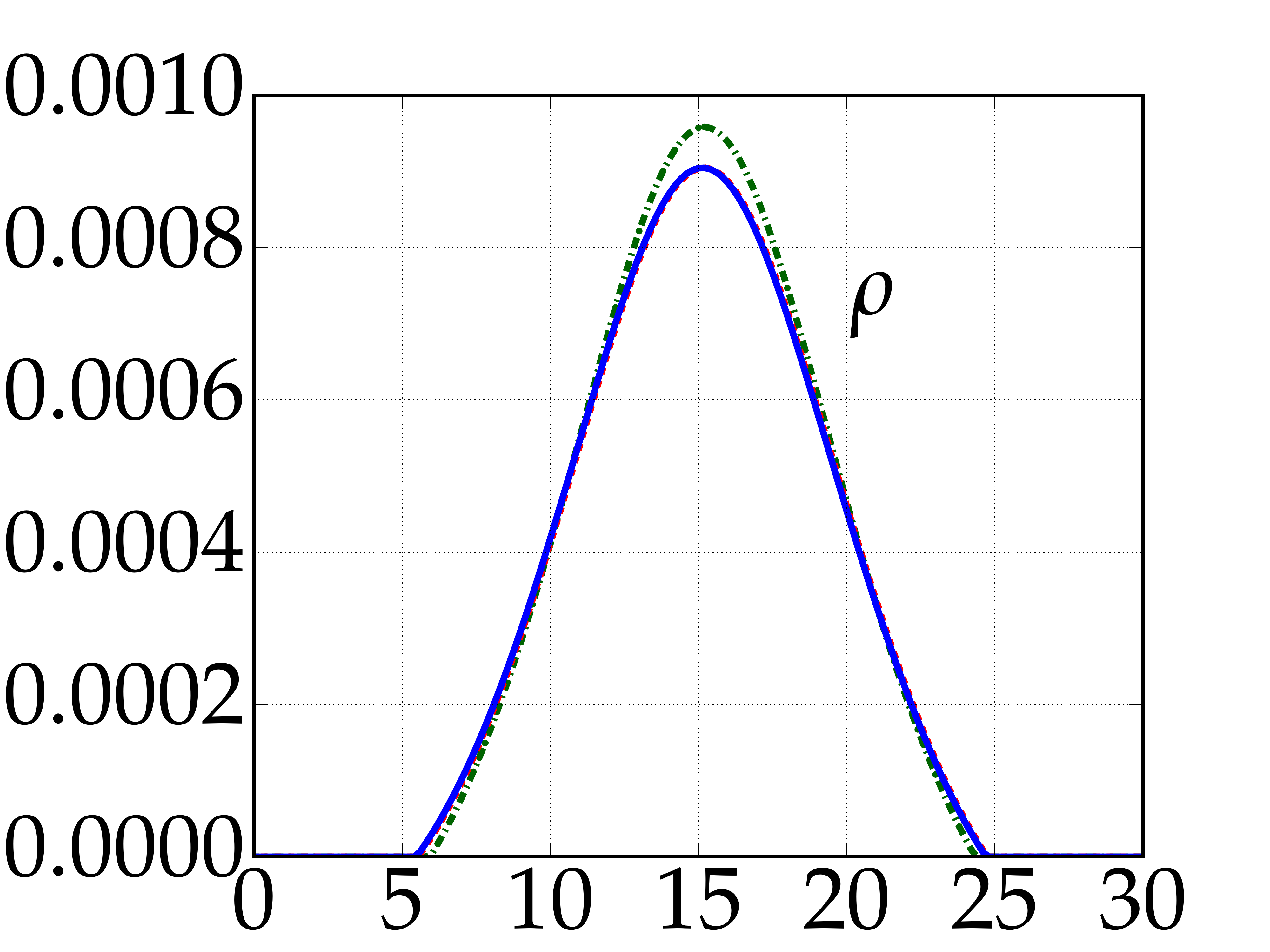}
\includegraphics[width=0.48\columnwidth]{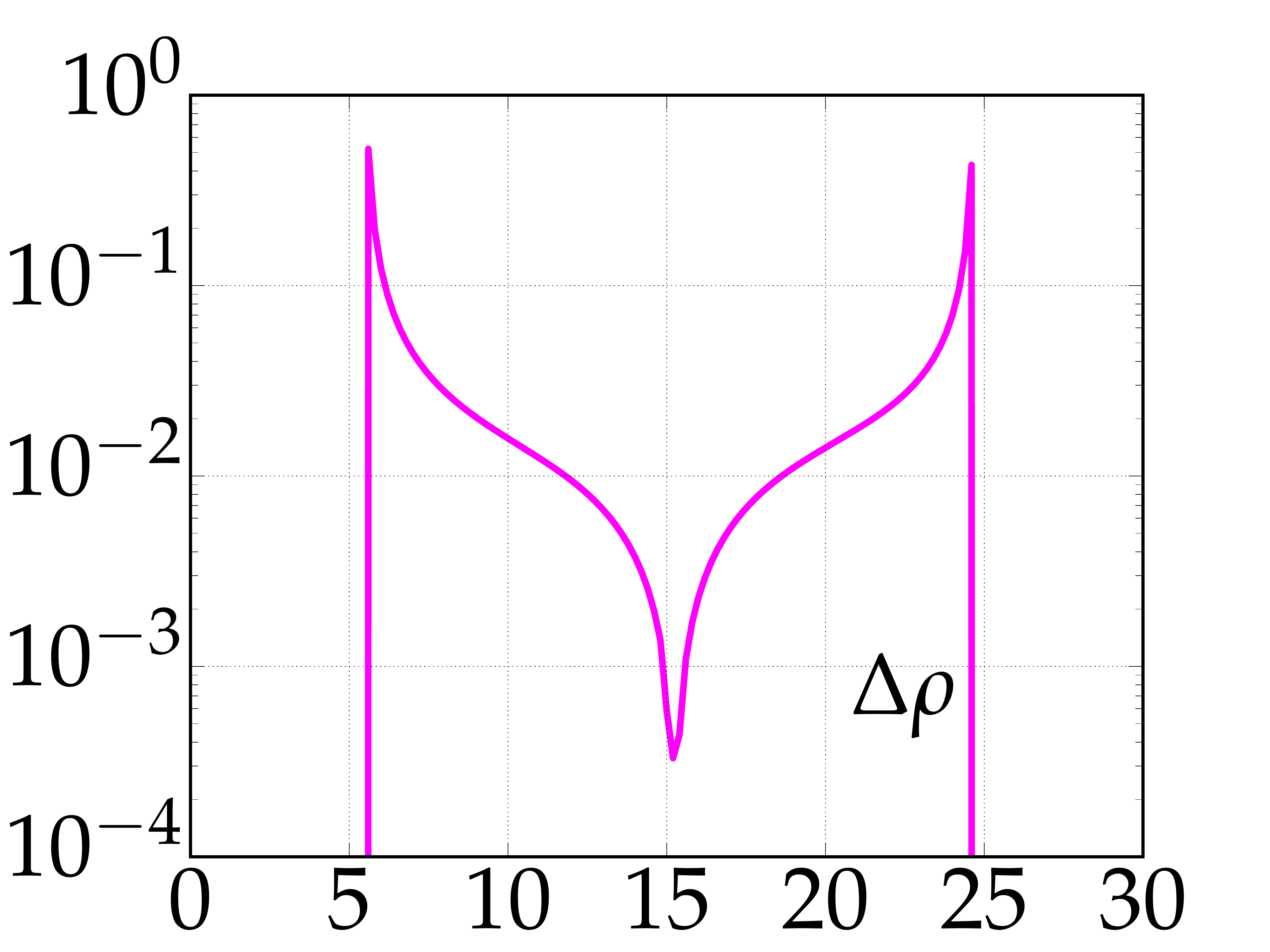}
\end{center}
\begin{center}
\includegraphics[width=0.48\columnwidth]{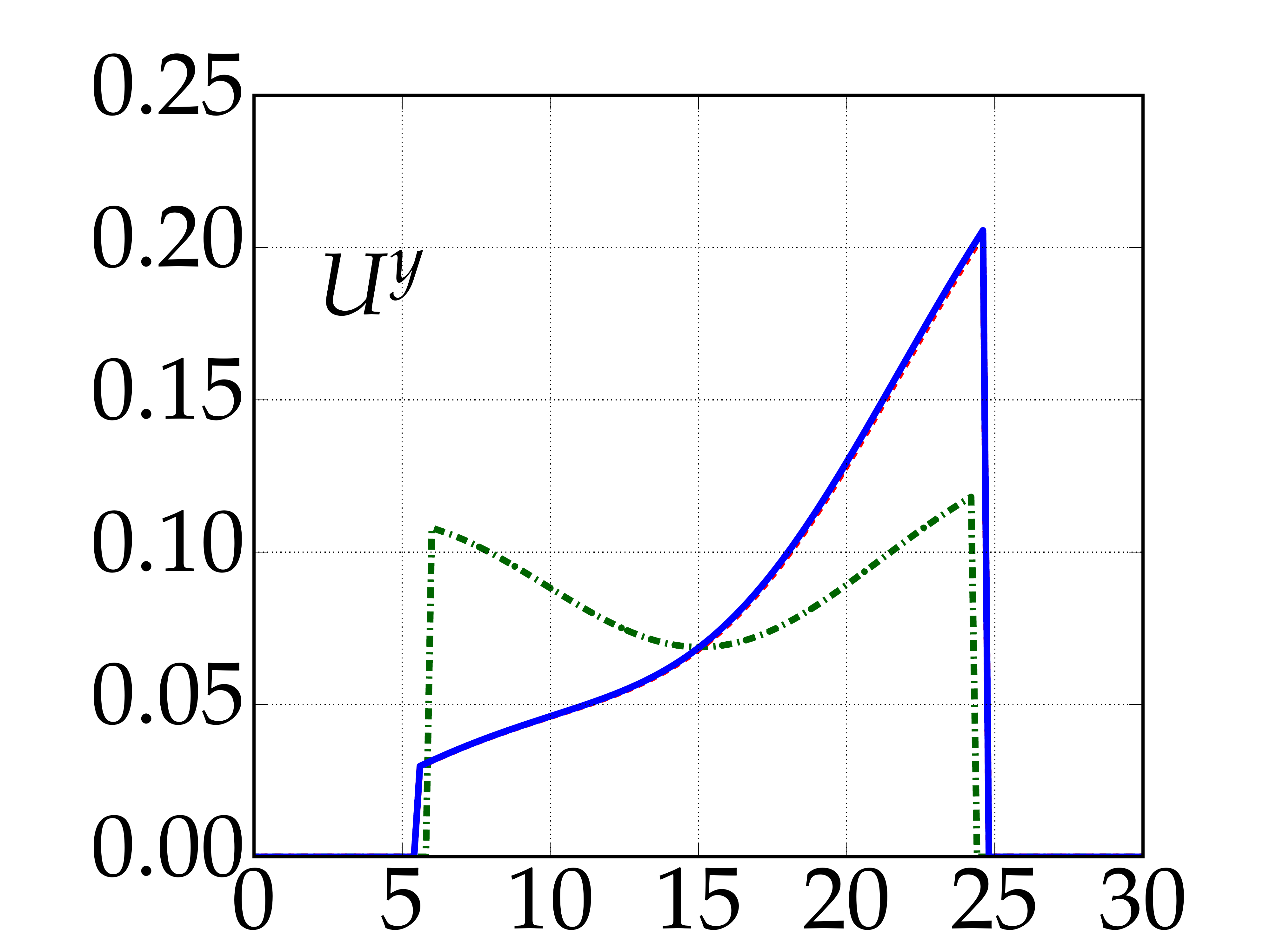}
\includegraphics[width=0.48\columnwidth]{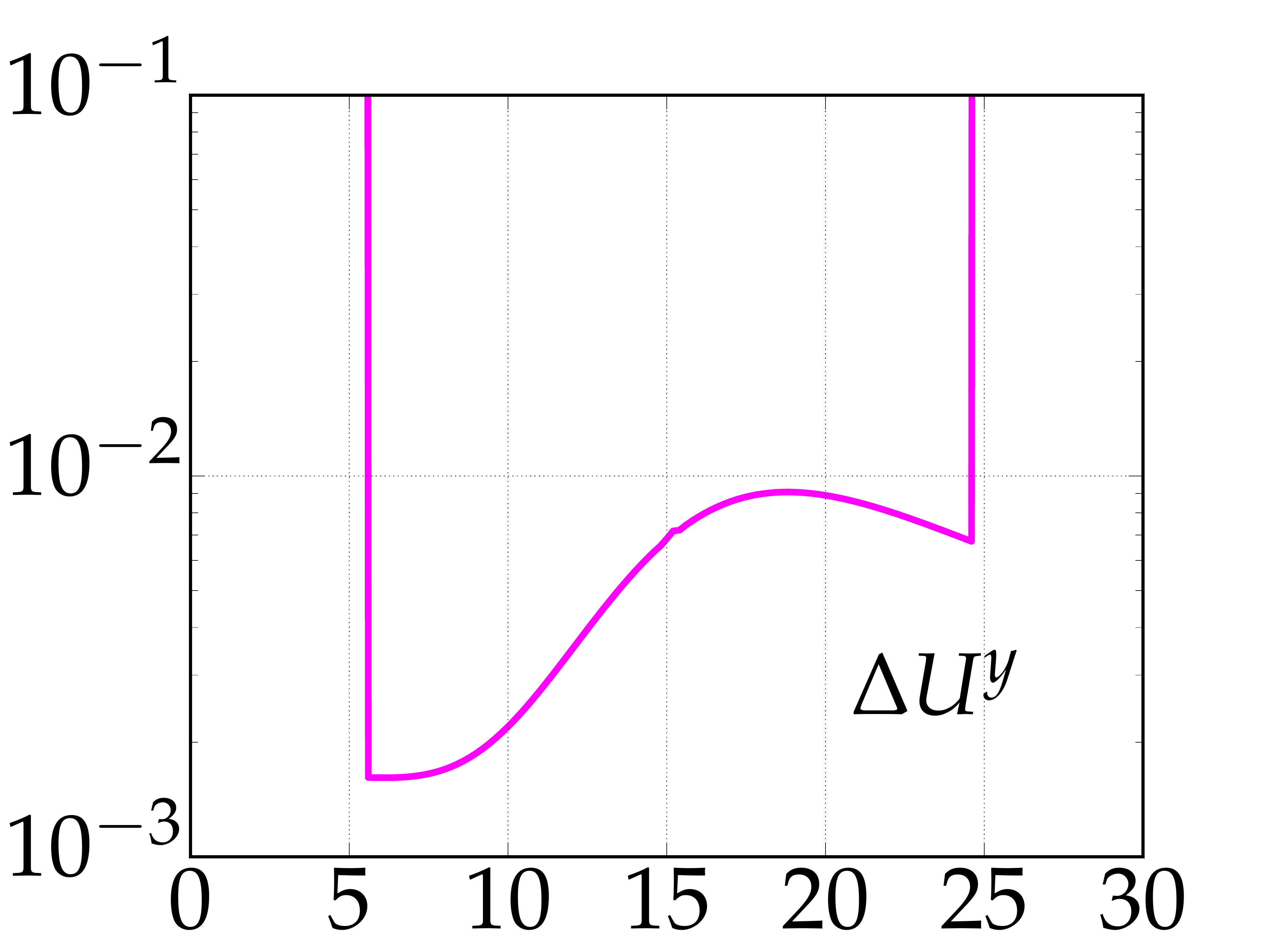}
\end{center}
\begin{center}
\includegraphics[width=0.48\columnwidth]{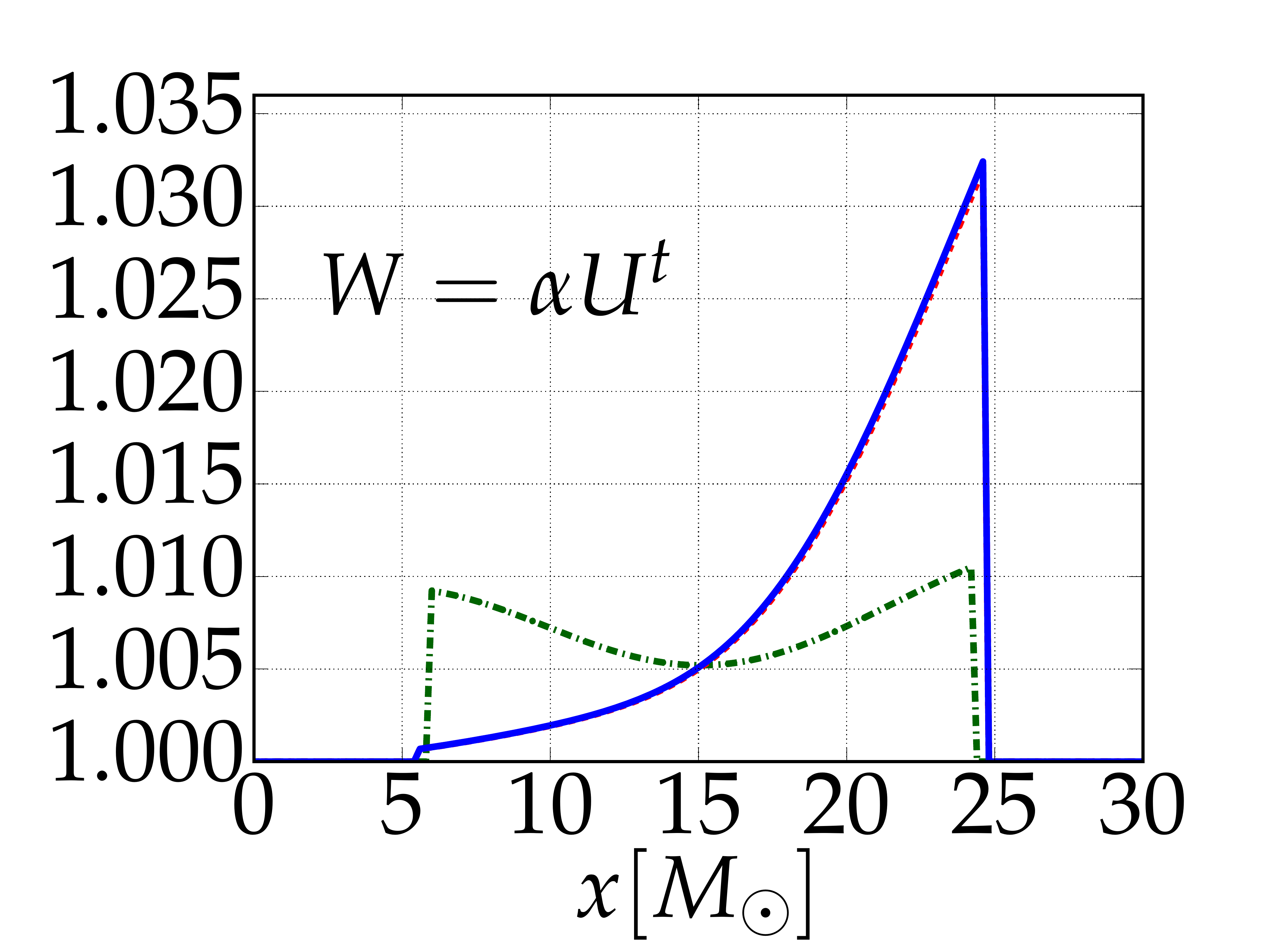}
\includegraphics[width=0.48\columnwidth]{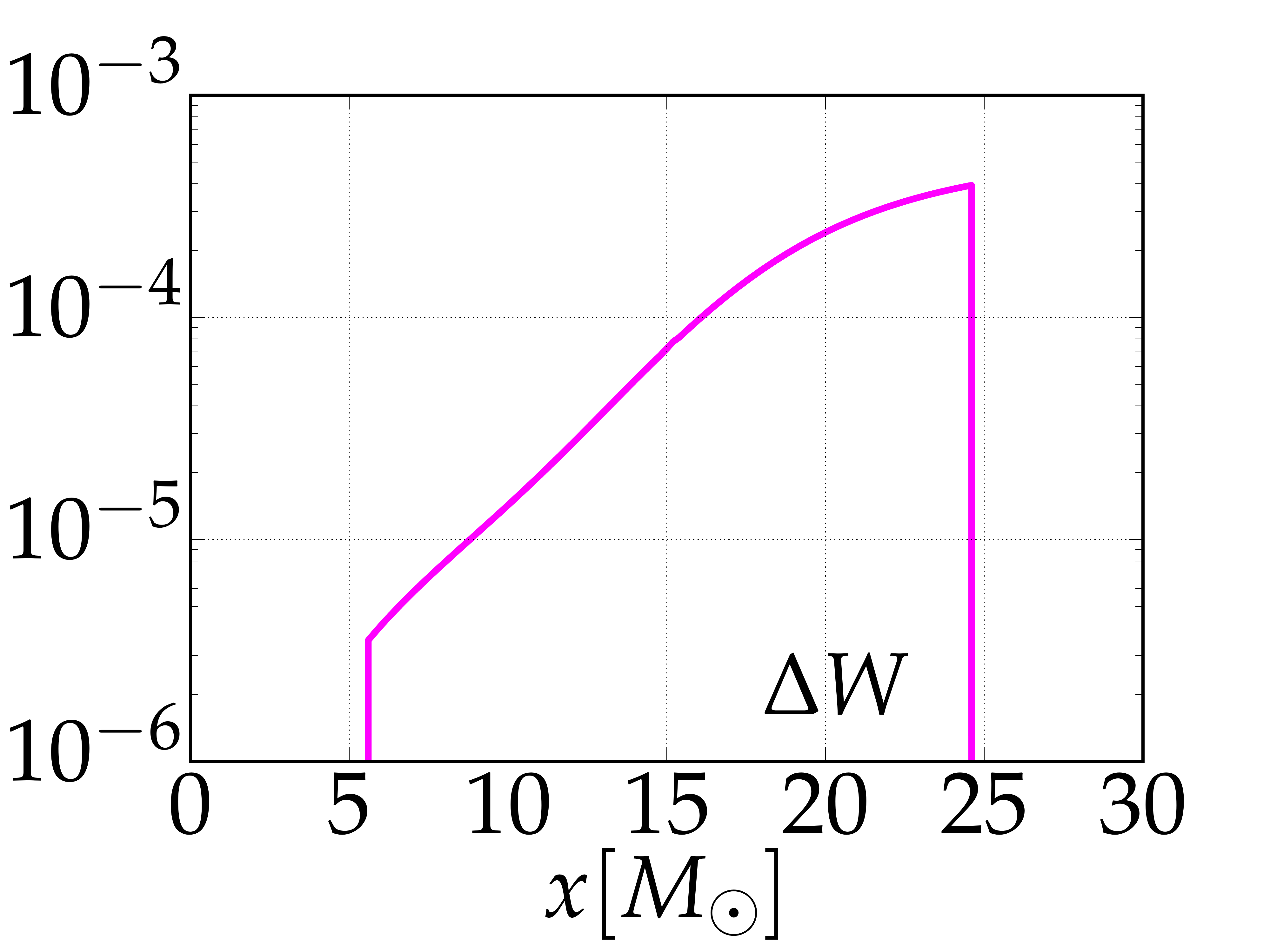}
\end{center}
\caption{The same as in Fig. \ref{fig:irrot_var1} but a corotating
  binary. The dashed green line refers to the irrotational solution in
  Fig. \ref{fig:irrot_var1}, and which has a very similar mass (\cf
  Tables \ref{tab:loco45} and \ref{tab:loco45corot}).}
\label{fig:corot_var1}
\end{figure}

Although corotating solutions are not considered as physically realistic
because the shear viscosity in neutron stars is too small to guarantee
that this tidal coupling takes place \cite{Kochanek92,Bildsten92}, in
this appendix we calculate a corotating neutron-star binary at $45\,{\rm
  km}$ and compare our solutions pointwise with a solution calculated
from \lorene{}. The reason is that corotating binaries are easier to
calculate, since the fluid rotates at the same angular velocity as the
binary, and hence they can be considered as a benchmark for error
estimation in binary calculations. Also, since they represent the
simplest spinning-binary configuration, they provide insight for the
magnitude of the error introduced by more complicated arbitrary spinning
solutions.

\begin{figure}
\begin{center}
\includegraphics[width=\columnwidth]{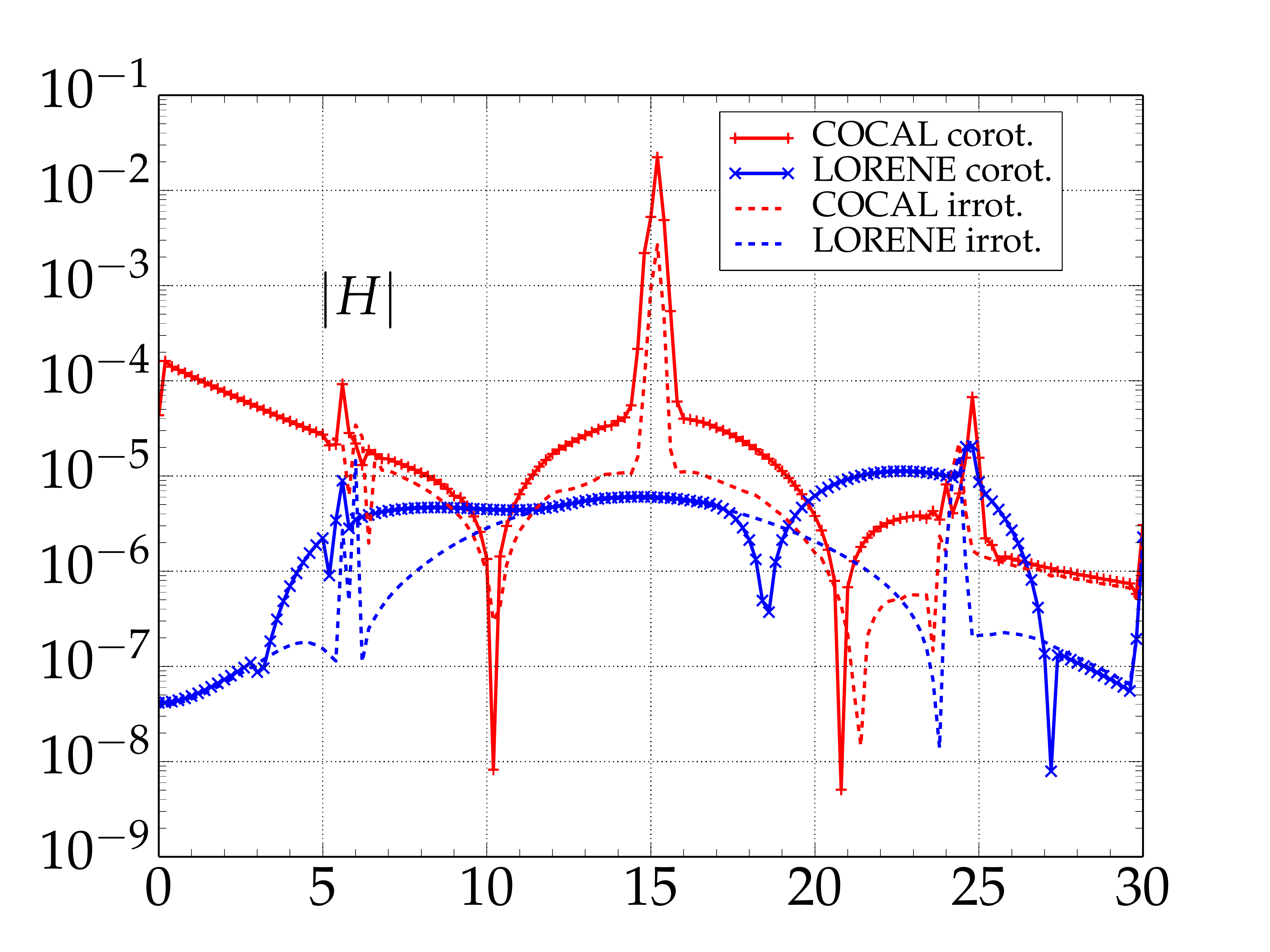}
\includegraphics[width=\columnwidth]{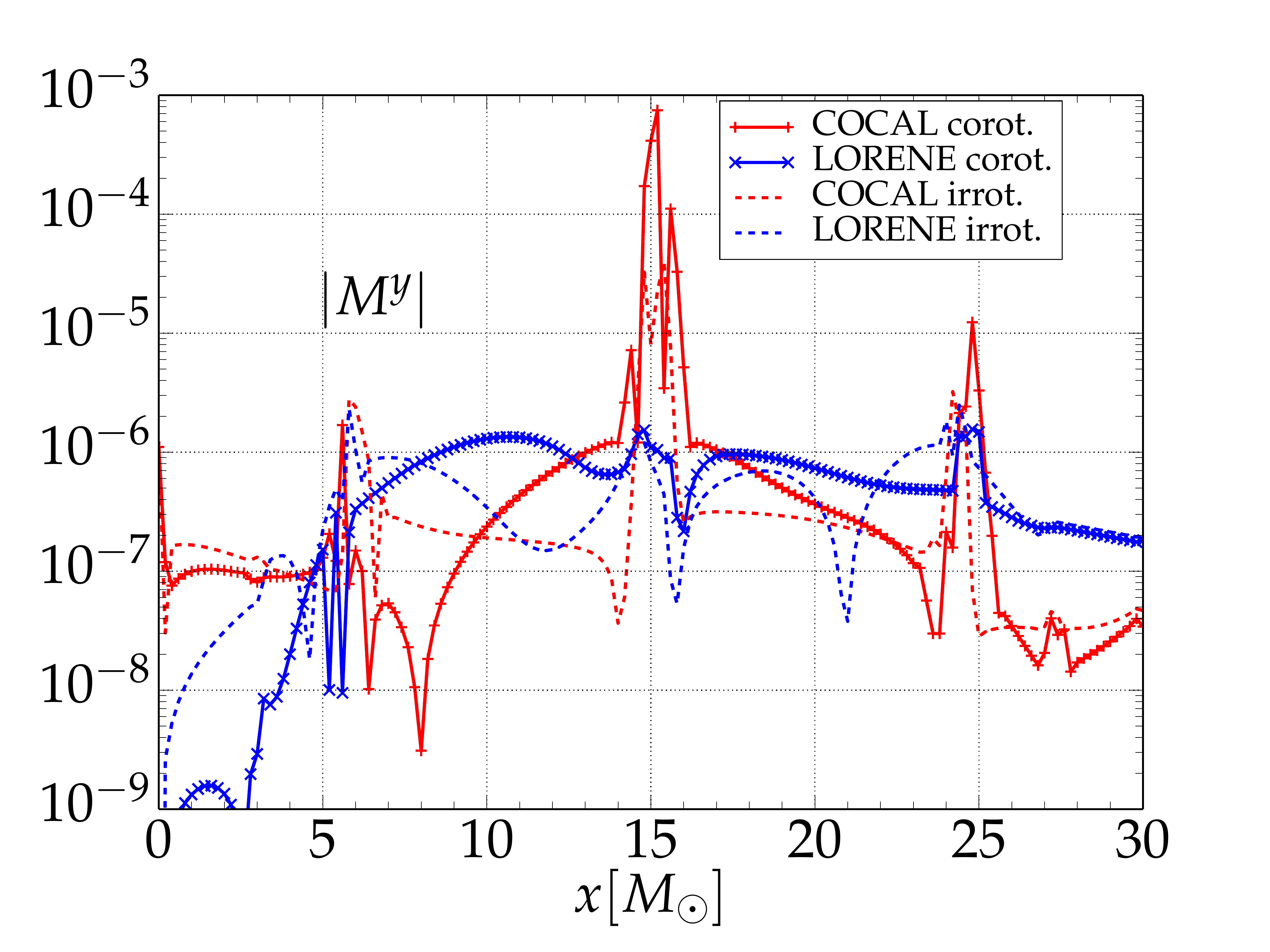}
\caption{Hamiltonian (top) and momentum violations (bottom) for the
  $y$-component of the shift ($\GB^y$) along the $x$-axis for corotating
  (solid lines) neutron-star binaries. Dashed lines are the corresponding
  irrotational \cocal{} and \lorene{} violations as appear in
  Fig.~\ref{fig:xaxis_cv}. The origin $x=0$ corresponds to the center of
  mass of the binary with the surface of the star to be located at
  $x\approx 6\,M_{\odot}$, and at $x\approx 24\,M_{\odot}$. Grid parameters
  used in \cocal{} are those of $\texttt{Hs3.0d}$.}
\label{fig:corot_xaxis_cv}
\end{center}
\end{figure}

To enforce corotation, we set $V^\GA=0$ and the Eulerian velocity is then
given by 
\begin{equation}
 U^i =\frac{\GO^i}{\GA} \,.
\label{eq:corot_vy}
\end{equation}
We only consider the \texttt{Hs3.0d} resolution and the main physical
quantities for both \cocal{} and \lorene{} are reported in
Table~\ref{tab:loco45corot}. Note that the central rest-mass density is
smaller than for the irrotational binary, while the ADM mass and
angular momentum being slightly larger. This is simply due to the stellar
rotation, that tends to stabilize the binary by including rotational
kinetic energy.
 
In Fig.~\ref{fig:corot_var1} we plot along the positive $x$-axis the
conformal factor, the lapse function $\alpha$, the $xy$-component of the
extrinsic curvature, the rest-mass density $\rho$, the $y$-component of
the velocity and the Lorentz factor for both \cocal{} (red lines) and
\lorene{} (blue lines) solutions. As in Fig.~\ref{fig:irrot_var1}, $x=0$,
corresponds to the center of mass of the system. Also plotted with a
dashed green line is the corresponding irrotational solution as reported
in Fig. \ref{fig:irrot_var1}. A rapid inspection shows that the conformal
factor and the lapse are slightly smaller inside the star, while the
extrinsic curvature increases (decreases) towards the outer (inner) part
of the star. Also the velocity
profile has much larger values in the outer parts of the star (\ie those
farther away from the center of mass) and this is the an obvious
manifestation of the large spin component introduced by the corotation
and that is reflected in the Lorentz factor too. Overall, and as for the
irrotational case, also here the differences between the two datasets are
$\lesssim 1\%$.

In Fig.~\ref{fig:corot_xaxis_cv} we plot the constraint violations as we
have done in Fig.~\ref{fig:xaxis_cv} for the irrotational binaries. Only
one resolution for \cocal{}, the \texttt{Hs3.0d}, is plotted, together
with the corresponding violations from the irrotational solutions (shown
with dashed lines; \cf Fig.~\ref{fig:xaxis_cv}), that are shown for
comparison. 

The comparison with the results from \lorene{} shows a very similar
behaviour to the one already discussed for the irrotational case: the
Hamiltonian violations are larger but the violations of the momentum
constraint smaller. Comparing instead the \cocal{} irrotational data with
the the corotating cases, we see that the violations are larger in the
corotating binary. Hence, although the fluid formulation is significantly
more complicated in the case of irrotational binaries, the large rotation
present in corotating binaries induces a small amount of extra violations
for both finite-difference and spectral-method codes. We expect that a
similar behavior will be shown also by neutron-star binaries with
arbitrary spins.


\begin{table}
\begin{tabular}{l|rr}
\hline
\hline
  & \lorene{} & \cocal{}  \\
\hline
\hline
$M_0$                      & $1.62504$ & $1.62505$  \\
$M_{\rm ADM}$                & $3.00274$ & $3.00275$  \\
$\MK         $             & $-$       & $3.00243$  \\
$\GR_c\ (\times 10^{-4})$   & $9.04601$ & $9.04969$  \\
$J_{\rm ADM}$                & $9.76287$ & $9.75909$  \\
$\Omega\  [{\rm rad/sec}]$ & $1857.82$ & $1848.84$   \\
$d_s\ [{\rm km}]$          & $44.731$  & $44.736$   \\
$R_{\rm eq}\ [{\rm km}]$     & $14.193$  & $14.181$  \\
\hline
\end{tabular}
\caption{Physical parameters for a corotating binary computed with either
  \cocal{} or \lorene{} (see Table \ref{tab:loco45} for a description of
  the various quantities). The resolution used for \cocal{} is
  $\texttt{Hs3.0d}$ of Table \ref{tab:grid_param}, except for parameter
  $r_s=0.7925$ in order to create a binary at separation $44.7\,{\rm
    km}$.}
\label{tab:loco45corot}
\end{table}

\bibliographystyle{apsrev4-1}
\bibliography{aeireferences}



\end{document}

%% file: kigou.tex
\newcommand{\Madm}{M_{\rm ADM}}
\newcommand{\MK}{M_{\rm K}}

\newcommand{\beq}{\begin{equation}} 
\newcommand{\eeq}{\end{equation}} 
\newcommand{\beqn}{\begin{eqnarray}} 
\newcommand{\eeqn}{\end{eqnarray}}

\newcommand{\Kab}{K^a{}\!_b}

\newcommand{\zD}{{\raise1.0ex\hbox{${}^{\ \circ}$}}\!\!\!\!\!D}
\newcommand{\alone}{{\raise0.5ex\hbox{${}^{\ 1}$}}\!\!\!\!\alpha}

\newcommand{\Dl}{\Delta}

\newcommand{\nalam}{\mathrel{\raise0.9ex\hbox{$^\lambda$}\mkern-14mu
\lower0.0ex\hbox{$\nabla$}}}

\newcommand{\Nrf}{{N_r^{\rm f}}}
\newcommand{\Nrm}{{N_r^{\rm m}}}

\newcommand{\zeroD}{{\raise1.0ex\hbox{${}^{\ \circ}$}}\!\!\!\!\!D}

\newcommand{\zLap}{{\raise1.0ex\hbox{${}^{\ \circ}$}}\!\!\!\!\Delta}
\newcommand{\zna}{{\raise1.0ex\hbox{${}^{\ \circ}$}}\!\!\!\!\!\nabla}
\newcommand{\zS}{{\raise1.0ex\hbox{${}^{\ \circ}$}}\!\!\!\!\!S}

\newcommand{\carpet}{\textsc{carpet}}
\newcommand{\cocal}{\textsc{cocal}}
\newcommand{\coctocac}{\textsc{coc$2$cac}}
\newcommand{\lorene}{\textsc{lorene}}
\newcommand{\kadath}{\textsc{kadath}}
\newcommand{\scrid}{\textsc{scrid}}
\newcommand{\cactus}{\textsc{cactus}}
\newcommand{\bam}{\textsc{bam}}
\newcommand{\spec}{\textsc{spec}}
\newcommand{\etoolkit}{\textsc{Einstein Toolkit}}
\newcommand{\whisky}{\textsc{Whisky}}
\newcommand{\whiskythc}{\textsc{WhiskyTHC}}
\newcommand{\mclachlan}{\textsc{McLachlan}}